\documentclass[12pt]{article}
\usepackage{epsfig}

\newcommand{\postscript}[2]
 {\setlength{\epsfxsize}{#2\vsize}
\setlength{\epsfysize}{#2\vsize}
  \centerline{\epsfbox{#1}}}

\newcommand{\be}{\begin{equation}}
\newcommand{\ee}{\end{equation}}
\newcommand{\bea}{\begin{eqnarray}}
\newcommand{\eea}{\end{eqnarray}}

\newcommand{\Sigb}{{\overline\Sigma}}
\newcommand{\oot}{\overline {126}}
\newcommand{\boot}{\bf\oot}
\newcommand{\nnu}{\nonumber\\}
\def\blfootnote{\xdef\@thefnmark{}\@footnotetext}
\newcommand{\rbtau}{ R_{b\tau/s\mu}}

\begin{document}
%%%%%%%%%%%%%%%%%%%%%%%%%%%%
%\begin{titlepage}
%%%%%%%%%%%%%%%%%%%%%%%%%%%%%
\vspace{4\baselineskip}
%%%%%%%%%%%%%%%%%%% TITLE %%%%%%%%%%%%%%%%%%
\vspace{4cm}
\begin{center}{\Large\bf NMSGUT-III: Grand Unification upended
   }

\end{center}
%%%%%%%%%%%%%%%% AUTHORS %%%%%%%%%%%%%%%%%%%%%%%
\vspace{2cm}
\begin{center}
{\large
 Charanjit  S.  Aulakh\footnote{E-Mail:  aulakh@pu.ac.in   }
     }
\end{center}
%%%%%%%%%%%%%%%%%%%%%%% AFFILIATION %%%%%%%%%%%%
\vspace{0.2cm}
\begin{center}
%\small
  {\it
Dept. of Physics, Panjab University,\\ Chandigarh, India.}
\end{center}

\centerline{\Large{\bf{Abstract}}}

\vspace{0.2cm} We show that matter  yukawa couplings of the New
Minimal Supersymmetric (SO(10)) GUT(NMSGUT) are  subject to  very
significant GUT scale threshold corrections.  Including these
threshold effects relaxes the constraint $ y_b-y_\tau\simeq
y_s-y_\mu$ operative in $\textbf{10} -\textbf{120} $
 plet generated tree level MSSM matter fermion yukawas $y_f$. We
find accurate fits of the MSSM fermion mass-mixing data in terms
of NMSGUT superpotential couplings and 5 independent   soft Susy
breaking parameters $M_0, M_{1/2}, A_0, M^2_{H,\bar{H}}$ at $M_X$.
The fits generally have elevated unification scale $M_X$ near
$M_{Planck}$, viable values of $\alpha_3(M_Z)$,
 and are  consistent with current limits on B violation,
$b\rightarrow s\gamma$, muon magnetic moment anomaly and Standard
Model $\rho$ parameter. The associated novel and distinctive
  soft Susy spectra have light
gauginos,  a \emph{normal} s-hierarchy and  Bino LSP. The Bino LSP
is accompanied by second and first
 generation right chiral sfermions light enough to mediate a consistent WIMP dark matter co-annihilation
cosmology and to be discoverable at LHC, while third  generation sfermions
 are in the LHC undiscoverable range of 3-50 TeV. The fits found require  $|\mu|,|A_0|\sim 100$ TeV   which imply both
  deep CCB/UFB minimae and stability of the MSSM standard vacuum  on cosmological time scales.
    Our results   indicate that  a consistent realistic phenomenology may
be specifiable in terms of SO(10) (NMS)GUT parameters at $M_X$
alone and  that a new viable sector of the soft supersymmetry
parameter space may exist if flavour violation constraints can be
satisfied in the 43 dimensional parameter space.
%\pacs{PACS: 12.10-g, 12.15.Ff, 14.60.Pq}

\vskip1pc

\section{  Introduction}. \hspace{0.5cm}

Supersymmetric Grand Unification based on the SO(10) gauge group
has received well deserved attention over the last 3 decades. Models proposed
 fall into two broad classes : those that preserve R-parity down to low energies
 \cite{aulmoh,ckn,abmsv,ag1,ag2,rparso10,blmdm}
 using as Higgs the special representations (${\bf{\oot}}$) of SO(10) that contain R-parity even SM singlets and
another large class of SO(10)   R-parity violating
models\cite{nRGUTs} that attempt to construct viable models using
sets of small SO(10) representations even after sacrificing the
vital distinction provided by  R-parity between matter and Higgs
multiplets.  Besides structural attractions, such as the automatic
inclusion of the conjugate neutrino fields necessary for neutrino
mass, SO(10) GUTs
  offer  a number of other  natural
features.  Among these are  third generation yukawa
unification\cite{3Gyukuni,hallrathempf}, automatic embedding of
minimal supersymmetric Left-Right models, natural R-parity
preservation\cite{rparso10} down to the weak scale and
consequently  natural LSP WIMP dark matter, economic and
explicitly soluble  symmetry breaking   at the GUT
scale\cite{abmsv}, explicitly calculable superheavy
spectra\cite{ag1,bmsv,fuku04,ag2}, interesting gauge unification
threshold effects\cite{ag2,gmblm,blmdm,nmsgutI,nmsgutII}  which
can lead to a  natural elevation of the unification scale to near
the Planck scale\cite{nmsgutI}, GUT scale threshold corrections to
the QCD coupling $\alpha_3(M_Z)$ of the required\cite{langpol}
sign and size\cite{precthresh} and a deep interplay between the
scales of Baryon and Lepton number violation as suggested by   the
neutrino oscillation measurements and the   seesaw formulae
connecting neutrino masses  to the B-L breaking scale.

  The   fascination of the MSSM RG flow at large $\tan\beta$
   stems from  the  tendency of third
 generation yukawa couplings  to converge, at the  MSSM unification scale
\cite{3Gyukuni,hallrathempf}, in a  manner   reminiscent of gauge
unification in  the MSSM  RG flow\cite{marsenj,amaldimssm}.
    For suitably large
$\tan\beta$ and for close to central input values of SM fermion
couplings at the Susy breaking scale  $ M_S \sim M_Z$, third
generation yukawas actually almost coincide at $M_X$.  On the
other hand, in SO(10) theories with only the simplest possible
fermion mass giving  Higgs content (a single \textbf{10}-plet),
when all the complications of threshold effects at $M_X\sim
10^{16} GeV$(not to speak of those at   seesaw scales
$M_{\bar\nu}\sim 10^7-10^{12}\, GeV$) are ignored, one does expect to
generate boundary conditions for the gauge and yukawa couplings
that are unified gauge group wise and (third generation) flavour
wise.

However, fitting the rest of the known fermion data (15 more
parameters) definitely requires other Fermion  Higgs multiplets
(more \textbf{10}-plets, $\mathbf{120,\oot}$ s etc). A principled
position (\emph{monoHiggism}?)  with regard to the choice of Higgs
multiplets responsible for fermion mass (FM Higgs) is to accept
\emph{only one} of each irrep present in the conjugate of the
direct product   of fermion representations. This principle may be
motivated by regarding the different Higgs representations as
characteristic ``FM channels'' through which the fermion mass (FM) is
transmitted  in structurally distinguishable ways. For example the
Georgi-Jarlskog mechanism distinguishes the \textbf{45} plet Higgs
in SU(5) ($\boot$  in SO(10)) from the $\mathbf{5+{\bar 5}}$
(\textbf{10} in SO(10)) due to their ability to explain the
quark-lepton mass relations in the second and third generations
respectively. Similarly the $\boot$ in SO(10) is peculiarly
suitable for implementing the Type I and Type II seesaw mechanisms
for neutrino mass. If one duplicates  the   Higgs multiplets
transforming as  the \emph{same} gauge group representation,  for
example by taking multiple \textbf{10}-plets in SO(10), then one
abandons  the quest for a structural explanation of the pattern of
fermion masses in favour of ``just so'' solutions.   Moreover the
many Higgs doublets typically present in the theory mix with the
\textbf{10}-plet derived Higgs doublets  to make up the light
doublets of the effective MSSM. Both these features dilute the
expectation of exact third generation  yukawa unification.

The technical difficulties of $b-\tau$ unification  are attendant
upon the fact that the GUT scale values of the third generation
Yukawa couplings, in sharp contrast to the three gauge couplings,
are highly infrared sensitive(see \cite{wells} for a cogent
discussion).  The accuracy of unification of third generation
Yukawa couplings after the MSSM RG flow    up to the 1-loop GUT
scale  $M_X^0(\sim 10^{16.25}$ GeV) is strongly dependent on the
low energy values of the yukawa couplings in the MSSM and thus
requires large $\tan\beta$. It is also strongly dependent on the
precise values of $y_b^{MSSM}(M_Z)$ and $y_t^{MSSM}(M_Z)$(both of
which can receive significant Susy threshold corrections) besides
$\alpha_3(M_Z)$ (which needs  significant GUT threshold
corrections to be consistent with experiment\cite{langpol}). The
large threshold corrections \cite{hallrathempf,hitanbetacorr} to
the relations between Standard Model(SM) and MSSM Yukawa couplings
for down type quarks precisely in the large $\tan\beta$ MSSM
variants that are in focus in a SO(10) context, combined with
infrared sensitivity of yukawa unification  casts  doubt on the
naturalness of exact $\mathbf{t-b-\tau}$ unification. As  these
yukawa couplings  have been measured more precisely over the same
period as the investigation of   third generation yukawa
unification in Susy SO(10) models,  the strain of  these models
has increased with time, and various just so mechanisms are
invoked to preserve the exact third generation unification. Yet it
would be strange if Susy SO(10) had no mechanisms to accommodate
not only this difficulty but even the full complexity of the
fermion spectrum.

   Many  detailed  investigations\cite{3Gyukuni,hallrathempf}
    of the yukawa unification    problem since the early days
     have focussed on models based on
   small Higgs representations($\mathbf{16,10,45}$)
       and  seek  to understand only the
   third generation fermion masses. These studies   assume  that no Higgs Multiplet other
    than the \textbf{10}-plet could contribute significantly to the third
   generation masses and use that to justify the assumption
        that high scale threshold corrections to the
     yukawa couplings are negligible.
  In the few studies  that   attempt to tackle   the full fermion spectrum,
   recourse is    had to a   subset of the many  possible
     non-renormalizable operators
  that contribute to fermion yukawa couplings in the effective theory
  below the unification scale.  Since the first two generation
  yukawas are so small relative to the third generation  it is
  possible   to build models using  small
  non-renormalizable  contributions that allow one to fit
  three generation charged  fermion  data\cite{andersonraby,lucasraby}
  and even keep the operator dimension five contributions to
  baryon violation within or near to experimental
  limits\cite{lucasraby}.  When other
     aspects, such as GUT scale spontaneous symmetry breaking,
       are to be addressed multiple Higgs \textbf{10,45,16}-plets are often
     introduced\cite{nRGUTs}. These then require  ad-hoc discrete
     symmetries to replace the    R/matter parity that is   broken in such
     theories.
     These symmetries also serve to prevent  interference between the assigned functions of the
     multiple copies of low dimensional Higgs representation
     introduced (along with appropriately limited
     non-renormalizable interactions that in effect play the role
     of the higher dimensional Higgs irreps that were disallowed at
     the outset).
  With  stringent unification  constraints at GUT scales and large
threshold corrections at the Susy breaking scale only narrow
regions of the Susy-GUT parameter space remain viable. Moreover
  specification  of the soft Susy spectrum and couplings to
ensure    simultaneous  cancellations to keep the threshold
corrections to  $m_b^{MSSM}(M_S)$ small and the $b\rightarrow
s\gamma$ rate consistent with current limits
\cite{raby,baer,wells} is needed.  In general  such investigations
have focussed on evaluating various    non minimal
  effects such as non-universal gaugino masses, D-term
contributions to scalar non universality etc, operating in
specific (tight) corners of the soft susy parameter space, to
evade the conflicting demands  of  infra red sensitivity, large
infrared threshold corrections and accurate yukawa unification. A much smaller
set of papers
\cite{aulmoh,ckn,rparso10,ag1,abmsv,gmblm,blmdm,bert3,nmsgutI}
focusses on renormalizable models without ad-hoc symmetries  and
confronts the full complexity of the spontaeous symmetry breaking
 problem conjointly with fitting the fermion spectra \cite{nmsgutI,nmsgutII}.

Renormalizable    SO(10) Susy GUTs\cite{aulmoh,ckn,abmsv}
  employ  not only the {\bf{10}}-plet but also the other two SO(10)-allowed
 fermion mass (FM) Higgs  representations i.e the
  $\mathbf{\oot}\,\,\mbox{and}\,\,{\bf{120}}$.
   Renormalizability of the GUT model is strictly maintained
   throughout  as a criterion for the structure of the
  superpotential. As a result these theories have a very parameter-economical
  structure\cite{aulmoh,ckn,abmsv,nmsgutI}. In 1992, just as
  yukawa unification came into focus it was proposed\cite{babmoh} that
  use of only the $\mathbf{10-\oot}$ FM Higgs might be sufficient to
  account for the entire fermion mass spectrum \emph{including} the
  neutrino masses and mixings in terms of their (15)  yukawa
  coupling parameters. This obviously attractive possibility was
  investigated as a \emph{generic} possibility in SO(10) (i.e
  without using the formulae in terms of the fundamental GUT
  parameters but only having the symmetries and (naive\cite{pinmsgut})parameter
  counting   implied by their FM Higgs  yukawa  structure). Although
  initially unsuccessful, increasingly refined studies based upon
  the increasingly well defined neutrino mass data ultimately
  showed\cite{TypeIfits,TypeIIfits} that the generic  $\mathbf{10}-\boot$
    parametrizations permitted accurate
 fits of all the known fermion mass data  using \textbf{10}-plet domination of
 third generation mass terms combined with Georgi-Jarlskog
  mechanism for the second generation
 and   either  the Type I or Type II seesaw  mechanisms or a
 combination thereof to fit the neutrino data.
 Perforce  these generic models always assumed  freedom to dial
  the relative strength of the
two seesaw mechanisms.These accurate fits were
found even while assuming that the precisely known SM
  yukawa coupling uncertainties  persevered unchanged over the  RG
  flow  up to the GUT scales. In the light of the
drastic susy threshold corrections at large $\tan\beta$ this
assumption was  naive and peculiarly uninformed of the attention
long paid\cite{hallrathempf,hitanbetacorr} to the role of these
corrections in the other strand of SO(10) GUTs.  The appearance of
\cite{antuschspinrath,rosserna} however reminded us of this
essential feature    and we have thenceforth adopted the much more
realistic estimates of theoretical and experimental uncertainties
advocated by \cite{antuschspinrath} in our subsequent
work\cite{nmsgutII}.

Just as the  intriguing results on the successful implementation
of the Babu-Mohapatra proposal in generic SO(10) GUTs  emerged it
was also shown\cite{gmblm,blmdm,bert3,pinmsgut} that : (i) The use
of MSGUT specific fermion mass formulae implied that neither
seesaw mechanism could account for neutrino masses while also
fitting the charged fermion data. (ii) Conclusions derived using
generic formulae could not be transferred to  concrete
and specific GUTs because of intrinsic difficulties in untangling highly nonlinear
constraints placed by the underlying fundamental GUT on the
generic parameters. (iii) In a specific scenario\cite{blmdm}
  inclusion of the remaining allowed  FM Higgs (i.e the
${\bf{120}}$-plet) could permit the generation of large enough
Type I seesaw masses due to suppressed $\mathbf{\oot}$ couplings
and charged fermion mass fits due to the $\mathbf{10-120} $
combination.

  The $\mathbf{10-120} $ FM Higgs combination proposed by us to
  tackle the charged fermion fits  (with the
  $\boot$ couplings too small to affect any but the first generation
  masses) was seemingly  shown\cite{grlav} to face a generic    difficulty in
providing large enough strange and down quark masses.  This was only to be expected in any scenario where the
Georgi-Jarlskog  or other mechanism  to  distinguish  second generation  down type
yukawas is  not implemented. This difficulty appears in a quite
different light when one considers\cite{nmsgutII} that   Susy
threshold corrections at large $\tan\beta$
 tend \cite{hallrathempf,hitanbetacorr} to drastically modify  the effective value of the down and
  strange quark yukawas. Thus we proposed\cite{nmsgutII} that the excellent fits
  obtainable  for the other (16) fermion data (besides
  $y_{d,s}$) justified searching for the susy threshold corrections to
implement the required corrections to lower $y_{d,s}$  and  this
could be achieved not only with free soft susy parameters at $M_S$
(as in version 1 of this paper\cite{nmsgutII})  but even with just
5 GUT compatible $N=1$ supergravity(SUGRY), Non Universal Higgs
Masses(NUHM), type soft parameters $ ( m_{
\frac{1}{2}},m_0,A_0,m^2_{H,\bar{H}})$ at $ m_Z$ specified at
$M_X$\cite{nmsgutII}
 and two more $(|\mu|, B\sim m_A^2)$ determined by electro-weak spontaneous symmetry breaking  and run back up to $ M_X$ to provide
7 parameters at $ M_X$ coding all the SUGRY-NUHM information.

 These  developments have put detailed consideration
of the role of ${\mathbf{120-\oot}}$ FM Higgs representations
firmly on the agenda of Susy SO(10) yukawa unification. In this
paper we argue that the assumption that third generation fermion
yukawas are protected from large GUT scale threshold corrections
associated with non-\textbf{10} -plet FM Higgs irreps is facile.
We show that it is belied in the class of Susy SO(10)
theories\cite{aulmoh,ckn,abmsv,blmdm,nmsgutI,nmsgutII} which
actually face up to the task of accounting for \emph{all} the
available fermion mass data in a fully specified model  without
invoking uncontrollable higher dimensional operators or ad-hoc
symmetries and rely solely on SO(10) gauge symmetry,
supersymmetry, and structural(parameter counting) minimality  as
guiding principles. In particular the NMSGUT with a large
 $\mathbf{120}$-plet coupling (the $\mathbf{120}$-plet
yukawa  is antisymmetric and hence has two eigenvalues that are
equal in magnitude  and one that is zero) \emph{requires} evaluation of
the GUT scale threshold corrections to the fermion Yukawa
couplings. The large number of fields,  \textbf{120}-plet
couplings  comparable to the \textbf{10}-plet couplings,   and the
fact that these threshold corrections arise from chiral
supermultiplet  wave function normalization(so that any field that
couples to a matter field or a Higgs doublet can run in the self
energy loop) raises the possibility that these corrections may be
far from negligible\cite{dixitsher}. In this paper we show that
this is in fact the case.

The particular importance of the wave function  corrections for the fermion data fitting program
 is that they can relax the  stringent constraint
$y_b-y_\tau\simeq y_s-y_\mu$ that we found \cite{core,msgreb}
operative at $M_X$ in SO(10) models with a \textbf{10-120} FM
Higgs system. Thus at least any model that employs the
$\mathbf{120}$-plet must be sufficiently specified to allow the
calculation of these very significant corrections which can
drastically change the yukawa couplings at $M_X$. Such a
specification involves   solution of the spontaneous symmetry
breaking, calculation of the GUT spectrum and couplings and a RG
analysis of threshold effects based thereon. So far, to our
knowledge, such a complete calculation is possible  only for the
SO(10)  NMSGUT \cite{blmdm,nmsgutI,nmsgutII}for which we give the
dominant 1-loop contributions( certain sub-dominant mixing effects
are postponed to a future calculation). The calculation of the
(wave function) renormalization that gives rise to the threshold
effects modifying the matter fermion Yukawa couplings highlights
the issue of perturbative consistency. One finds that while the
corrections to the matter fermion lines of the mass generating
Higgs-Yukawa vertices are no more than 25\%, the wave function
corrections($\Delta_{H,\bar H}$) to the (light) MSSM Higgs lines
can be much larger.   The most serendipitous scenario would be if
searches that restricted these corrections to be less than 1 were
able to find fits that are also consistent with Baryon decay
limits. This is not what we have observed in our searches so far :
accurate fits with $|\Delta_{H,\bar H}| <1$ are obstructed in
achieving small values ($<< 10^{-18} GeV^{-1} $) of the Baryon
violating dimension 5 operator Wilson coefficients and hence give
baryon lifetimes $\sim 10^{27}$ years. On the other hand  if one
allows large values of $\Delta_{H,\bar H}$ the searches achieve
lifetimes that are a million times or more  larger. Thus
 although threshold corrections to gauge couplings turned out to be mild due to cancellations
\cite{ag2,blmdm} in spite of early alarms\cite{dixitsher} the wave
function renormalization effects at one loop are indeed  large.
Investigation of higher loop contributions is thus be called for
to see if the perturbation theory is sensible. In this paper we
take the view that it is necessary to pursue the investigation of
the realistic features of the SO(10) while including the large
1-loop threshold effects even though full perturbative convergence
may take long to prove( or indeed may never be possible : as for
instance in the most precise known theory QED).

In Section \textbf{2} we briefly review the structure of the
NMSGUT\cite{blmdm,nmsgutI,nmsgutII} to establish the notation  for
  presentation of our results  on    threshold effects in
Section \textbf{3} and   Appendix \textbf{A}. In
 Section \textbf{4} we present illustrative examples to underline the
 significance of the GUT scale threshold effects and the need to include
 them. In Section \textbf{5} we present  examples of improved
fits(specially with respect to
  acceptable $d=5$ operator Baryon violation rates).  In Section \textbf{6}   we discuss the broad features of
   the emerging phenomenology of the NMSGUT. In Section \textbf{7}
  we summarize   our conclusions  and discuss the   various   improvements
    in the fitting, RG flows and searches   and    that are called
    for. Appendix \textbf{A} contains details of the calculation of  threshold
    effects  at $M_X$.  Appendix \textbf{B} contains a discussion of the two loop RGE flow at large $A_0,|M^2_{H,\bar H}|$
    values which results in the unconventional susy spectra with normal s-hierarchy and
 gaugino masses   not even close to the $1:2:7$ ratio found at one loop in  Susy GUTs with universal gaugino masses.
        Appendix \textbf{C} contains
    tables of parameter values for   additional example fits
     to provide a wider view of the possibilities and as inputs
      for phenomenological explorations of the viability of our results. .

\section{NMSGUT recapitulated }

 The    NMSGUT \cite{nmsgutI}  is  a
  renormalizable  globally supersymmetric $SO(10)$ GUT
 whose Higgs chiral supermultiplets  consist of AM(Adjoint Multiplet) type   totally
 antisymmetric tensors: $
{\bf{210}}(\Phi_{ijkl})$,   $
{\bf{\overline{126}}}({\bf{\Sigb}}_{ijklm}),$
 ${\bf{126}} ({\bf\Sigma}_{ijklm})(i,j=1...10)$ which   break the $SO(10)$ symmetry
 to the MSSM, together with Fermion mass (FM)
 Higgs {\bf{10}} (${\bf{H}}_i$) and ${\bf{120}}$($O_{ijk}$).
  The  ${\bf{\overline{126}}}$ plays a dual or AM-FM
role since  it also enables the generation of realistic charged
fermion   and    neutrino masses and mixings (via the Type I
and/or Type II Seesaw mechanisms);  three  {\bf{16}}-plets
${\bf{\Psi}_A}(A=1,2,3)$  contain the matter  including the three
conjugate neutrinos (${\bar\nu_L^A}$).
 The   superpotential   (see\cite{abmsv,ag1,bmsv,ag2,blmdm,nmsgutI,nmsgutII} for
 comprehensive details ) contains the  mass parameters
 \bea
 m: {\bf{210}}^{\bf{2}} \quad ;\quad  M : {\bf{126\cdot{\overline {126}}}}
 ;\qquad M_H : {\bf{10}}^{\bf{2}};\qquad m_O :{\bf{120}}^{\bf{2}}
\eea

and trilinear couplings corresponding to the superfield chiral
invariants indicated :
  \bea
 \lambda &:& {\bf{210}}^{\bf{3}} \qquad ; \qquad  \eta   :
 {\bf{210\cdot 126\cdot{\overline {126}}}}
 ;\qquad\qquad \rho :{\bf{120\cdot 120 \cdot{  { 210}}}}
\nnu k &:& {\bf{ 10\cdot 120\cdot{ {210}}}} \qquad;\gamma \oplus
{\bar\gamma}  : {\bf{10 \cdot 210}\cdot(126 \oplus{\overline
{126}}}) \nnu \zeta \oplus {\bar\zeta}  &:& {\bf{120 \cdot
210}\cdot(126 \oplus {\overline {126}}})
  \eea

In addition   one has two   symmetric matrices $h_{AB},f_{AB}$ of
Yukawa couplings of the the $\mathbf{10,\oot}$ Higgs multiplets to
the $\mathbf{16_A .16_B} $ matter bilinears and one antisymmetric
matrix $g_{AB}$ for the coupling of the ${\bf{120}}$ to
 $\mathbf{16_A .16_B} $. One of the complex symmetric matrices can be
made real and diagonal by a choice of SO(10) flavour basis. Thus
 initially complex FM Yukawas    contain 3 real and 9 complex
parameters. Five overall phases (one for each Higgs), say those of
 $m,M, \lambda ,\gamma,\bar\gamma$, can be set by fixing phase conventions. One(complex
parameter)  out of the rest of the superpotential
parameters i.e
   $m,M_H,M,m_o,\lambda,\eta,\rho,k,\gamma,\bar\gamma,\zeta,\bar\zeta$ , say
   $M_H$,  can be fixed by the fine tuning condition to keep two doublets
   light so that the effective theory is the MSSM. After removing un-
   physical phases  this leaves 23 magnitudes and
15 phases as parameters : still in the lead out of any theories
aspiring to do as much\cite{abmsv}. As explained in\cite{abmsv,ag1,ag2}
   the fine tuning fixes the Higgs fractions i.e the composition
   of the massless electroweak doublets in terms of the (6 pairs of suitable)
    doublet fields in  the GUT.  A subtle point here is that even
    if the other parameters are taken real  the fine tuned $M_H$
   (which does not itself enter into the low energy lagrangian)
     will be complex. Thus strictly speaking one cannot
     justify the use of only real superpotential
     parameters  by invoking `spontaneity' of CP violation
     and we will not do so.

The GUT scale vevs and therefore the mass spectrum are all
expressible\cite{abmsv,ag2,bmsv} in terms of a single complex
parameter $x$ which is a solution of the cubic equation

\be 8 x^3 - 15 x^2 + 14 x -3 +\xi (1-x)^2=0 \label{cubic} \ee
where $\xi ={{ \lambda M}\over {\eta m}} $.

In our programs we find it convenient to scan over the ``three for
a buck"\cite{bmsv06}  parameter $x$ and determine $\xi$ therefrom . Then the
phase of $\lambda$ is adjusted to be that implied by $x$ and the
relation $\xi ={{ \lambda M}\over {\eta m}} $ and is not itself
scanned over independently.
 It is a convenient fact that the 492
fields in the Higgs sector fall into precisely 26 different types
of SM gauge representations which  can hence be naturally labelled
by the 26 letters of the English alphabet\cite{ag2}.
  The decomposition of SO(10) in terms of the labels  its ``Pati-Salam''
maximal subgroup $SU(4)\times SU(2)_R\times SU(2)_L$
provided\cite{ag1} a   translation manual from SO(10) to unitary
group labels.  The complete GUT scale spectrum and couplings of
this theory have been given in \cite{ag2,nmsgutI}.

 The tree level fermion yukawa  couplings and
neutrino masses of  the effective
 MSSM arising from this GUT below the GUT scale
 after fine tuning to keep one pair of Higgs multiplets  light
 are given  in \cite{nmsgutI,nmsgutII}.

As mentioned, in the NMSGUT the conjugate(i.e ``right handed'')
neutrino Majorana masses are 4 or more orders of magnitude smaller
than the GUT scale due to very small $\boot$ couplings. Therefore
for purposes of calculating the threshold corrections to the
Yukawa couplings at $M_X$ we can consistently treat the conjugate
neutrinos as light particles on the same footing as the other 15
fermions of each SM family. These fermion mass formulae, after
correcting for threshold effects, are to be confronted with the
RG-extrapolated data (from $Q=M_Z $ to $Q=M_X^0=10^{16.25}$ GeV
 including neutrino masses and mixing angles). The
calculation of $\Delta_X$ also fixes the scale $ {m}$ of the high
scale symmetry breaking\cite{blmdm,bert3,nmsgutI}. The stringent
simultaneous requirements of of a common unification-seesaw scale,
gauge unification (including the right high scale gauge RG
threshold corrections to shift the GUT prediction of
$\alpha_3(M_Z)$ down to acceptable values\cite{precthresh}), third
generation yukawa unifcation,  {as well as} fits to all the other
fermion masses and mixing matrices, are effective in singling out
characteristic and suggestive GUT parameters (including Susy
breaking parameters at $M_X$).

\section{ GUT scale Yukawa threshold corrections}

The technique of\cite{wright} for calculating high scale threshold
corrections to yukawa couplings,  generalizes  the
Weinberg-Hall\cite{weinberghall} method  for calculating threshold
corrections to gauge couplings, and has long been available  but
has not been exploited much; possibly due to the assumption that
such effects are always negligible. In supersymmetric theories the
superpotential parameters are renormalized only due to wave
function correction  and this is easy -if tedious- to calculate
for the large number of MSSM submultiplets  in the
$\mathbf{120}$-plet which couple to the light fermions and   MSSM
Higgs at an SO(10) yukawa vertex.  There is also a contribution
from heavy gauge field couplings to the light(i.e MSSM) fields at
the matter yukawa  vertices. The calculation involves going to a
basis in which the heavy field supermultiplet mass matrices are
diagonal. This basis is easily computable given the complete set
of mass matrices and trilinear coupling decompositions given in
\cite{ag1,ag2,nmsgutI}. For a generic heavy field type $\Phi$ the
mass terms in the superpotential diagonalize as  :
 \bea{\overline{\Phi }}= U^{\Phi}{\overline{\Phi' }} \quad ;
 \qquad {\Phi } =  V^{\Phi}\Phi'\quad \Rightarrow \quad
 {\overline{\Phi}}^T M \Phi ={\overline{\Phi'}}^T M_{Diag}
 \Phi'     \eea

Threshold correction to a Yukawa coupling matrix (which occurs in
the superpotential as $W={\bar f}^T Y_f f H_\pm$) then have the
form
  \bea Y_f=Y_f+  {\Delta}_{\bar f}^T \cdot Y_f +  \Delta_f \cdot Y_f +
    \Delta_{H^{\pm}}
  Y_f\eea

  where the  $\pm$ refers to the $Y=\pm 1 $ Higgs multiplets
  appropriate to give mass to
   $ T_3=\pm {1\over 2}$ fermions and $\Delta_{f,{\bar f}},  \Delta_{H^{\pm}}$ are
    the 1-loop wave function correction factors.
  For a generic interaction superpotential
   $W= {1\over 6} \sum_{ijk}Y_{ijk} \Phi^i\Phi^j\Phi^k $,
   the quantities $\Delta$, at the renormalization scale $Q$,
     have the form ($\Delta=-{1\over
   2} K $ in the notation of \cite{wright})

  \bea \Delta_i^{ j}(Q)= {1\over {32 \pi^2}}(- 2 g_{10}^2 \sum_{k,A} F_1(m_A,m_k,Q)
  I^A_{ik} I^A_{kj}+{1\over 2}\sum_{kl} Y_{ikl}Y^*_{jkl}
  F_1(m_k,m_l,Q))\eea

  We have used $Q=M_X^0  =10^{-\Delta_X}M_X$ where $\Delta_X$ is the
  shift (away from the one loop MSSM unification value $Log_{10}M_X^0=16.25$)
    due to  loop and threshold effects.
  Here $A$ is a generic gauge field(adjoint)  index and $i,j,k$   are generic
  chiral field indices. The gauge couplings and generators $g_{10},I^A$
  of SO(10) are related to the usual ($SU(5)$normalization)   gauge coupling
   and generators  $g_5=g,T^A$
  by $g_{10}=g/{\sqrt{2}},I^A={\sqrt{2}} T^A$.
    When both the fields running in the loop are heavy
  fields ($F_{1}$ is a symmetric  Passarino-Veltman function)
  $F_1$ should be taken to be

\bea F_{12}(M_A,M_B,Q)={1\over {(M_A^2- M_B^2)}}(  M_A^2 ln
{M_A^2\over Q^2} -M_B^2 ln {M_B^2\over Q^2} )- 1 \eea which
reduces to just \bea F_{11}(M_A,Q)=F_{12}(M_A,0,Q)=    ln
{M_A^2\over Q^2}  - 1 \eea when one field is light
($M_B\rightarrow 0)$.
  When   one   of   the heavy   fields in the loop has MSSM doublet type
   $G_{321}$ quantum numbers $[1,2,\pm 1]$ (so that one eigenvalue
   is light while the other \emph{five}\cite{nmsgutI}  are heavy)
    care should be taken to avoid summing over light-light loops since
      that calculation belongs to the MSSM radiative corrections.

   In the NMSGUT fitting scenario where
  $|f_{AB}|< 10^{-5}$  it is an excellent approximation  to ignore
  the  contributions of the $\mathbf{\oot}$ to the high scale threshold
  corrections.  Moreover  while calculating the wave function renormalization of the
  Higgs line in the matter fermion-antifermion-MSSM Higgs vertex, we shall assume that it is an good
  approximation to take the MSSM higgs
 to be   dominantly made up of the $\mathbf{10}$-plet (which dominance
   we know is required in order to
   account for top-bottom-$\tau$ (near) unification which is an intrinsic part of
    the large $\tan\beta$
   scenario) so that we can ignore the admixture of the other 5 MSSM type Higgs
    doublets pairs present in the theory.
    Hence, for the non-dominant i.e non \textbf{10}-plet derived light Higgs components,
       the   contributions to the  wavefunction
    renormalization of the light Higgs doublets
    in the theory will   all be suppressed by additional small (in keeping with
  the \textbf{10}-plet dominance of light Higgs composition) factors of
   $|\alpha_{i}|^2; i=2....6$ (for $H_0[1,2,1]$)
  or $|\bar\alpha_i|^2,i=2....6$ (for $\bar{H}_0 [1,2,-1]$). That
  is to say there will be a suppression of the contribution by
  $|\alpha_i|^2$ or  $|\bar\alpha_i|^2$
  unless the external doublets come from the \textbf{10}-plet
  in the $k \mathbf{10.120.210}$ or $\mathbf{10.210.(\gamma 126 + \bar\gamma\oot)}$
  terms of the superpotential.   We can
  therefore sidestep--for the moment-- the elaborate  calculation required for
  calculating the wave function renormalization of the external  light Higgs
  when all its 6 possible  components (from the
  $\mathbf{10,210,126,\oot,120}$ (two pairs) )
    are corrected by wave function renormalization.
This is an good ( but not perfect ) approximation in practice as
will be seen from the values of the Higgs
fractions\cite{abmsv,ag2,nmsgutI}. For example in case I-1(Table\ref{I-1-b} the values
are : \bea |{\vec \alpha}|^2 &=& \{0.689 , 0.0044, 0.0038 ,
0.1838 , 0.0302 , 0.0886\}\nnu |{\vec {\bar \alpha}}|^2 &=&
\{0.792, 0.0068 , 0.0037, 0.1020 , 0.0052 , 0.0903\} \eea
where the numbering of the components is\cite{nmsgutI} seen from :
\bea[1,2,-1](\bar{h}_1,\bar{h}_2,\bar{h}_3,\bar{h}_4,\bar{h}_5,\bar{h}_6)\oplus
[1,2,1](h_1,h_2,h_3,h_4,h_5,h_6)\equiv \\
(H^{\alpha}_{\dot{2}},\bar\Sigma_{\dot{2}}^{(15)\alpha},
\Sigma_{\dot{2}}^{(15)\alpha},\frac
{\Phi_{44}^{\dot{2}\alpha}}{\sqrt{2}},O^{\alpha}_{\dot{2}},O_{\dot{2}}^
{(15)\alpha}\hspace{2mm}) \oplus (H_{\alpha
\dot{1}},\bar\Sigma_{\alpha \dot{1}}^{(15)},\Sigma_{\alpha
\dot{1}}^{(15)},\frac{\Phi_{\alpha}^{44\dot{1}}}{\sqrt{2}},O_{\alpha\dot{1}}
,O_{\alpha\dot{1}}^{(15)})\nonumber\eea

   The decomposition of SO(10) invariant terms in the superpotential
and gauge terms yields\cite{ag1,ag2,nmsgutI}  a  large
number($\sim 100$ ) of vertices. It then  requires a tedious but
straightforward calculation to determine the threshold corrections
explicitly. The explicit  expressions are  given in  Appendix {\bf{A}}.

Heretofore such threshold  corrections have mostly been argued to
be negligible($< 1\%$)   although at least one paper \cite{baer}
faced with the  difficulties of literal third generation yukawa
unification   has considered the possibility, without any explicit
model which permitted calculation, that the third generation
yukawa unification relations must necessarily be subject to
threshold corrections of up to $50\%$.  In which case it was found
that  the various
 stratagems invoked to permit precise  3 generation
Yukawa unification  could    become redundant. We shall see  that
 the calculation of the GUT scale 1-loop Yukawa threshold
effects  in the NMSGUT  can actually change the naive(i.e pure
\textbf{10}-plet)  unification relations $y_t=y_b=y_\tau$
significantly. Furthermore the $ \mathbf{10-120}$ plet fermion
fits have been shown (  in the absence of GUT scale threshold
effects) to require a close equality $ | y_b-y_\tau/ ( y_s-
y_\mu)| \approx 1 $ at $M_X $ which is very constricting when
searching for fits. The fits we exhibited in
\cite{nmsgutI,nmsgutII} were all of this type. However in the
present case the fits we obtain can deviate significantly   from $
{\frac{y_b-y_\tau}{y_s- y_\mu}}\simeq 1 $ which was obeyed by both
the fits presented in \cite{nmsgutII} where no threshold
corrections were applied to the yukawas  at $M_X$.
 The large values of $\Delta_{H,{\bar H} }$ can   radically reduce the
    magnitudes of SO(10) yukawas required to reproduce the MSSM couplings at $M_X$,
     thus threshold effects can help in loosening
 this constriction of fitting freedom. Moreover the changes
are such as to help in finding  fits with a   slower B violation
rate. We note again that one must  study the higher  loop
threshold corrections    and the steps necessary to define a
consistent perturbative expansion (possibly involving some
variants of large N resummation and use of the exact SO(10) Susy
gauge beta functions\cite{vainshif}) to see if the 1-loop results
we find are stable.

\section{Numerical fits to the  fermion data    and   threshold effects}

To appreciate the significance of the threshold corrections at
$M_X$ for the matter fermion yukawas it is sufficient to consider
the values of the matrices $\Delta_{f,{\bar{f}}}$  for the various
MSSM fields when the formulae derived in the appendix are
evaluated using parameters from the examples of fits (found
ignoring GUT scale threshold corrections) given in
\cite{nmsgutI,nmsgutII}. The NMSGUT is , to our knowledge, the
\emph{only} SO(10) model where the spontaneous symmetry breaking
and the consequent spectra and Higgs fractions have been
explicitly calculated so that the Yukawa threshold corrections can
be   evaluated for specific GUT parameter based fits.

The example  fits from \cite{nmsgutII} are of the 18  known
fermion data namely the yukawa couplings
$y_{t,b,\tau,c,s,\mu,u,d,e}$, the CKM angles and phase
$\theta^q_{12,13,23},\delta^q$, the neutrino mass squared
differences $\Delta m^2_{21,32}$ and Leptonic mixing angles
$\theta^L_{12,13,23}$ in terms of the NMSGUT hard and soft
parameters. We search  assuming normal neutrino hierarchy, a very
light ($\leq1~ meV$) electron neutrino, and a small neutrino
mixing angle $\theta^L_{13}<5^\circ$\footnote{ After the calculations for this paper were completed
 the T2K and MINOS results on the first measurements of $\theta^L_{13}$ using reactor muon neutrinos directed at distant neutrino detectors  were announced\cite{T2K,MINOS}. These results indicate that the likely range is  $\theta^L_{13}\sim 10^\circ \pm 5 ^\circ$.
 Since the search found fits with $\theta_{13}^L \sim 5^\circ $   we anticipate no problem if the target is raised by 1-5 degrees.  }.  The fits are found by a
 random search based on  the downhill simplex method which requires
the definition of a $\chi^2$ function formed from the difference
of the GUT implied and target (i.e RG extrapolated from $M_Z$)
values of the fermion parameters normalized by the uncertainties
in these parameters\cite{antuschspinrath}. An important point is
that heretofore fitting in this class of renormalizable SO(10)
models assumed that the uncertainties involved were merely the
(very small) error estimates  for the SM extrapolated to the GUT
scale. The complexities induced by the uncertainties due to the
strong threshold effects at the Susy and GUT thresholds were never
given any shrift in this context. Recently the papers
\cite{antuschspinrath,rosserna} motivated us to use the threshold
effects to evade the difficulty of the failure of the
Georgi-Jarlskog mechanism in the NMSGUT by lowering the couplings
$y_{d,s}$ via threshold corrections due to Susy partners. Thus in
\cite{nmsgutII} we used  the more realistic values
of\cite{antuschspinrath} for the error estimates  and eschewed the
spurious precision of  most previous efforts in the context of
renormalizable MSGUTs.

The first example of an accurate fit  presented in \cite{nmsgutI}
was able to achieve a $\chi_X=0.0538$ for a fit of the 18 fermion
data at the scale $M_X$ (accompanied by $\chi_Z=0.027
  $ fit at the scale $M_Z$ for matching the run down
charged fermion yukawas to the Standard Model results after
inclusion of the Susy threshold corrections). When the same data
are used with the threshold corrections switched \emph{on} one
gets $\chi_X=773.6 $ and the unification parameters are
$\Delta_{X,G,3}$ are unacceptable.  Clearly the threshold
corrections can make a great deal of difference ! The
corresponding changes for the second solution are $
 \chi_X=286364.7 $ and the unification parameters are
$\Delta_{X,G,3}$ are unacceptable.  It is also worth noting that
even though the fermion yukawas generated from the NMSGUT formulae
change radically when one inserts the threshold corrections (using
couplings determined by the fits found ignoring them), one finds
that the accuracy of satisfaction of $y_b-y_\tau=y_s-y_\mu$ as
measured by the ratio $R_{b\tau/s\mu}= (y_b-y_\tau)/y_s-y_\mu $
remains good with the ratios for these two solutions changing as :
$( 1.084,0.97 ) \rightarrow ( 1.16 ,0.96 )$. \clearpage

\begin{table}
 $$
 \begin{array}{|c|c|c|c| }
 \hline
 && NTH-1& \\
   \hline
      Eigenvalues(\Delta_{\bar u})&   0.010085&   0.031557&   0.032194\\
  Eigenvalues(\Delta_{\bar d})&   0.013965&   0.037585&   0.038028\\
Eigenvalues(\Delta_{\bar \nu})&   0.015536&   0.057369&   0.057938\\
  Eigenvalues(\Delta_{\bar e})&   0.015275&   0.039068&   0.040438\\
       Eigenvalues(\Delta_{Q})&   0.007961&   0.024766&   0.025598\\
       Eigenvalues(\Delta_{L})&   0.005748&   0.038287&   0.039675\\
    \Delta_{\bar H},\Delta_{H}&        3.007406   &        1.661375    &{} \\
 \hline
 && NTH-2& \\
 \hline
  Eigenvalues(\Delta_{\bar u})&   0.069710&   1.007792&   1.033000 \\
  Eigenvalues(\Delta_{\bar d})&   0.066965&   1.024073&   1.053104 \\
Eigenvalues(\Delta_{\bar \nu})&   0.078188&   1.049538&   1.074279 \\
  Eigenvalues(\Delta_{\bar e})&   0.086420&   0.999649&   1.027649 \\
       Eigenvalues(\Delta_{Q})&   0.063482&   1.058937&   1.085331 \\
       Eigenvalues(\Delta_{L})&   0.077448&   1.075814&   1.104087 \\
    \Delta_{\bar H},\Delta_{H}&        9.309859   &        4.858437    &{} \\

  \hline
 \end{array}
 $$
  \caption{\small{NTH12: Eigen values of the wave function correction matrices calculated using coupling values found in
 \cite{nmsgutII} \emph{without} incorporating threshold corrections. Third generation values are in the first column.
  Note the significant  threshold effects on the Higgs lines and the twofold  degeneracy among the first
  two generations as well as suppression of the third generation corrections due to cancelation between
  the large gauge and third generation yukawa contributions. Due to reduction of  typical SO(10)
  \textbf{16}-plet yukawas both these features are modified in the fits found after incorporating threshold corrections at $M_X$. }}
\label{NTH12} \end{table}

Typical values for the fermion line dressing coefficients due to
yukawa couplings of the first two generations  are of a few
percent or smaller as seen in Table \ref{NTH12} where we
 exhibit eigenvalues of the fermion/Higgs line dressing matrices calculated
  for the two fits given in \cite{nmsgutII} which did \emph{not} incorporate
  threshold corrections. Note that values for the Higgs lines can add up and achieve   large values
 of up to several 100\% !   If we switch off these corrections to $\Delta_{H,\bar H}$ then the
changes in the fits are much smaller  but still  by no means
negligible.  These numbers make it clear that the light  fermion
lines  and specially the light  Higgs lines suffer very
significant  threshold corrections. Thus realistic  GUT theories
must face up to the task of specifying themselves sufficiently
explicitly so that the threshold corrections may be calculated,
tree level estimates are likely to be only rough pctures or even
completely misleading. It is in this sense that we speak of Grand
Unification ``upended''.

\section{ Realistic fits with threshold corrections included}

If we use our search programs to find fits after including the
threshold effects we can impose   strict perturbativity in
the sense that no threshold correction may exceed  1  i.e \bea
|\Delta_{f,\bar f,H,\bar H}| < 1 \label{strictpert}\eea The search
programs\cite{nmsgutII} do find solutions (quite far from the
examples of \cite{nmsgutII} in that some couplings, such as $\eta$
underwent major changes) which satisfied this constraint and still
provided accurate unification and accurate fits of the fermion
mass data. However when one evaluates the rate of Baryon number
violation (in the dominant $B\rightarrow Meson~+ \nu $ channels)
one finds (as in the case of the fits in \cite{nmsgutII}) that
typical solutions predict lifetimes of $10^{27}$ years or smaller.
This is   6 orders of magnitude below the current experimental
limits \cite{superKBlimits}. This problem is  an extension   of
that exhibited by  the solutions found without threshold effects\cite{nmsgutII}.

Given the vast parameter space it is natural to ask whether there
exist solutions where there is a suppression of the $d=5$ operator
mediated Baryon decay.  Such fits  can be found by instructing(via
a $\chi^2$ penalty for rapid baryon number violation) the `amoeba'
(i.e  is the search engine of the downhill simplex method for
nonlinear fitting\cite{pressteukolsky}) to look for fits that have
sufficiently low B-decay rates.  An exhaustive statistical
characterization of the parameter space and its possibilities
requires the marshalling of considerable (super)computing
resources which we shall eventually accomplish. For the moment we
avoid over-determining an already excruciating search  for viable
fits  by not also demanding such strict perturbativity, in the
above sense, in addition to all the other demands of a sensible
supersymmetric phenomenology  that we must anyway impose.

 Since the complete programs for calculating B-decay rates (based upon the
formulae provided in \cite{lucasraby,gotonihei}) are large and
  time consuming it would have slowed down both our
search engine and the present computations too much to interface
the complete Baryon violation  programs (which include
renormalizing some 447 variables from $M_X$ to $M_Z$, as well as a
time consuming evaluation of the Baryon decay amplitudes)  with
our search programs at this still exploratory stage. Therefore we
adopted the expedient of computing the the maximal absolute
magnitude $Max(O^{(4)})$ of the LLLL and RRRR coefficients in the
$d=5,\Delta B\neq 0$ effective superpotential for the
NMSGUT\cite{nmsgutI,nmsgutII}. For the solutions found so far this
quantity was found to be typically of order $10^{-18} $ to
$10^{-16} \,GeV^{-1}$ corresponding to the fast baryon decay rates
   $\sim 10^{-27}\, yr^{-1}$ obtained. Thus a quick fix to the problem of limiting the
   B-decay rate while searching for accurate fermion fits is to
   limit($\tilde{O}$  is the dimensionless operator in units od $|m/\lambda|$)
    $Max(\tilde{O}^{(4)})< 10^{-5 } GeV^{-1}$. This produced fits with proton lifetimes
     above $10^{36}$ yrs so we also relaxed the limit in some searches to just  $Max(\tilde{O}^{(4)})< 10^{-4 } GeV^{-1}$. When
   this condition was imposed simultaneously with the requirement
   of strict perturbativity  (\ref{strictpert}) above, we were
   unable to find any accurate fits  so far. On the other hand if
   one removes the condition of strict perturbativity   but
   limits $Max(O^{(4)})< 10^{-21} \, Gev^{-1}$ or $Max(O^{(4)})< 10^{-22} \, GeV^{-1}$
     then it was possible to find accurate fits
   that gave lifetimes in excess of $10^{33} yrs$.
   In addition to the above mentioned penalties we also required
   the fits to  satisfy the following consistency/phenomenological constraints :

   \begin{itemize}

   \item As already explained in detail in \cite{nmsgutI} the gauge
   unification RG flow is constrained so that perturbation theory in the
    gauge coupling at unification remains valid, the unification
     scale is less than $M_{Planck}$  and the GUT threshold
     contributions to $\alpha_3(M_ZX)$ are in the right range\cite{langpol,precthresh,nmsgutI} :
\bea
-22.0\leq \Delta_G &\equiv&  \Delta  (\alpha_G^{-1}(M_X))  \leq 25 \nonumber \\
3.0 \geq  \Delta_X &\equiv &\Delta (Log_{10}{M_X}) \geq - 0.3\nonumber \\
-.017< \Delta_{3} &\equiv & \alpha_3(M_Z)  < -.004\label{criteria}
\eea

\item We constrain  the $|\mu(M_Z)|, |(A_0)_{ii}|(i=1,2,3)$
parameters to  be smaller than  150 TeV. Typically these
parameters emerge in the range $\sim 50-90 $ TeV while the gaugino
masses $M_i$  are driven to the lower limits imposed(since it is
the ratios $|\mu(M_Z)|/M_i,  (A_0)/M_i$ which cointrol the
efficacy of the large tan$\beta$ corrections for our purposes.
This  is the price one must pay to correct the fermion yukawas to
achievable values in the NMSGUT. Large values of $A_0$ are well
known to lead to deep charge and colour breaking (CCB)
minimae\cite{CCB}  or unbounded from below (UFB)
potentials\cite{UFB}. However it is also
established\cite{kuslangseg} that the metastable standard vacua
that we are considering (with all mass squared parameters of
charged or coloured or  sneutrino scalar fields \emph{positive}
i.e at a local minimum which preserves colour,charge and R-parity)
can well be stable on times scales ($\sim 10 $ Giga-years)   of
order the age of the universe. Thus individual cases need to be
re-examined in detail before dismissing any  otherwise viable fit
out of hand. We consider the situation further in the next
section.

\item In accordance with experimental constraints \cite{pdg2010}
we also constrain lightest chargino (essentially wino $\tilde
W^\pm$ ) masses to be greater than 110 GeV.
All the charged  sfermions  as well as the charged
 Higgs are constrained to lie above 110 GeV and the uncharged Higgs($h^0$) above 105 GeV.

 \item Since the  susy threshold corrections to $y_{d,s,b}$ \emph{necessary} for the survival
 of the NMSGUT as a viable theory of fermion masses    depend on logarithms of ratios of soft
 susy breaking parameters, our scenario is obviously incapable of
 \emph{predicting} the all important susy breaking scale.  On the other hand this is also a theory that counts
  providing exact unbroken R-parity  down to the lowest scales (so that the LSP is stable and  a good Dark
  Matter candidate) as one of its main virtues. Thus
 it behooves us to search for fits constrained by requiring the mass of the LSP(which is purely Bino due
 to   the large value $\mu\sim 50 TeV$  that emerges) in various  ranges best motivated from a LSP dark matter scenario point of view,
 such  as $>101\,GeV$(range I),  $5-50\, GeV$ (Range II) and $50-101\, GeV $ (Range III),  to get an
 initial glimpse of whether and how the effective Dark Matter scale  is linked to the
 pattern of superpartner masses. Thus in Tables \ref{I-1-a} to \ref{III-1-f} we provide
    examples of fits found with the LSP/Bino mass in each of these ranges.
    In addition, in the Appendix we provide 8 additional examples of fits in these categories.
     We emphasize that these examples are only indicative and that a thorough investigation
     should have a finer grain in the LSP mass ranges chosen  and must also incorporate loop
      corrections, specially for sfermion   masses besides various improvements discussed below.
      It must also consider the statistics of fits found by  much larger scale systematic searches
      than those we have so far been able to mount : due to the scale of computer resources required.
       We note that the Type II and III fits are
       variations on those in I found by changing just the constraint on the LSP/Bino mass.
        One of the cases namely II-2 did not yield a satisfactorily exact fit  in spite of
        extensive running, therefore it is omitted from the Appendix.

   \end{itemize}

 In Tables of Type X-a ($X=I,II,III$) we give the complete set of NMSGUT parameters defined at the one loop unification scale $M_X^0$
 together with the values of the soft Susy breaking parameters ($\{m_0,m_{1/2},A_0,B, M^2_{H,\bar H}\} $) together
with   Supersymmetric parameter $\mu $. Thus we have tried out only a N=1 Supergravity GUT motivated scenario with
  relaxation permitting  Non universal Higgs masses(SUGRY-NUHM) (this is justified from a GUT point of view since
   the light doublets are a mixture of doublets from several sources in different SO(10) representations).
    The procedure followed for finding the fits, specially the use of Susy threshold correction  to correct
    the yukawas $y_{d,s}$ to values consistent with the \textbf{10-120} FM Higgs structure, as well as to
    raise the effective $y_b$ in the MSSM
so as relieve the well known tension requiring this SM value of this  parameter to lie several standard deviations
 above its experimental value in order to achieve $b,\tau$ unification\cite{king},  is described in detail
 in the second paper of this series \cite{nmsgutII}. Besides these parameter values of the   SUGRY-NUHM NMSGUT
  we also give the mass spectrum of superheavy heavy fields including the right handed neutrinos and the Type I
  and Type II neutrino seesaw masses as well as the unification parameters $\Delta_{X,G,3}$ described in detail in \cite{nmsgutI}.

In tables of Type X-b we have given the values of the target(i.e two loop RGE extrapoled Susy threshold corrected MSSM yukawas and Susy Weinberg operator coefficients)
 fermion parameters and their uncertainties (estimated a la \cite{antuschspinrath}) together with the achieved values and pulls.  The reader may check that the fits are all excellent with typical fractional errors O(0.1\%). We remark that we found it tedious and meaningless to push our program to further narrow the fit in view of the large uncertainties and the numerous corrections still to be incorporated in our calculations (see below). We also give the eigenvalues of the GUT scale yukawa vertex  threshold correction operators. We note that Tables X-a  show that there is a significant lowering of the size of the SO(10) fermion yukawas so that the universal gauge corrections dominate and make the corrections to all three generations almost equal specially when the lowering is pronounced. We give also the values of the the ``Higgs fractions \cite{abmsv,ag2,nmsgutI} $\alpha_i, {\bar \alpha}_i$ crucial for determining the fermion mass formulae \cite{ag2,gmblm,blmdm,nmsgutII}.
These parameters are determined as a consequence of the GUT scale symmetry breaking and the fine tuning to preserve a light pair of MSSM Higgs doublets. They   distill the influence of the SO(10) GUT (and its spinorial clebsches determined appositely for this purpose in \cite{ag1})  on the low energy fermion physics. The reader may use them together with the formulae given in \cite{nmsgutI} to check the fits even without entering into the details of our GUT scale mass spectra. We note that the first components of the $\alpha_i, {\bar \alpha}_i$ were chosen real by convention \cite{nmsgutI}(see Appendix \textbf{C}).

In Tables of type X-c values of the SM masses at $M_Z$ are compared with those of masses from  the run down yukawas achieved
in the NMSGUT both before and after large $\tan\beta$ driven radiative corrections.  Note that the central value of
$m_b(M_Z)=2.9$ GeV becomes prima facie acceptable in contrast to small $A_0$ scenarios where the need for $m_b(M_Z)>3.1$ GeV, ie
more than one standard deviation from the experimental central value, has   been a principal source of tension and anxiety for
small $A_0$ models \cite{king}  and should perhaps motivate  acceptance of exploration of the large $A_0$ parameter space which
 seems almost unexplored so far.

  In Tables of type X-d values of the soft supersymmetry breaking parameters which are the most crucial and remarkable
   output of this study  since they tie the  survival of the NMSGUT to a specific
  type of  soft Susy  spectrum  with large
  $\mu,A_0,B$ and third  generation sfermion masses generation in the 10-100 TeV range. Remarkably and in sharp
   contrast to received (small $A_0,M^2_{H,{\bar H}}$) wisdom the third s-generation is much \emph{heavier} than the first two
    generations, which however are themselves not very light \emph{except}
for the \emph{right chiral} smuon/selectron , sstrange/sdown  and scharm/sup  \emph{right handed}   charged sfermions  which can
 actually descend close to their experimental lower limits.   In doing so they keep alive the  effectiveness
  of the pure Bino LSP (and pure Wino lightest chargino and next to lightest Neutralinos)  as candidate
   dark matter by providing co-annihilation channels of the sort  a light $stau$ is often enlisted for
    in standard Susy GUT scenarios. Also remarkable is   the friability of the  familiar   $1:2:7$ ratio
    of the gaugino masses $M_1,M_2,M_3$ that we discover repeatedly in our calculations.
    This ratio  is almost  fixed in stone  by  one loop RGE and GUT mandated gaugino mass universality
      for small $A_0$   (and   often provokes baroque elaborations of the  gauge invariance principle designed to avoid it).
       Here however the large influence of the $A_0$ parameters can change the RGE flow to the point
        where the gluino is often \emph{lighter} than the Winos which are sometimes as much thrice the mass of the Bino.
         Note also that except for the very heavy third generation the actual sizes of the sfermion trilinear
          couplings are rather modest since they are the product of the $A_0$ parameters and the Yukawas.
           For the third sgeneration the   trilinears are roughly the same size as the masses themselves
           thus preserving naturality. We remind the reader that a diagonal two loop  RGE flow from $M_X$ to $M_Z$ was
            used to determine these soft susy parameters via the Susy threshold effects since only the diagonal
             formulae were easily accessible and seemed justified in view of our limited expectations of overall
              accuracy  of sfermion spectra which we have so far evaluated only at \emph{tree
              }level. In view of the importance of the susy   spectra we discuss in detail in  Appendix
\textbf{B} the  features of the two loop RGE flow which result in
the unconventional spectra noted above.

Finally Tables of type X-e,f  give Susy particle determined using two loop RGEs  with and without generation mixing
switched on. They are so similar as to justify the use of the diagonal values for estimating the Susy threshold
 corrections. For the case of the lightest sfermions however the corrections are sometimes as large as 10-30\%.
 This again sounds a note of caution regarding the exact numerical values of the lighter sfermion masses we provide.
  However even after incorporation of Loop corrections in addition to these effects we certainly expect that the
  broad division into LHC discoverable particles lighter than say  2.5 TeV  (the  LHC beam energy available per
   parton which sets the upper limit of discovery potential at the LHC) and those heavier to have
   some cogency regarding the limits of the discoverable. Thus we have provided
    an additional Table \ref{LHCdetectables}  that collects all superparticles with masses less than this limit.
    We also note that the Higgs masses were calculated using the 1-loop corrected electroweak symmetry breaking
    conditions and  1-loop effective potential using a subroutine\cite{porod}  based on\cite{loophiggs}.
   The wary reader fearful of loop corrections destroying the whole  scenario may be reassured at least on that count.

\begin{table}
 $$
 \begin{array}{|c|c|c|c|}
 \hline
 {\rm Parameter }&Value &{\rm  Field }&\hspace{10mm} Masses\\
 &&{\rm}[SU(3),SU(2),Y]&\hspace{10mm}( Units\,\,of 10^{16} GeV)\\ \hline
       \chi_{X}&  0.0464           &A[1,1,4]&    645.64 \\ \chi_{Z}&
    0.0148
                &B[6,2,{5/3}]&            0.3633\\
           h_{11}/10^{-6}&  0.0212         &C[8,2,1]&{     35.72,    325.28,    339.02 }\\
           h_{22}/10^{-4}&  0.0344    &D[3,2,{7/ 3}]&{     35.03,    349.70,    362.74 }\\
                   h_{33}&  0.0026     &E[3,2,{1/3}]&{      0.58,     26.33,     26.33 }\\
 f_{11}/10^{-6}&
  0.0781-  0.1368i
                      &&{    28.661,    393.61,    441.49 }\\
 f_{12}/10^{-6}&
 -1.9955-  0.0830i
          &F[1,1,2]&      6.15,      6.15
 \\f_{13}/10^{-5}&
  0.0580+  0.0517i
                  &&     25.31,    325.28  \\
 f_{22}/10^{-5}&
  6.6036-  4.9627i
              &G[1,1,0]&{     0.091,      0.72,      0.72 }\\
 f_{23}/10^{-4}&
  2.0080+  2.3459i
                      &&{     0.718,     30.69,     30.92 }\\
 f_{33}/10^{-3}&
 -1.0051+  0.4427i
              &h[1,2,1]&{     1.437,     20.71,     34.27 }\\
 g_{12}/10^{-4}&
  0.0605+  0.1232i
                 &&{    541.46,    563.22 }\\
 g_{13}/10^{-5}&
 -0.0460+  1.8407i
     &I[3,1,{10/3}]&      1.26\\
 g_{23}/10^{-4}&
  6.3251+  5.7460i
          &J[3,1,{4/3}]&{     1.387,     14.31,     14.31 }\\
 \lambda/10^{-2}&
 -0.8601-  1.4974i
                 &&{     44.05,    383.45 }\\
 \eta&
-10.3248+  2.5325i
   &K[3,1, {8/ 3}]&{     50.91,    468.83 }\\
 \rho&
  0.7042-  2.2528i
    &L[6,1,{2/ 3}]&{     24.18,    752.08 }\\
 k&
  0.0151-  0.0805i
     &M[6,1,{8/ 3}]&    761.55\\
 \zeta&
  1.6200+  0.5400i
     &N[6,1,{4/ 3}]&    757.79\\
 \bar\zeta &
  1.0084+  0.4594i
          &O[1,3,2]&   1454.74\\
       m/10^{16}GeV&    0.04    &P[3,3,{2/ 3}]&{     14.50,   1130.98 }\\
          m_o/10^{16}GeV& -21.047e^{-iArg(\lambda)}     &Q[8,3,0]&     1.041\\
             \gamma&    3.71        &R[8,1, 0]&{      0.40,      1.55 }\\
              \bar\gamma& -2.8691     &S[1,3,0]&    1.7528\\
 x&
  0.9397+  0.6629i
         &t[3,1,{2/ 3}]&{      1.15,     19.66,     47.99,     78.65        }\\\Delta_X&      1.16 &&{    252.51,    337.25,   7050.35 }\\
              \Delta_{G}&   4.863           &U[3,3,{4/3}]&     1.480\\
 \Delta\alpha_{3}(M_{Z})&  -0.013               &V[1,2,3]&     1.046\\
    \{M^{\nu^c}/10^{11}GeV\}&{    0.01,   14.65,  613.85    }&W[6,3,{2/ 3}]&            877.20  \\
 \{M^{\nu}_{ II}/10^{-12}eV\}&  0.4997,    927.38,          38856.88               &X[3,2,{5/ 3}]&     0.353,    28.201,    28.201\\
                  M_\nu(meV)&{    2.17,    7.63,   42.34    }&Y[6,2, {1/3}]&              0.44  \\
  \{\rm{Evals[f]}\}/ 10^{-7}&{    0.15,  282.68,11844.25         }&Z[8,1,2]&              1.54  \\
 \hline
 \mbox{Soft parameters}&{\rm m_{\frac{1}{2}}}=
           -88.707
 &{\rm m_{0}}=
          4198.698
 &{\rm A_{0}}=
         -1.1832 \times 10^{   5}
 \\
 \mbox{at $M_{X}$}&\mu=
          9.4206 \times 10^{   4}
 &{\rm B}=
         -5.9399 \times 10^{   9}
  &{\rm tan{\beta}}=           50.0000\\
 &{\rm M^2_{\bar H}}=
         -7.1782 \times 10^{   9}
 &{\rm M^2_{  H} }=
         -6.7789 \times 10^{   9}
 &
 {\rm R_{\frac{b\tau}{s\mu}}}=
  2.6935
  \\
 Max(|L_{ABCD}|,|R_{ABCD}|)&
          6.1149 \times 10^{ -23}
  {\,\rm{GeV^{-1}}}&& \\
 \hline\end{array}
  $$
  \caption{\small{I-1-a : Column 1 contains values   of the NMSGUT-SUGRY-NUHM  parameters at $M_X$
  derived from an  accurate fit to all 18 fermion data and compatible with RG constraints.
 Unificaton parameters and mass spectrum of superheavy and superlight fields are  also given.
 The values of $\mu(M_X),B(M_X)$ are determined by RG evolution from $M_Z$ to $M_X$
 of the values determined by the EWRSB conditions.}}\label{I-1-a}\end{table}
 \begin{table}
 $$
 \begin{array}{|c|c|c|c|c|}
 \hline
 &&&&\\
 {\rm  Parameter }&Target =\bar O_i &Uncert.= \delta_i    &Achieved= O_i &Pull =(O_i-\bar O_i)/\delta_i\\
 \hline
    y_u/10^{-6}&  2.098620&  0.801673&  2.098242& -0.000472\\
    y_c/10^{-3}&  1.022984&  0.168792&  1.023751&  0.004546\\
            y_t&  0.383855&  0.015354&  0.383875&  0.001336\\
    y_d/10^{-5}&  6.826388&  3.979784&  6.843968&  0.004417\\
    y_s/10^{-3}&  1.294453&  0.610982&  1.301389&  0.011351\\
            y_b&  0.460136&  0.238811&  0.460919&  0.003278\\
    y_e/10^{-4}&  1.201757&  0.180264&  1.202499&  0.004117\\
  y_\mu/10^{-2}&  2.464838&  0.369726&  2.468141&  0.008935\\
         y_\tau&  0.527874&  0.100296&  0.523894& -0.039688\\
             \sin\theta^q_{12}&    0.2210&  0.001600&    0.2210&           -0.0005\\
     \sin\theta^q_{13}/10^{-4}&   29.1027&  5.000000&   29.1272&            0.0049\\
     \sin\theta^q_{23}/10^{-3}&   34.2424&  1.300000&   34.2402&           -0.0017\\
                      \delta^q&   60.0207& 14.000000&   60.0550&            0.0024\\
    (m^2_{12})/10^{-5}(eV)^{2}&    5.3580&  0.567948&    5.3569&           -0.0019\\
    (m^2_{23})/10^{-3}(eV)^{2}&    1.7330&  0.346597&    1.7347&            0.0050\\
           \sin^2\theta^L_{12}&    0.2887&  0.057748&    0.2883&           -0.0080\\
           \sin^2\theta^L_{23}&    0.4620&  0.138613&    0.4639&            0.0132\\
       \theta^L_{13}(degrees) &3.7&          3.7&              4.57                    &\\
 \hline
  Eigenvalues(\Delta_{\bar u})&   0.048796&   0.048800&   0.048805&\\
  Eigenvalues(\Delta_{\bar d})&   0.046619&   0.046623&   0.046629&\\
Eigenvalues(\Delta_{\bar \nu})&   0.052548&   0.052552&   0.052558&\\
  Eigenvalues(\Delta_{\bar e})&   0.059078&   0.059082&   0.059087&\\
       Eigenvalues(\Delta_{Q})&   0.046281&   0.046284&   0.046289&\\
       Eigenvalues(\Delta_{L})&   0.054386&   0.054390&   0.054394&\\
    \Delta_{\bar H},\Delta_{H}&       71.935369   &       62.415435    &{}&\\
 \hline
 \alpha_1 &
  0.8302+  0.0000i
 & {\bar \alpha}_1 &
  0.8899+  0.0000i
 &\\
 \alpha_2&
  0.0644+  0.0157i
 & {\bar \alpha}_2 &
  0.0516+  0.0644i
 &\\
 \alpha_3 &
 -0.0425-  0.0442i
 & {\bar \alpha}_3 &
 -0.0571-  0.0204i
 &\\
 \alpha_4 &
 -0.3970+  0.1618i
 & {\bar \alpha}_4 &
  0.3186-  0.0237i
 &\\
 \alpha_5 &
  0.1534+  0.0818i
 & {\bar \alpha}_5 &
  0.0709+  0.0141i
 &\\
 \alpha_6 &
  0.1116-  0.2759i
 & {\bar \alpha}_6 &
  0.1187-  0.2760i
 &\\
  \hline
 \end{array}
 $$
 \caption{\small{I-1-b: Fit   with $\chi_X=\sqrt{ \sum_{i=1}^{17}
 (O_i-\bar O_i)^2/\delta_i^2}=
    0.0464
 $. Target values,  at $M_X$ of the fermion yukawa
 couplings and mixing parameters, together with the estimated uncertainties, achieved values and pulls.
 The eigenvalues of the wavefunction renormalization increment  matrices $\Delta_i$ for fermion lines and
 the factors for Higgs lines are given, assuming the external Higgs is 10-plet
 dominated.  Note the close similarity of the eigenvalues which
 suggests that the small values of the SO(10) yukwas utilized when threshold corrections are in play  lead
 to gauge dominated corrections which are the same for all three
 generations.
 The Higgs fractions $\alpha_i,{\bar{\alpha_i}}$ which control the MSSM fermion yukawa couplings  are also
 given. Notice the dominance of the first components $\alpha_1,{\bar{\alpha_1}}$
 consistently with  the assumption made. Right handed neutrino threshold  effects   have been ignored.
  We have truncated numbers for display although all calculations are done at double
 precision.}}
 \label{I-1-b} \end{table}
 \begin{table}
 $$
 \begin{array}{|c|c|c|c|}
 \hline &&&\\ {\rm  Parameter }&SM(M_Z) & m^{GUT}(M_Z) & m^{MSSM}=(m+\Delta m)^{GUT}(M_Z) \\
 \hline
    m_d/10^{-3}&   2.90000&   0.63332&   2.92341\\
    m_s/10^{-3}&  55.00000&  12.04268&  55.60296\\
            m_b&   3.00000&   3.07778&   3.00662\\
    m_e/10^{-3}&   0.48657&   0.47249&   0.48691\\
         m_\mu &   0.10272&   0.09694&   0.10270\\
         m_\tau&   1.74624&   1.73514&   1.73675\\
    m_u/10^{-3}&   1.27000&   1.09930&   1.27007\\
            m_c&   0.61900&   0.53635&   0.61968\\
            m_t& 172.50000& 149.05874& 172.47396\\
 \hline
 \end{array}
 $$

 \caption{\small{I-1-c: Values of standard model
 fermion masses in GeV at $M_Z$ compared with the masses obtained from
 values of GUT derived  yukawa couplings  run down from $M_X^0$ to
 $M_Z$  both before and after threshold corrections.
  Fit with $\chi_Z=\sqrt{ \sum_{i=1}^{9} (m_i^{MSSM}- m_i^{SM})^2/ (m_i^{MSSM})^2} =
0.0148$.}}
  \label{I-1-c}\end{table}
  \begin{table}
 $$
 \begin{array}{|c|c|c|c|}
 \hline
 {\rm  Parameter}  &Value&  {\rm  Parameter}&Value \\
 \hline
                       M_{1}&            124.94&   M_{{\tilde {\bar {u}}_1}}&           5392.74\\
                       M_{2}&            328.39&   M_{{\tilde {\bar {u}}_2}}&           5392.01\\
                       M_{3}&            569.92&   M_{{\tilde {\bar {u}}_3}}&          17882.28\\
     M_{{\tilde {\bar l}_1}}&           1101.73&               A^{0(l)}_{11}&         -75168.69\\
     M_{{\tilde {\bar l}_2}}&            165.22&               A^{0(l)}_{22}&         -75083.12\\
     M_{{\tilde {\bar l}_3}}&          11100.23&               A^{0(l)}_{33}&         -47868.83\\
        M_{{\tilde {L}_{1}}}&           6400.13&               A^{0(u)}_{11}&         -86227.11\\
        M_{{\tilde {L}_{2}}}&           6353.87&               A^{0(u)}_{22}&         -86226.53\\
        M_{{\tilde {L}_{3}}}&          10210.78&               A^{0(u)}_{33}&         -43721.21\\
     M_{{\tilde {\bar d}_1}}&           2917.32&               A^{0(d)}_{11}&         -75501.89\\
     M_{{\tilde {\bar d}_2}}&           2916.57&               A^{0(d)}_{22}&         -75501.25\\
     M_{{\tilde {\bar d}_3}}&          26552.10&               A^{0(d)}_{33}&         -32031.46\\
          M_{{\tilde {Q}_1}}&           4928.60&                   \tan\beta&             50.00\\
          M_{{\tilde {Q}_2}}&           4927.98&                    \mu(M_Z)&          76666.76\\
          M_{{\tilde {Q}_3}}&          22679.30&                      B(M_Z)&
          9.6672 \times 10^{   8}
 \\
 M_{\bar {H}}^2&
         -6.0301 \times 10^{   9}
 &M_{H}^2&
         -6.3249 \times 10^{   9}
 \\
 \hline
 \end{array}
 $$
  \caption{ \small {I-1-d: Values (GeV) in  of the soft Susy parameters  at $M_Z$
 (evolved from the soft SUGRY-NUHM parameters at $M_X$).
 The  values of soft Susy parameters  at $M_Z$
 determine the Susy threshold corrections to the fermion yukawas.
 The matching of run down fermion yukawas in the MSSM to the SM   parameters
 determines  soft SUGRY parameters at $M_X$. Note the  heavier third
 sgeneration.  The values of $\mu(M_Z)$ and the corresponding soft
 susy parameter $B(M_Z)=m_A^2 {\sin 2 \beta }/2$ are determined by
 imposing electroweak symmetry breaking conditions. $m_A$ is the
 mass of the CP odd scalar in the in the Doublet Higgs. The sign of
 $\mu$ is assumed positive. }}
 \label{I-1-d}\end{table}

 \begin{table}
 $$
 \begin{array}{|c|c|}
 \hline {\mbox {Field } }&Mass(GeV)\\
 \hline
                M_{\tilde{G}}&            569.92\\
               M_{\chi^{\pm}}&            328.38,          76666.85\\
       M_{\chi^{0}}&            124.94,            328.38,          76666.81    ,          76666.82\\
              M_{\tilde{\nu}}&          6399.784,          6353.520,         10210.565\\
                M_{\tilde{e}}&           1102.65,           6400.31,            160.56   ,           6354.33,          10105.12,          11196.70  \\
                M_{\tilde{u}}&           4928.29,           5392.61,           4927.63   ,           5391.92,          17876.33,          22684.87  \\
                M_{\tilde{d}}&           2917.43,           4928.98,           2916.66   ,           4928.38,          22663.07,          26566.04  \\
                        M_{A}&         219898.08\\
                  M_{H^{\pm}}&         219898.09\\
                    M_{H^{0}}&         219898.06\\
                    M_{h^{0}}&            111.45\\
 \hline
 \end{array}
  $$
\caption{\small{I-1-e: Spectra of supersymmetric partners calculated ignoring generation mixing effects.
 Inclusion of such effects   changes the spectra only marginally. Due to the large
 values of $\mu,B,A_0$. The LSP and light chargino are  essentially pure Bino and Wino($\tilde W_\pm $).
   The light  gauginos and  light Higgs  $h^0$, are accompanied by a light smuon and  sometimes  selectron.
 The rest of the sfermions have multi-TeV masses. The mini-split supersymmetry spectrum and
 large $\mu,A_0$ parameters help avoid problems with FCNC and CCB/UFB instability\cite{kuslangseg}.
 The sfermion masses  are ordered by generation not magnitude. This is useful in understanding the spectrum
  calculated including generation mixing effects. The mass of the Higgs particles($M_A,M_{h^0},M_{H^\pm}$)  are calculated by
  incorporating one loop contributions to the Electroweak  symmetry breaking i.e  to the effective potential\cite{loophiggs,porod,nmsgutII}.
  }} \label{I-1-e}\end{table}

 \begin{table}
 $$
\begin{array}{|c|c|}
 \hline {\mbox {Field } }&Mass(GeV)\\
 \hline
                M_{\tilde{G}}&            570.03\\
               M_{\chi^{\pm}}&            328.43,          76664.05\\
       M_{\chi^{0}}&            124.95,            328.43,          76664.02    ,          76664.02\\
              M_{\tilde{\nu}}&           6353.56,           6399.82,         10211.527\\
                M_{\tilde{e}}&            133.31,           1102.50,           6353.86   ,           6400.35,          10109.89,          11195.65  \\
                M_{\tilde{u}}&           4926.11,           4928.47,           5392.09   ,           5392.79,          17873.85,          22683.48  \\
                M_{\tilde{d}}&           2916.90,           2917.69,           4926.85   ,           4929.16,          22661.71,          26564.36  \\
                        M_{A}&         219973.32\\
                  M_{H^{\pm}}&         219973.33\\
                    M_{H^{0}}&         219973.30\\
                    M_{h^{0}}&            111.74\\
 \hline
 \end{array}
 $$
 \caption{\small{I-1-f : Spectra of supersymmetric partners calculated including  generation mixing effects.
 Inclusion of such effects   changes the spectra only marginally. Due to the large
 values of $\mu,B,A_0$ the  LSP and light charginos are  essentially pure Bino and Wino($\tilde W_\pm $).
  Note that the ordering of the eigenvalues in this table follows their magnitudes,
 comparison with the previous table is necessary to identify the sfermions}}\label{I-1-f}\end{table}

 \begin{table}
 $$
 \begin{array}{|c|c|c|c|}
 \hline
 {\rm Parameter }&Value &{\rm  Field }&\hspace{10mm} Masses\\
 &&{\rm}[SU(3),SU(2),Y]&\hspace{10mm}( Units\,\,of 10^{16} GeV)\\ \hline
       \chi_{X}&  0.0133           &A[1,1,4]&    622.89 \\ \chi_{Z}&
    0.0386
                &B[6,2,{5/3}]&            0.4399\\
           h_{11}/10^{-6}&  0.0224         &C[8,2,1]&{     34.59,    332.64,    344.97 }\\
           h_{22}/10^{-4}&  0.0283    &D[3,2,{7/ 3}]&{     33.75,    322.29,    340.48 }\\
                   h_{33}&  0.0026     &E[3,2,{1/3}]&{      0.70,     26.06,     26.06 }\\
 f_{11}/10^{-6}&
  0.0782-  0.1347i
                      &&{    27.899,    377.48,    429.89 }\\
\hline\end{array}
 $$
 \caption{\small{ II-1-a: Column 1 contains values   of the NMSGUT-SUGRY-NUHM  parameters at $M_X$
  derived from an  accurate fit to all 18 fermion data and compatible with RG constraints.
  See caption to Table 2 for explanation.}} \label{II-1-a}\end{table}
 \begin{table}
 $$
 \begin{array}{|c|c|c|c|c|}
 \hline
 &&&&\\
 {\rm  Parameter }&Target =\bar O_i &Uncert.= \delta_i    &Achieved= O_i &Pull =(O_i-\bar O_i)/\delta_i\\
 \hline
    y_u/10^{-6}&  2.141027&  0.817872&  2.141050&  0.000028\\
    y_c/10^{-3}&  1.043728&  0.172215&  1.043795&  0.000391\\
            y_t&  0.386577&  0.015463&  0.386580&  0.000221\\
    y_d/10^{-5}&  6.891765&  4.017899&  6.919378&  0.006873\\
    y_s/10^{-3}&  1.311160&  0.618867&  1.311543&  0.000620\\
            y_b&  0.451549&  0.234354&  0.453520&  0.008411\\
\hline
 \end{array}
 $$
  \caption{\small{II-1-b: Fit   with $\chi_X=\sqrt{ \sum_{i=1}^{17}
 (O_i-\bar O_i)^2/\delta_i^2}= 0.0133$. Target values,  at $M_X$ of the fermion yukawa
 couplings and mixing parameters, together with the estimated uncertainties, achieved values and pulls.
 See caption to Table 3 for explanation.  }}
 \label{II-1-b}\end{table}
 \begin{table}
 $$
 \begin{array}{|c|c|c|c|}
 \hline &&&\\ {\rm  Parameter }&SM(M_Z) & m^{GUT}(M_Z) & m^{MSSM}=(m+\Delta m)^{GUT}(M_Z) \\
 \hline
    m_d/10^{-3}&   2.90000&   0.64294&   2.90511\\
    m_s/10^{-3}&  55.00000&  12.18662&  54.90038\\
            m_b&   3.00000&   3.05467&   3.02134\\
    m_e/10^{-3}&   0.48657&   0.48198&   0.48770\\
         m_\mu &   0.10272&   0.10177&   0.10289\\
         m_\tau&   1.74624&   1.74369&   1.74266\\
    m_u/10^{-3}&   1.27000&   1.11799&   1.23689\\
            m_c&   0.61900&   0.54503&   0.60300\\
            m_t& 172.50000& 149.51960& 172.26015\\
 \hline
 \end{array}
 $$

 \caption{\small{II-1-c: Values of standard model
 fermion masses in GeV at $M_Z$ compared with the masses obtained from
 values of GUT derived  yukawa couplings  run down from $M_X^0$ to
 $M_Z$  both before and after threshold corrections.
  Fit with $\chi_Z=\sqrt{ \sum_{i=1}^{9} (m_i^{MSSM}- m_i^{SM})^2/ (m_i^{MSSM})^2} =
0.0386$.}}
  \label{II-1-c }\end{table}

\begin{table}
 $$
 \begin{array}{|c|c|c|c|}
 \hline
 {\rm  Parameter}  &Value&  {\rm  Parameter}&Value \\
 \hline
                       M_{1}&             35.00&   M_{{\tilde {\bar {u}}_1}}&           2490.81\\
                       M_{2}&            113.01&   M_{{\tilde {\bar {u}}_2}}&           2490.90\\
                       M_{3}&             83.09&   M_{{\tilde {\bar {u}}_3}}&          16928.60\\
     M_{{\tilde {\bar l}_1}}&           1045.62&               A^{0(l)}_{11}&         -39522.44\\
     M_{{\tilde {\bar l}_2}}&           1148.28&               A^{0(l)}_{22}&         -39473.12\\
     M_{{\tilde {\bar l}_3}}&          14376.75&               A^{0(l)}_{33}&         -25104.91\\
        M_{{\tilde {L}_{1}}}&           3699.46&               A^{0(u)}_{11}&         -44729.39\\
        M_{{\tilde {L}_{2}}}&           3714.57&               A^{0(u)}_{22}&         -44729.08\\
        M_{{\tilde {L}_{3}}}&          10786.67&               A^{0(u)}_{33}&         -22492.46\\
     M_{{\tilde {\bar d}_1}}&            320.40&               A^{0(d)}_{11}&         -39448.33\\
     M_{{\tilde {\bar d}_2}}&            323.59&               A^{0(d)}_{22}&         -39447.98\\
     M_{{\tilde {\bar d}_3}}&          19724.24&               A^{0(d)}_{33}&         -16948.89\\
          M_{{\tilde {Q}_1}}&           2422.85&                   \tan\beta&             50.00\\
          M_{{\tilde {Q}_2}}&           2423.11&                    \mu(M_Z)&          48104.14\\
          M_{{\tilde {Q}_3}}&          18383.64&                      B(M_Z)&
          3.0799 \times 10^{   8}
 \\
 M_{\bar {H}}^2&
         -2.3660 \times 10^{   9}
 &M_{H}^2&
         -2.5187 \times 10^{   9}
 \\
 \hline
 \end{array}
 $$
 \caption{ \small {II-1-d: Values (GeV) in  of the soft Susy parameters  at $M_Z$
 (evolved from the soft SUGRY-NUHM parameters at $M_X$).
    See caption to Table 5 for explanation.}}
  \label{II-1-d}\end{table}

 \begin{table}
 $$
 \begin{array}{|c|c|}
 \hline {\mbox {Field } }&Mass(GeV)\\
 \hline
                M_{\tilde{G}}&             83.09\\
               M_{\chi^{\pm}}&            113.01,          48104.28\\
       M_{\chi^{0}}&             35.00,            113.01,          48104.23    ,          48104.24\\
              M_{\tilde{\nu}}&          3698.868,          3713.977,         10786.469\\
                M_{\tilde{e}}&           1046.59,           3699.78,           1147.01   ,           3715.55,          10777.59,          14383.72  \\
                M_{\tilde{u}}&           2422.23,           2490.54,           2422.10   ,           2491.00,          16922.19,          18390.64  \\
                M_{\tilde{d}}&            321.46,           2423.62,            324.40   ,           2423.91,          18355.16,          19750.86  \\
                        M_{A}&         124120.40\\
                  M_{H^{\pm}}&         124120.43\\
                    M_{H^{0}}&         124120.39\\
                    M_{h^{0}}&            130.18\\
 \hline
 \end{array}
 $$
\caption{\small{II-1-e: Spectra of supersymmetric partners calculated ignoring generation mixing effects.
  See caption to Table 6 for explanation.
  }} \label{II-1-e}\end{table}

 \begin{table}
 $$
 \begin{array}{|c|c|}
 \hline {\mbox {Field } }&Mass(GeV)\\
 \hline
                M_{\tilde{G}}&             83.15\\
               M_{\chi^{\pm}}&            113.03,          48102.54\\
       M_{\chi^{0}}&             35.01,            113.03,          48102.49    ,          48102.50\\
              M_{\tilde{\nu}}&           3698.88,           3713.99,         10786.642\\
                M_{\tilde{e}}&           1046.52,           1144.84,           3699.80   ,           3715.67,          10778.23,          14383.76  \\
                M_{\tilde{u}}&           2422.27,           2438.70,           2490.58   ,           2491.10,          16922.33,          18388.56  \\
                M_{\tilde{d}}&            321.58,            324.46,           2423.67   ,           2440.53,          18353.16,          19750.82  \\
                        M_{A}&         124163.88\\
                  M_{H^{\pm}}&         124163.90\\
                    M_{H^{0}}&         124163.87\\
                    M_{h^{0}}&            130.20\\
 \hline
 \end{array}
 $$
 \caption{\small{II-1-f: Spectra of supersymmetric partners calculated including  generation mixing effects.
   See caption to Table 7 for explanation.}}\label{II-1-f}\end{table}
\clearpage

 \begin{table}
 $$
 \begin{array}{|c|c|c|c|}
 \hline
 {\rm Parameter }&Value &{\rm  Field }&\hspace{10mm} Masses\\
 &&{\rm}[SU(3),SU(2),Y]&\hspace{10mm}( Units\,\,of 10^{16} GeV)\\ \hline
       \chi_{X}&  0.0038           &A[1,1,4]&    756.12 \\ \chi_{Z}&
    0.0030
                &B[6,2,{5/3}]&            0.4240\\
           h_{11}/10^{-6}&  0.0211         &C[8,2,1]&{     39.13,    387.00,    401.03 }\\
           h_{22}/10^{-4}&  0.0329    &D[3,2,{7/ 3}]&{     38.83,    385.83,    402.57 }\\
                   h_{33}&  0.0026     &E[3,2,{1/3}]&{      0.68,     29.71,     29.71 }\\
 f_{11}/10^{-6}&
  0.0774-  0.1363i
                      &&{    31.591,    443.85,    505.56 }\\
\hline\end{array}
 $$
  \caption{\small{III-1-a : Column 1 contains values   of the NMSGUT-SUGRY-NUHM  parameters at $M_X$
  derived from an  accurate fit to all 18 fermion data and compatible with RG constraints.
   See caption to Table 1 for explanation.}}\label{III-1-a}\end{table}
 \begin{table}
 $$
 \begin{array}{|c|c|c|c|c|}
 \hline
 &&&&\\
 {\rm  Parameter }&Target =\bar O_i &Uncert.= \delta_i    &Achieved= O_i &Pull =(O_i-\bar O_i)/\delta_i\\
 \hline
    y_u/10^{-6}&  2.116326&  0.808437&  2.116304& -0.000027\\
    y_c/10^{-3}&  1.031618&  0.170217&  1.031400& -0.001282\\
            y_t&  0.386527&  0.015461&  0.386531&  0.000285\\
    y_d/10^{-5}&  6.872020&  4.006388&  6.880189&  0.002039\\
    y_s/10^{-3}&  1.303124&  0.615075&  1.301971& -0.001874\\
            y_b&  0.468198&  0.242995&  0.468608&  0.001688\\
\hline
 \end{array}
 $$
  \caption{\small{III-1-b: Fit   with $\chi_X=\sqrt{ \sum_{i=1}^{17}
 (O_i-\bar O_i)^2/\delta_i^2}=
    0.0038
 $. Target values,  at $M_X$ of the fermion yukawa
 couplings and mixing parameters, together with the estimated uncertainties, achieved values and pulls.
  See caption to Table 2 for explanation. }}
\label{III-1-b} \end{table}
 \begin{table}
 $$
 \begin{array}{|c|c|c|c|}
 \hline &&&\\ {\rm  Parameter }&SM(M_Z) & m^{GUT}(M_Z) & m^{MSSM}=(m+\Delta m)^{GUT}(M_Z) \\
 \hline
    m_d/10^{-3}&   2.90000&   0.63252&   2.90621\\
    m_s/10^{-3}&  55.00000&  11.96937&  55.00879\\
            m_b&   3.00000&   3.09207&   3.00111\\
    m_e/10^{-3}&   0.48657&   0.47228&   0.48641\\
         m_\mu &   0.10272&   0.09698&   0.10252\\
         m_\tau&   1.74624&   1.74550&   1.74567\\
    m_u/10^{-3}&   1.27000&   1.10652&   1.27009\\
            m_c&   0.61900&   0.53927&   0.61899\\
            m_t& 172.50000& 149.30372& 172.48084\\
 \hline
 \end{array}
 $$

 \caption{\small{III-1-c: Values of standard model
 fermion masses in GeV at $M_Z$ compared with the masses obtained from
 values of GUT derived  yukawa couplings  run down from $M_X^0$ to
 $M_Z$  both before and after threshold corrections.
  Fit with $\chi_Z=\sqrt{ \sum_{i=1}^{9} (m_i^{MSSM}- m_i^{SM})^2/ (m_i^{MSSM})^2} =
0.0030$.}}
  \label{III-1-c}\end{table}

 \begin{table}
 $$
 \begin{array}{|c|c|c|c|}
 \hline
 {\rm  Parameter}  &Value&  {\rm  Parameter}&Value \\
 \hline
                       M_{1}&            100.07&   M_{{\tilde {\bar {u}}_1}}&           4539.23\\
                       M_{2}&            270.77&   M_{{\tilde {\bar {u}}_2}}&           4538.61\\
                       M_{3}&            428.43&   M_{{\tilde {\bar {u}}_3}}&          17182.81\\
     M_{{\tilde {\bar l}_1}}&            906.34&               A^{0(l)}_{11}&         -65947.70\\
     M_{{\tilde {\bar l}_2}}&            142.19&               A^{0(l)}_{22}&         -65871.94\\
     M_{{\tilde {\bar l}_3}}&          11976.39&               A^{0(l)}_{33}&         -41513.30\\
        M_{{\tilde {L}_{1}}}&           5564.41&               A^{0(u)}_{11}&         -75901.65\\
        M_{{\tilde {L}_{2}}}&           5528.49&               A^{0(u)}_{22}&         -75901.14\\
        M_{{\tilde {L}_{3}}}&          10194.21&               A^{0(u)}_{33}&         -38216.99\\
     M_{{\tilde {\bar d}_1}}&           2291.12&               A^{0(d)}_{11}&         -66155.83\\
     M_{{\tilde {\bar d}_2}}&           2290.56&               A^{0(d)}_{22}&         -66155.26\\
     M_{{\tilde {\bar d}_3}}&          24659.75&               A^{0(d)}_{33}&         -27514.18\\
          M_{{\tilde {Q}_1}}&           4205.32&                   \tan\beta&             50.00\\
          M_{{\tilde {Q}_2}}&           4204.83&                    \mu(M_Z)&          68576.15\\
          M_{{\tilde {Q}_3}}&          21289.22&                      B(M_Z)&
          7.4868 \times 10^{   8}
 \\
 M_{\bar {H}}^2&
         -4.7966 \times 10^{   9}
 &M_{H}^2&
         -5.0786 \times 10^{   9}
 \\
 \hline
 \end{array}
 $$
  \caption{ \small {III-1-d: Values (GeV) in  of the soft Susy parameters  at $M_Z$
 (evolved from the soft SUGRY-NUHM parameters at $M_X$).
  See caption to Table 4 for explanation. }}
 \label{III-1-d}\end{table}

 \begin{table}
 $$
 \begin{array}{|c|c|}
 \hline {\mbox {Field } }&Mass(GeV)\\
 \hline
                M_{\tilde{G}}&            428.43\\
               M_{\chi^{\pm}}&            270.76,          68576.25\\
       M_{\chi^{0}}&            100.07,            270.76,          68576.21    ,          68576.22\\
              M_{\tilde{\nu}}&          5564.015,          5528.094,         10193.999\\
                M_{\tilde{e}}&            907.46,           5564.62,            136.00   ,           5529.05,          10149.71,          12014.32  \\
                M_{\tilde{u}}&           4204.96,           4539.08,           4204.40   ,           4538.53,          17176.99,          21294.85  \\
                M_{\tilde{d}}&           2291.27,           4205.76,           2290.68   ,           4205.29,          21272.06,          24674.64  \\
                        M_{A}&         193516.73\\
                  M_{H^{\pm}}&         193516.75\\
                    M_{H^{0}}&         193516.72\\
                    M_{h^{0}}&            117.18\\
 \hline
 \end{array}
 $$
\caption{\small{III-1-e: Spectra of supersymmetric partners calculated ignoring generation mixing effects.
  See caption to Table 6 for explanation.
  }} \label{III-1-e }\end{table}

 \begin{table}
 $$
 \begin{array}{|c|c|}
 \hline {\mbox {Field } }&Mass(GeV)\\
 \hline
                M_{\tilde{G}}&            428.53\\
               M_{\chi^{\pm}}&            270.80,          68573.70\\
       M_{\chi^{0}}&            100.09,            270.80,          68573.66    ,          68573.67\\
              M_{\tilde{\nu}}&           5528.13,           5564.05,         10194.739\\
                M_{\tilde{e}}&            105.36,            907.31,           5528.77   ,           5564.65,          10152.37,          12014.38  \\
                M_{\tilde{u}}&           4205.11,           4205.54,           4538.67   ,           4539.24,          17175.20,          21293.36  \\
                M_{\tilde{d}}&           2290.89,           2291.51,           4205.91   ,           4206.41,          21270.60,          24673.37  \\
                        M_{A}&         193585.06\\
                  M_{H^{\pm}}&         193585.07\\
                    M_{H^{0}}&         193585.04\\
                    M_{h^{0}}&            117.39\\
 \hline
 \end{array}
 $$
\caption{\small{III-1-f : Spectra of supersymmetric partners calculated including  generation mixing effects.
 See caption to Table 7 for explanation.}} \label{III-1-f}\end{table}
\clearpage

\section{Structural features and Phenomenology}

As mentioned the threshold corrections to the Higgs lines in a
light fermion- light higgs vertex can obtain a thick wave function
dressing from GUT scale particles leading to an amplification of
the effective SO(10) yukawa couplings and making much weaker
SO(10) couplings capable of fitting the fermion data. Moreover the
threshold corrections of the light fermion yukawa couplings are
highly nonlinear, and as such there is no reason to expect that
the constraint ${y_b-y_\tau}\simeq{y_s-y_\mu}$ found on the basis
of the tree level coupling of the \textbf{10,120}-plets to the
\textbf{16}-plets of fermions\cite{core,msgreb}continues to be effective.  The influence of
these large corrections  can be seen in the value of   $ \rbtau=
 \frac{y_b-y_\tau}{y_s-y_\mu}  $ given in the tables. We see that
 only in cases $I-2,III-2,III-4$ is $\rbtau$ still approximately
 unity while in the rest it is typically $\sim 3$:  which is the
 magnitude of the ratio when all threshold corrections to SM
 values are ignored and the couplings are run up using MSSM RGEs.
This  shows that the tree level constraint can be evaded;
and with profit since the magnitudes are no longer mismatched. Note
that the cases with $\rbtau\sim 1$ typically have
$\Delta_{H,{\overline {H}}}$ smaller by a factor of about 5 than
the cases where $\rbtau\sim 3$.

The ultra small values of the $\mathbf{\oot} $ couplings ensure
that they make little difference to the 2nd and 3rd generation
charged fermion yukawas but ensure that $ M_{\nu^c}$ are light and
thus Type I neutrino masses are viable\cite{blmdm}. Note that for
the same reason in all fits the Type II neutrino masses  are
completely negligible.  The other superpotential couplings are
unremarkable except that $\eta$ is somewhat large. However one
should recall   that it occurs in the superpotential divided by
$ 4!=24$. Actual coefficients of the radiative corrections will
typically be powers of $(\sim\frac{|\eta|}{4 \pi })$.

As explained in Section \textbf{1,2} the unification scale
$M_X=M_X^0 10^{\Delta_X}$, (typically $\sim  10^{17.5} $ GeV)   is
identified with the mass of the $X[3,2,\pm\frac{5}{3}]$ gauge
sub-multiplet ( exchange of which gives rise to d=6 operators for
B decay ) and determines the scale parameter $m$ of the
superpotential via \be m=\frac{|\lambda| 10^{16.25 + \Delta_X}}{
\tilde{M}_X} \, GeV\ee where \[ \tilde{M}_X=
g_r\sqrt{4|\tilde{a}+\tilde{\omega}|^2+2|\tilde{
p}+\tilde{\omega}|^2 }\] and \[ g_r=\sqrt{2 \pi ( 25.6 +
\Delta_G)^{-1}} \] is the corrected SO(10) gauge coupling. We see
that the unification scale is generally  elevated(
$10^{16.7}-10^{19}$ GeV) but the SO(10) coupling at unification is
still perturbative though sometimes only marginally so.

 The right handed neutrino masses are important for Leptogenesis
  and for lepton RGE flows  at intermediate scales. We find  $M_{\nu^c}$   is generically in the range
  $ 10^9-10^{13} $ GeV (with normal hierarchy);
     which is also  the preferred   range for Leptogenesis. It is
     determined by the   necessarily (for viable neutrino spectra)
   ultra small $\mathbf{\oot}$ couplings $f\sim
10^{-8}-10^{-3}$. It is interesting to note that the threshold
corrections may also weaken the influence of the   right handed
neutrino thresholds on yukawa unification.  The interplay of the
GUT scale corrections and the $M_{{\nu^c}}$ thresholds will be
interesting to evaluate, specially since the latter are
known\cite{nucandyukuni} to lead to tension  for yukawa
unification. It may be that as in the case of the tension
regarding the value of $m_b(M_Z)$ acceptable for yukawa
unification\cite{king}, which is relieved in our model  by the threshold corrections at
$M_S=M_Z$, so also GUT scale threshold corrections  may help with
relieving the effect of right handed neutrino thresholds.

The super heavy masses lie in the range    $10^{14}-10^{20}$  GeV
 with a few multiplets sometimes having an uncomfortably large
 mass even  greater than the Planck mass. Overall the unification
parameter and the spectra of the Fits with the large $ \Delta_{H,
\bar{H}}\sim 10^2 $ and $\rbtau \sim 3$ seem   more palatable than
those for  Case I-2  with $ \rbtau \sim 1 $, $\Delta_{H,
\bar{H}}\sim 10 $.

Let us turn next to the conjectured Soft Spectra determined by
requiring EWRSB at tan $ \beta \sim 50 $ and large threshold
corrections to lower $ y_{d,s}$ by a factor of 5 or so. The
required values of the $|\mu| , |A_0|$ parameters turn out to be
so huge $(\sim 10^2 $  TeV)  that they will incite controversy
driven by concerns regarding deep CCB minima. For the moment we
take the pragmatic attitude that we have checked the local
stability by ensuing the positivity of all scalar mass squared
parameters. The stability against tunneling to CCB minima on
cosmological time scales calls for further investigation after
loop corrections to scalar masses have been included. However the
literature\cite{kuslangseg,riotroulet} supports the pragmatic
attitude we adopt regarding the viability of metastable minima.
The seminal and clear investigations of\cite{kuslangseg} regarding
meta-stability in the parameter region of ultra large $\mu, A_0$
are so encouraging that we cannot resist quoting them verbatim.
Firstly they note that (our interpolations in square brackets)``
\emph{the height of the barrier separating the
[\textrm{metastable}] minimum from the CCB minimum is roughly
proportional to $1/y_{min}^2$, where $y_{min}$ is the smallest
Yukawa coupling associated with the fields that acquire non-zero
vev in the CCB minimum.  The corresponding tunnelling rates are
greatly suppressed for small $y$.} "  Thus since   it is only the
third generation of matter sfermion fields that have appreciable
yukawa couplings the violation of the CCB and UFB
bounds\cite{CCB,UFB} is not likely to be a matter of concern for
the first two generations. Moreover they note ``\emph{the most
stringent constraints come from the small ${\rm tan} \beta $
region, where the top Yukawa coupling is larger}". Whereas we are
in the large ${\rm tan} \beta \sim 50 $ regime.

\begin{figure}
\postscript{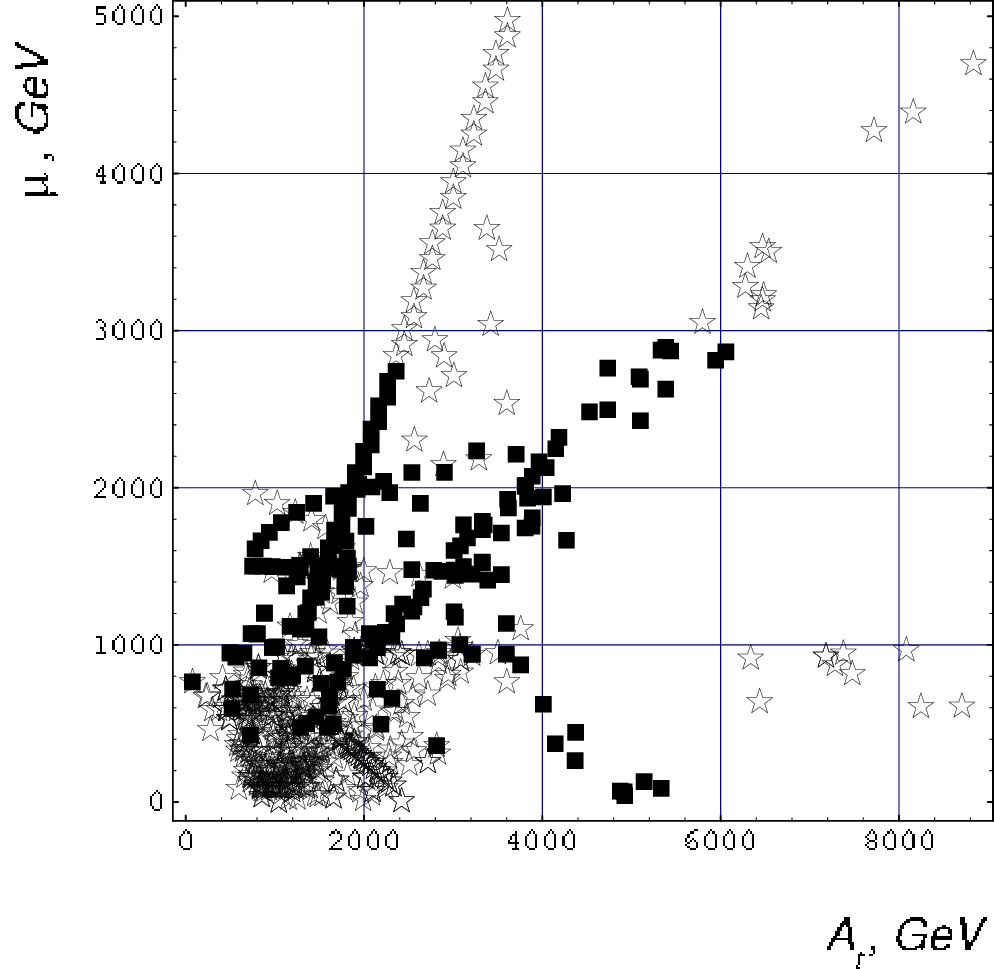}{0.5} \caption{``\emph{Tunneling
probability for unphysically large values of {$A_t$} and {$\mu$}.
As the CCB minimum moves farther away, it becomes ``less
dangerous''. As before, the stars mark the points with {$S>400$},
while the boxes depict those with {$S<400$}.}''.  The points
marked with stars correspond to MSSM ``standard/realistic vacua''
that are long lived on the scale of the age of the universe. From
\cite{kuslangseg} with permission. } \label{amufig}
\label{kusfig}\end{figure}

         The investigations of \cite{kuslangseg} focussed on the  region
         $|\mu|,|A_0|< 4$  TeV.  Their findings  confirmed that in this region
         ``\emph{the larger the trilinear coupling the more dangerous is the
         corresponding CCB minimum}".  However they had the
         prescience to realize that understanding the behaviour in
         the regions with much larger $|\mu|,| A_0|$ would
         illuminate the dynamics of CCB and tunnelling  and
         clarify  the operation of decoupling arguments which
         seem violated by the above tradecraft maxim  but are always
         crucial to establish an intuitive grip on field theoretic dynamics.
           Thus they note
``\emph{it is instructive to examine what happens to the
tunnelling probability in the limit of very large $\mu$ and $A_t$
(and large enough squark mass terms to ensure the existence of the
[\textrm{metastable}] minimum). In that limit, as the CCB minimum
moves away from the SML [\textrm{standard}] minimum, the barrier
separating the two becomes thicker, and the false vacuum should
become more stable. This is, in fact, what happens....  As
expected, the tunnelling probability diminishes for very large
values of $A_t$ and $\mu$, and $m_{\tilde{t}_{_L}}$ and
$m_{\tilde{t}_{_R}}$. To summarize, if the global CCB minimum is
nearly degenerate with the local SML minimum (thin-wall limit),
then the tunnelling probability is extremely small.  As the
trilinear couplings increase, the false vacuum decay rate
increases because the escape point of the bounce moves out of the
flat vicinity of the global minimum into the region in which the
gradient of the potential is significant. \textbf{However, a
further increase in the size of the trilinear couplings, as well
as the consequent increase in the squark mass terms, makes the
barrier thicker and pushes the escape point away from the
[\textrm{metastable}] minimum.  This eventually causes a decrease
in the tunneling rate. In accordance with one's intuition, the
low-energy physics is unaffected by the physics at the very high
energy scales.} }"  These effects are clearly visible in the Fig.
\ref{kusfig} and indicate that $|\mu|,|A_0|>10 $ TeV  should be
utterly safe as regards CCB/UFB issues.

The above  arguments justify our retention of the parameter sets
which we have found and which prima facie suffer from   CCB
problems. Our very large values of $\tan\beta, |\mu|,|A_0|$ and
$B$ militate for metastability on cosmic time scales. However for
the dubious we note two further points :  we have not used the
full freedom in to choose trilinears  at $M_X$, contenting
ourselves with a single $A_0(M_X)$ parameter. The requirement of
viable metastability could itself be incorporated as a search
criteria in future versions of our search routines. In sum it
would be premature to dismiss the novel   scenario emerging from
an apparently well motivated fundamental theory on the basis of
technical grounds of CCB instability whose very efficacy in
choosing vacua has long been in doubt. We note that recently
arguments have been advanced that cast doubt on even the very
possibility of tunnelling from a Poincare invariant vacuum to a
lower energy state\cite{gia}. Should it be the case that a
phenomenology package assumes some version of CCB constraints
without calculating metastability then it is the package that must
be improved not necessarily  our proposed parameters that need to
be considered as discredited.

 The mass spectra obey a normal s-hierarchy  (third
sgeneration heavier than degenerate first two generations) coupled
with a mini split supersymmetry ($   m_{\tilde{f}}\gg M_i$) with
pure Bino LSP, Wino(${\tilde{W_\pm}}$) light charginos, and next
to  lightest neutralino(${\tilde{W_3}}$), and Higgsino heavy
neutralinos and chargino. What is  remarkable and interesting from
the point of view of the Dark Matter Cosmology is that the quasi
inert Bino LSP,  which could serve as an ideal form of Cold Dark
Matter, is generally accompanied by a light Right sfermion
 of the first or second generation (often a smuon).  This is in sharp contrast to most
Susy GUT spectra which predict the stau, stop and sbottom    as
the lightest sfermions because their masses are driven to lower values by the effect of the
their large yukawas. Here however the additional presence of large negative Higgs mass terms
  drives the the third sgeneration to large masses (see Appendix \textbf{B}).
 Thus our model is marked out from other GUT models by a distinctive low energy spectrum that puts it
 in a different  and novel universality class of models.
  The LHC at 14 TeV will provide about 1-2
TeV per colliding parton and so, as a rule of thumb, particles
lighter than 2.5 TeV and with reasonably large couplings to SM
particles may be detectable. Besides the light Bino which is very
weakly coupled and the light Higgs $h^0$, the other MSSM gauginos
$ {\hat W}_a,{\tilde g}$, and from the sfermions some from among
${\tilde{\bar {\mu}}},{\tilde{\bar {e}}},{\tilde{\bar
{d}}},{\tilde{\bar {s}}}, {\tilde{\bar {u}}},{\tilde{\bar {c}}}$
are   lighter than 2 TeV  and should be discoverable at LHC. In
Table 20 we give the "detectable spectra" for each case.
\begin{table}
 $$
 \begin{array}{|c|c|c|c|c|c|}
 \hline
 {\rm Particle\Rightarrow }&LSP &Winos& Gluino&Higgs& Sfermions   \\ \hline
 {\rm Case \Downarrow }&(\tilde B ) &({\tilde W}^{\pm,0})&\tilde g  &h^0&{\tilde { \bar f}}, \tilde {F} \\ \hline
  I-1 &0.125&0.328&0.570&0.111&1.1({\tilde { \bar e}}),0.16({\tilde { \bar \mu}})\\
 I-2 &0.105&0.354&0.269&0.113&1.89({\tilde { \bar e}}),0.3({\tilde { \bar \mu}}),0.82({\tilde { \bar d}}),0.83({\tilde { \bar s}})  \\
I-3 &0.147&0.406&0.599&0.115&1.63 ({\tilde { \bar e}}),0.39 ({\tilde { \bar \mu}})   \\
I-4&0.104&0.302&0.351&0.128&0.12({\tilde { \bar e}}),0.89({\tilde { \bar \mu}}),1.7({\tilde { \bar d}}),1.7({\tilde { \bar s}})  \\

II-1 &0.035&0.113&0.83&0.130&1.05({\tilde { \bar e}}),1.14
({\tilde { \bar \mu}}),0.32({\tilde { \bar d}}),0.32({\tilde {
\bar s}}) ,2.49({\tilde { \bar u}}),2.49({\tilde { \bar c}}),2.49({\tilde  Q_{1,2} })\\
II-3 & .044& .144&0.09 & .124&2.15({\tilde { \bar
e}}),2.07({\tilde { \bar \mu}}),0.46({\tilde { \bar
d}}),0.46({\tilde { \bar s}}),0.20 ({\tilde { \bar u}}),0.20({\tilde { \bar c}}),2.45({\tilde  Q_{1,2}} )\\

II-4 &0.035&0.110&0.082&0.127&1.67({\tilde { \bar
e}}),1.68({\tilde { \bar \mu}}),0.364({\tilde { \bar
d}}),0.36 ({\tilde { \bar s}}),1.36({\tilde { \bar u}}),1.36({\tilde { \bar c}}),2.11({\tilde  Q_{1,2}} )\\

III-1 &0.100&0.271&0.429&0.117&0.105({\tilde { \bar
e}}),0.91({\tilde { \bar \mu}}),2.29 ({\tilde { \bar
d}}),2.29 ({\tilde { \bar s}}) \\

III-2 &0.99&0.342&0.232&0.113&1.87({\tilde { \bar e}}),0.34
({\tilde { \bar \mu}}),0.72({\tilde { \bar
d}}),0.71({\tilde { \bar s}}) \\

III-3 &0.98&0.273&0.387&0.118&1.06({\tilde { \bar e}}),0.14
({\tilde { \bar \mu}}),2.15({\tilde { \bar
d}}),2.15({\tilde { \bar s}}) \\

III-4 &0.94&0.284&0.269&0.128&.09({\tilde { \bar e}}),0.11({\tilde
{ \bar \mu}}),1.17({\tilde { \bar
d}}),1.17 ({\tilde { \bar s}}) \\

 \hline\end{array}
 $$
 \caption{\small{Table of nominally discoverable particles.
 Mass values (in TeV)  below 2.5 TeV, rounded off to two decimal places,
  calculated at tree level(except the Higgs which include one loop corrections) using two loop
 RGE equations \emph{including} generation mixing.
  The principal component of the mass  eigenstate is indicated in
  brackets after the mass value. The eigenstates are quite pure.}}
\label{LHCdetectables} \end{table}

The cosmology of the Bino LSP Dark matter would  be determined by
the co-annihilation through the sfermion channels (the
pseudo-scalar Higgs is unavailable for the purpose being very
heavy). The combination of a Bino LSP and very light sfermion is
thus   ideal   for Bino-LSP WIMP DM and it is interesting to note
that this feature emerged naturally from completely independent
considerations.   Note that-as is often the case - the smuon is
the lightest sfermion then  such annihilation could lead to a
excess of charged leptons relative to nucleons from DM decay. Such
signals have indeed been reported recently by a number of
experiments\cite{pamela,atic,fermilat,hess}. We will return to
detailed examination of the rates of the co-annihilation and
charged particle production therefrom elsewhere.

In our fits typical gluino masses lie in the $ 200- 600 $ GeV
range. It is interesting that the ratio of gaugino  masses can
diverge significantly from the $ M_1 : M_2 : M_3 :: 1:2:7$ ratio
 dictated by the 1-loop RG invariance of $M_i/g_i^2$  if one  begins from universal
gaugino masses at $M_X$ : as is simple and plausible in a GUT
context. However we find that the gluino can be quite light (  as in Fit I-2 where it is lighter than
even the Wino). Further phenomenological analysis to revise the
gluino mass bounds in the special soft Susy parameter region of
high $\tan\beta$, and multi TeV $\mu,B,A_0$,  which has emerged from this analysis as a distinct, novel   possibly viable region of soft Susy parameter space,  is required before the
viability of such fits can be decided.    This unusual feature like many
others in this scenario, is also due to the effects of the large
values of $A_0$ that we have been required to consider by the
 down and strange quark   fitting requirements of the NMSGUT. The large values
of $M_{H ,{\bar{H}}}^2, |A_0|$ required by the fermion fitting
 lead to correspondingly large values of required $|\mu(M_Z)| ,  B(M_Z)$
 (and therefore also of the additional pseudoscalar mass  $M_A$
 which emerges much to heavy to play any role in Dark Matter cosmology)   through   the EWRSB
conditions that tie them together.   The large(negative) values of
$M_{H ,{\bar{H}}}^2$ have a dramatic consequence : the one loop RG
equations for the evolution of sfermion masses contain terms
proportional to $M_{H}, M_{\bar{H}} ^2$ times the yukawa couplings
squared  $Y_f^\dagger Y_f$. These terms dominate the RGEs   for
the third generation sfermions and drive their masses far above
the masses of the first two s-generations independently of the
value of $A_0$. These  important features of the RGE are  shown
graphically in Appendix \textbf{B}.

 Baryon decay via $ d=5$ operators is, as usual\cite{lucasraby,gotonihei},
  dominated by the chargino mediated channels.
The heavy sfermions help with suppressing B-decay. The dominant
channels are  $ Baryon \rightarrow Meson  + neutrino $. We
emphasize that the  flavour violation required by $d=5$ B
violation is supplied  entirely by the rundown values of the (off
diagonal) SuperCKM values determined by the fitting of the fermion
yukawas at $M_X$ by the SO(10)   light fermion yukawa
formulae\cite{ag1,ag2,abmsv,blmdm,nmsgutI}.   Using the formulae
given in \cite{nmsgutI}  and adapting the formalism
of\cite{lucasraby,gotonihei},  the proton decay decay rates in the
six   dominant channels for the 11 Fits we present in this paper
are given in Table \ref{BDEC}. These lifetimes are enhanced by up
to 8 orders of magnitude relative to those calculated for generic
fits where\cite{nmsgutII} no attempt was made to suppress the
coefficient of the Baryon violating $d=5$ operators or to take
account of threshold corrections to the yukawa couplings. If one
tries to trace how this has been accomplished one sees that the
minimum value of the masses of  standard B violating triplets
$[3,1,\pm {2/3}](t\oplus\bar{ t})$ (which anyway dominate -being
lighter- the novel $[3,3,\pm 2/3](P \oplus \bar{P}),  [3,1,\pm
8/3](K \oplus \bar{K})$  triplets) is raised by some two orders of
magnitude and in addition the   SO(10) yukawa couplings are also
reduced significantly relative to those that were necessary to
reproduce the MSSM fermion yukawas in the theory without threshold
corrections. A point that requires further investigation is the
effect of wave function renormalizations on the Baryon decay
operators. As matters stand we have not applied such corrections
since the external lines are all light fermion lines and the
corrections on these lines are typically smaller than the
systematic errors in the B- violation calculation(see tables of
type X-b). It is another matter that dimension six operators
containing external light Higgs vevs may be enhanced by wave
function renormalization. However, the   additional dimensional
suppression may well keep these operators subdominant.

 \begin{table}
 $$
 \begin{array}{|c|c|c|c|c|c|}
 \hline
 {\rm Case }&\tau_p(M^+\nu)  &\Gamma(p\rightarrow \pi^+\nu) & BR( p\rightarrow
 \pi^+\nu_{e,\mu,\tau})&\Gamma(p\rightarrow K^+\nu) & BR( p\rightarrow K^+\nu_{e,\mu,\tau})\\ \hline
 I-1
 &
                   2.4 \times 10^{  36}
 &
                   6.2 \times 10^{ -38}
 &
 \{
                 3.6  \times 10^{  -7}
 ,
                 0.082
 ,
                 0.918
 \}&
                   3.5 \times 10^{ -37}
 &\{
                 2.8  \times 10^{  -5}
 ,
                 0.119
 ,
                 0.881
 \} \\
 I-2
 &
                   6.7 \times 10^{  34}
 &
                   2.5 \times 10^{ -36}
 &
 \{
                 3.8  \times 10^{  -5}
 ,
                 0.06
 ,
                 0.94
 \}&
                   1.3 \times 10^{ -35}
 &\{
                 1.1  \times 10^{  -4}
 ,
                 0.111
 ,
                 0.889
 \} \\
 I-3
 &
                   5.7 \times 10^{  36}
 &
                   2.4 \times 10^{ -38}
 &
 \{
                 3.2  \times 10^{  -7}
 ,
                 0.100
 ,
                 0.900
 \}&
                   1.5 \times 10^{ -37}
 &\{
                 2.3  \times 10^{  -5}
 ,
                 0.139
 ,
                 0.861
 \} \\
 I-4
 &
                   5.7 \times 10^{  34}
 &
                   1.7 \times 10^{ -36}
 &
 \{
                 7.4 \times 10^{  -5}
 ,
                 0.052
 ,
                 0.948
 \}&
                   1.6 \times 10^{ -35}
 &\{
                 9.1 \times 10^{  -5}
 ,
                 0.046
 ,
                 0.954
 \} \\
 II-1
 &
                   1.5 \times 10^{  36}
 &
                   9.7 \times 10^{ -38}
 &
 \{
                 1.8 \times 10^{  -6}
 ,
                 0.114
 ,
                 0.886
 \}&
                   5.9 \times 10^{ -37}
 &\{
                 3.6  \times 10^{  -5}
 ,
                 0.170
 ,
                 0.830
 \} \\
 II-3
 &
                   1.7 \times 10^{  36}
 &
                   7.4 \times 10^{ -38}
 &
 \{
                 1.9  \times 10^{  -6}
 ,
                 0.153
 ,
                 0.847
 \}&
                   5.1 \times 10^{ -37}
 &\{
                 2.7 \times 10^{  -5}
 ,
                 0.205
 ,
                 0.795
 \} \\
 II-4
 &
                   6.2 \times 10^{  33}
 &
                   1.6 \times 10^{ -35}
 &
 \{
                 5.8 \times 10^{  -5}
 ,
                 0.071
 ,
                 0.929
 \}&
                   1.5 \times 10^{ -34}
 &\{
                 8.0 \times 10^{  -5}
 ,
                 0.069
 ,
                 0.931
 \} \\
 III-1
 &
                   2.3 \times 10^{  36}
 &
                   6.5 \times 10^{ -38}
 &
 \{
                 5.7 \times 10^{  -7}
 ,
                 0.088
 ,
                 0.912
 \}&
                   3.7 \times 10^{ -37}
 &\{
                 2.5  \times 10^{  -5}
 ,
                 0.128
 ,
                 0.872
 \} \\
 III-2
 &
                   5.0 \times 10^{  34}
 &
                   3.3 \times 10^{ -36}
 &
 \{
                 3.3  \times 10^{  -5}
 ,
                 0.050
 ,
                 0.950
 \}&
                   1.7 \times 10^{ -35}
 &\{
                 9.2  \times 10^{  -5}
 ,
                 0.093
 ,
                 0.907
 \} \\
 III-3
 &
                   2.2 \times 10^{  36}
 &
                   6.2 \times 10^{ -38}
 &
 \{
                 5.9  \times 10^{  -7}
 ,
                 0.103
 ,
                 0.897
 \}&
                   4.0 \times 10^{ -37}
 &\{
                 2.0 \times 10^{  -5}
 ,
                 0.141
 ,
                 0.859
 \} \\
 III-4
 &
                   6.2 \times 10^{  33}
 &
                   1.6 \times 10^{ -35}
 &
 \{
                 5.8 \times 10^{  -5}
 ,
                 0.071
 ,
                 0.929
 \}&
                   1.5 \times 10^{ -34}
 &\{
                 8.0 \times 10^{  -5}
 ,
                 0.069
 ,
                 0.931
 \} \\
 \hline\end{array}
 $$
\caption{\small{ of $d=5$ operator mediated proton lifetimes $\tau_p$(yrs), decay rates
 $ \Gamma ( yr^{-1} )$  and Branching ratios in the dominant Meson${}^++\nu$ channels. }} \label{BDEC}\end{table}

  We see that we have been able to suppress the B decay
rates to lie comfortably within the  current limits. Thus the
search criteria may even be loosened without conflict with
experiment. We note that our  programs can already calculate the
rates in other channels   driven by Gluino, Neutralino, Higgsino
etc exchange. However we defer a presentation of the results for
the subdominant channels till the various corrections and
improvements still needed (see below)  needed have been
implemented.  Our aim was to show that the NMSGUT is quite
compatible with the stability of the proton to the degree it has
been tested, and even beyond. Firm predictions will ensue only
once the susy spectrum is anchored in reality by a discovery of a
supersymmetric particle.

 The very heavy third sgeneration masses
  indicate  that the rate $\Gamma( b\rightarrow s \gamma)$,
is likely to be acceptable and uniform among the fits. The Susy contribution to
 muon (g-2) $\Delta a_\mu=\Delta(g-2)_\mu/2$ may vary considerably since
  the  mass of the smuon in the loop within which the photon couples
   is quite variable and generally quite low compared to other sfermions.
    Finally change in the $\rho $ parameter $\Delta\rho$  could also in
    principle be appreciable due to the 6-8 light superparticle present
    in most cases.    We   plugged our susy spectra into the ( tree
level) Spheno\cite{porod} routines to obtain    the contributions
shown in  Table \ref{LWENRGYCNSTRTS}
\begin{table}
 $$
 \begin{array}{|c|c|c|c|}
 \hline
 {\rm Case }&B.R(b\rightarrow s\gamma) &\Delta a_{\mu}&  \Delta \rho \\ \hline
 I-1
 &
                 3.294 \times 10^{  -4}
 &
                 5.796 \times 10^{  -9}
 &
                 5.985 \times 10^{  -6}
 \\
 I-2
 &
                 3.293 \times 10^{  -4}
 &
                 5.471 \times 10^{  -9}
 &
                 2.397 \times 10^{  -5}
 \\
 I-3
 &
                 3.294 \times 10^{  -4}
 &
                 2.300 \times 10^{  -9}
 &
                 2.825 \times 10^{  -6}
 \\
 I-4
 &
                 3.293 \times 10^{  -4}
 &
                 7.238 \times 10^{  -9}
 &
                 6.064 \times 10^{  -7}
 \\
 II-1
 &
                 3.290 \times 10^{  -4}
 &
                 1.360 \times 10^{ -10}
 &
                 2.503 \times 10^{  -6}
 \\
 II-3
 &
                 3.287 \times 10^{  -4}
 &
                 1.035 \times 10^{ -10}
 &
                 3.385 \times 10^{  -6}
 \\
 II-4
 &
                 3.278 \times 10^{  -4}
 &
                 1.043 \times 10^{ -10}
 &
                 3.612 \times 10^{  -6}
 \\
 III-1
 &
                 3.293 \times 10^{  -4}
 &
                 8.058 \times 10^{  -9}
 &
                 3.718 \times 10^{  -6}
 \\
 III-2
 &
                 3.293 \times 10^{  -4}
 &
                 6.824 \times 10^{  -9}
 &
                 2.105 \times 10^{  -5}
 \\
 III-3
 &
                 3.295 \times 10^{  -4}
 &
                 8.689 \times 10^{  -9}
 &
                 3.743 \times 10^{  -6}
 \\
 III-4
 &
                 3.294 \times 10^{  -4}
 &
                 7.452 \times 10^{  -9}
 &
                 5.989 \times 10^{  -7}
 \\
 \hline\end{array}
 $$
 \caption{\small{Table Low energy constraints from the limits on the branching
 ratio for $b\rightarrow s \gamma, \Delta a_\mu$ and $\Delta \rho$.}}
 \label{LWENRGYCNSTRTS}\end{table}

  The $ b\rightarrow s
\gamma$  branching ratio values are right in the centre of the
region $(3-4\times 10^{-4})\pm 15\%$  determined by measurements
at CLEO, BaBar and Belle\cite{pdg2010,cleo,babar,belle}. The
current difference between experiment and theory for the muon
magnetic moment anomaly is is $\Delta a_\mu=255(63)(49)\times
10^{-11}$\cite{pdg2010}. The results in Table \ref{LWENRGYCNSTRTS}
are thus certainly in the right ball park and we may well begin to
use the value of $\Delta a_\mu$  to discriminate between different
models provided one is confident that all instabilities in the
parameter determination process have been controlled by adequate
attention to loop and threshold effects. At the moment however we
simply note that there is no gross conflict. The predicted change
in the $\rho$ parameter is so small as to be insignificant
compared with the experimental uncertainties $\sim
.001$\cite{pdg2010}.

 The unification scale tends to be raised
above $M_X^0$ in the NMSGUT  i.e.  $\Delta_X >0$ . This is
especially true once we demand that $d=5$ operators mediating
proton decay be suppressed. In fact of the Fits  we have exhibited
here and in \cite{nmsgutI,nmsgutII} the values of $\Delta_X$ we
encounter are $-0.29, 1.82$ for the solutions without GUT scale
threshold  corrections (to fermion Yukawas) while  with threshold
corrections one gets $\{1.16, 2.82,1.28, 0.46\}$  for Cases I-1 to
I-4, $\{1.15,  1.36, 0.39\}$  for Cases II-1,3,4. $\Delta_G $, and
$\{1.22, 2.82,1.21, 0.43\}$ for Cases III-1 to I-4. Thus we see
that the unification scale-defined as the mass of the B-violating
gauginos of type $ X[3,2,\pm{5\over 3}]$
 is typically raised by one oorder of magintude or more. On the other hand the
 correction to the inverse value of the fine structure constant ($\Delta_G$) at
 the unification scale varies over a wide
 range from -20.0 to 8.1 so that  the value
of the  unification coupling may as well be weak as not. However
it remains true that above the new unification scale once we begin
to use the SO(10) RGE beta functions the gauge  coupling will
still explode\cite{trmin,tas} over an    energy scale range of only
about 5-10. Smaller $\alpha_G$ can only postpone this a little. An
ideal scenario is then that the theory is still weakly coupled at
the threshold corrected unification scale $M_X
> 10^{17.5}$ GeV but that thereafter the Susy GUT becomes strongly
coupled simultaneously with gravity. In that case the Planck scale
may be identified as a physical cutoff for the Susy NMSGUT where
it condenses as strongly coupled Supersymmetric gauge theory
described by  an appropriate sigma model. We envisage\cite{tas}
the possibility that gravity arises dynamically as an induced
effect of the quantum fluctuations of the Susy GUT  calculated in
a coordinate independent framework. This may be  realized as a
path integral over a background metric that begins to propagate
only at low energies leading to the near canonical  N=1 Supergravity perturbative
NMSGUT  as the effective theory  below $M_{Planck}$ that we assume
in our work.

    Besides the high(GUT)and low (Susy/Electroweak) scale
thresholds SO(10) theories are also typically subject to threshold
corrections and   RG flows associated with the couplings of the
righthanded neutrinos present in the theory. In fact   the NMSGUT
scenario makes essential use of  intermediate scale heavy
neutrinos($10^8-10^{13} $ GeV).  Such neutrinos are however not
inert: particularly as far as their effects on RG evolution of the
yukawa couplings of the light leptons  above the mass threshold
associated with the right handed neutrino masses are concerned.
The techniques for inclusion of these RG flows and threshold
effects are by now standard\cite{babupant,nuflow} and in
subsequent full analysis we will include also these effects.

   The NMSGUT also provides corrections to the QCD
coupling at $M_Z$ that are in the right range($-.017
<\Delta\alpha_3(M_Z)<-.004 $  to lower it as
required\cite{langpol,precthresh}.

  As  a direct consequence of the dominance of the Type I
seesaw mechanism, due to the low value of the $\mathbf{\oot}$
coupling, the right handed neutrinos emerge in just the range
$10^8- 10^{14} $ GeV required to implement
leptogenesis\cite{leptogenesis}. The influence of these threshold
on the RG evolution has not yet been factored in by us yet and may
have important implications for the Yukawa unification.  This is
straightforward to implement and of high priority for the next
round of improvement of the calculation.

    The  large   values of the crucial
 parameters $|\mu|,|A_0|,M_A, |M^2_{H,{\bar H}}| \sim 10^2\, TeV$ (where $M_A$ is the mass of the pseudo
 scalar Higgs remnant) play a crucial role in structuring the low energy phenomenology.
  Due to the  large value of $|\mu|$ the LSP is  essentially a pure Bino and the
  lighter chargino  is a Wino($\tilde W_\pm$) while the  heavier one ${\tilde H}^{\pm}$ is a pure Higgsino,  with mass set
 by the large $\mu$ parameter ( which also sets the scale for the
 two heavy neutralinos) while the next to lightest   neutralino is
 also an essentially pure Wino($\tilde W_3$).
  Since the required threshold corrections depend on ratios of scalar masses to gaugino masses
       there is actually a preference for light gaugino masses to enhance the ratios with
        the heavy masses. Running counter to this is only the constraint that the experimental
        lower mass limit on charged gauginos and exotic charged and coloured scalars generally
          is around 100 GeV. Thus we imposed a floor of 110 GeV for all such exotics. It is this
           chargino limit and the linked behaviour of the Bino and Wino masses that prevents the
           Bino mass from running to very low values( in our example fits the lowest Bino mass
           is 35 GeV). Also due to the link between the gaugino masses  have We have  light
           gluinos  below  500 GeV for the cases I-II where we allowed LSP/Binos in the range $5-150$ GeV.

   The parameters $\mu , m_A$ at $ M_Z$ are determined in terms of the run down
 Higgs mass parameters $M^2_{H,{\bar H}}$ by the    Electroweak symmetry breaking
 conditions which we implemented at the one-loop level by
 including Higgs tadpoles calculated using a subroutine from
 \cite{porod}   corresponding to the  formulae  given
 by \cite{piercebagger}.   $M_A$   sets the scale of the mass of the scalar Higgs apart
 from $h^0$ i.e $H^0, H^\pm,A$ all have masses close to $M_A$. The
 light scalar Higgs $h^0$ typically has a mass  in the range of
 $110-125 $ GeV. Once the values of $\mu$, $m_A$ ( or $B=m_A^2 Sin 2\beta/2$ ) at $M_Z$ are known
 we  run them back up to $ M_X$ since they do not interfere with the running of
 rest of parameters due to the modular structure of the RGE. Thus we give the  7 parameter  set $(m_{ \frac{1}{2}},
 \tilde{m}_0, A_0, M^2_{H,{\bar H}}, |\mu|, B )$ which together with the NMSGUT parameters completely
 specifies the theory at all scales and yields a distinctive scenario for the
 low energy supersymmetric accelerator phenomenology as well as for LSP Dark Matter  cosmology.

 The large but negative values of the Higgs mass-squared  parameters found by our search programs have
 a dramatic consequence that sets the type of Susy spectra obtained in the NMSGUT in marked contradiction
  with the generally accepted and used patterns of Sfermion masses derived from GUT boundary conditions
  at high scales. As shown  in Appendix \textbf{B} these large negative mass-squared parameters drive
  the third generation sfermions \emph{to be much heavier than the first two generations} even though
  we assume  a common mass squared for the sfermions in all three 16 plets at $M_X^0$. Thus third
   generation sfermion  generally lie in the range 5-50 TeV and are effectively decoupled from
    electroweak scale physics. However the NMSGUT susy spectra are \emph{not} of the split
    supersymmetry type since   the sfermions of the first two generations typically populate
    the mass band  between the light gauginos and the superheavy third sgeneration. Most of the
     sparticles emerge above the direct discovery limits but a couple of right chiral sfermions
     emerge( see Table \ref{LHCdetectables}) in the discoverable set; of which squarks are
      generally relatively heavy but right sleptons and specially the smuon can descend even
      to the Electroweak breaking scale. It remains to be seen if leptonic  flavour violation
       constraints in will actually permit such light sleptons. However the point to emphasize
       at this stage is that searches  based on the assumption of a light LSP inevitably lead
       to a characteristic Susy spectrum :
 \bea M_{\tilde B} < \{M_{\tilde W}, M_{\tilde g}, M_{\tilde {\bar f}_{1,2}}\} <<M_{\tilde F_{1,2}}<<M_{{\tilde{\bar  f}_3} ,{\tilde F}_3 }  <<  |\mu|,|A_0|,M_A, |M^2_{H,{\bar H}}|\eea
 It may be necessary to further constrain the search (and include 1-loop corrections to
  sfermion masses) to obtain consistency with flavour violation processes involving the first
  two generations (as can be seen from the Table \ref{LWENRGYCNSTRTS} the generally stringent
   limits due to $b\rightarrow s \gamma$  (specially at large $\tan\beta$ ) when the  third
   sgeneration is lightest are ineffective in our case). Thus whereas it may be premature to
    point to any one sfermion mass as a \emph{prediction} of the NMSGUT that can be verified
     at the LHC it seems safe to assert that that the NMSGUT does predict a light Bino and
      Chargino and more distinctively that the first sfermions discovered, at LHC or later,
      \emph{must} belong to the first two generations with some preference being exhibited
      for the smuon and or selectron (provided those cases are consistent with flavour
      violation constraints).

   \section{Conclusions and Outlook}
This paper is the third of  a series \cite{nmsgutI,nmsgutII}
  developing  the NMSO(10)GUT as a possibly  viable
and complete theory of particle physics.  The emphasis is to
develop the theory on the same lines and as explicitly as the the
Standard Model. We have shown that the theory is sufficiently
simple as to allow explicit calculation of the spontaneous
symmetry breaking, mass spectra and eigenstates  and allows a
computation of the RG flow in terms of the fundamental GUT
parameters to the point where one can attempt to actually fit the
low energy data, i.e the SM parameters together with the neutrino
mixing data,  in its entirety. In carrying out this project the
model meets a major obstacle in its inability to fit the
unmodified down and strange quark yukawas in the MSSM
(renormalized up to $M_X$), which it overcomes by staking its
viability on the operation of large $\tan\beta$ driven threshold
corrections -providentially known to be operative and important in
this context- which lower these yukawas from their SM values by a
factor of about 5. As a result the pattern- although unfortunately
not the scale- of the Susy breaking parameters tends to become
fixed by the need to preserve, or indeed somewhat raise, the b
quark yukawa while lowering the  $d,s$ quark yukawas when crossing
the SM-MSSM threshold(s) . Thus the major unknown for the model,
indeed for the whole field of Susy, remains the mass of at least
one of the light susy particles. Due to the tight
interrelationship of the susy spectra which have been generated
from just 5 soft susy parameters at $M_X$ and subjected to
multiple stringent  demands such as EW symmetry breaking, yukawa
modification, Baryon decay suppression, LSP consistency and so on,
it seems very plausible that input of any one of the susy particle
masses would   generate a prediction of all the susy  masses in
the context of a fit that was pinned to yield that particular mass
value along with satisfying the various requirements we have
mentioned. In the absence of Susy discovery data and the
difficulty of extending the mass ranges that can actually be
claimed to have been excluded without egregious assumptions,
 we can still call upon the the model to stand by it's claim to be a viable
 theory of Dark matter and yield its stable Bino LSP
in the $5-150$ GeV mass range preferred by Cold Dark matter WIMP scenarios.
 This requirement is sufficient to pin the prediction for the Susy
 spectrum to a specific and completely novel and distinctive \emph{type},
  if not yet to specific masses for specific particles. The model thus predicts
light gauginos and discoverable first or second generation right
chiral sfermions below 2 TeV and invisible third generation
sfermions. Thus we arrive at the attractively dangerous conclusion
that : \emph{If as all other GUT scenarios derive, the third
sgeneration emerges
 lightest the NMSGUT may be taken to be falsified.}

This remarkable model has thus added yet more feathers to the
already long  list of its attractions. Besides providing a natural
and minimal context for the supersymmetric seesaw  mechanism and
the implementation of R-parity as a part of the gauge structure of
unification, the theory   successfully accounts for the entire
available fermion mass data in terms of its own    parameters in
way consistent with all known phenomenology.  It also yields
insight into a necessary structure of it's susy breaking
parameters  and yields a viable and completely natural candidate
for the LSP combined with a surprising and novel candidate for the
scalar NLSP which can lead to an effective DM scenario.
Furthermore the theory naturally pushes the unification scale
towards the Planck scale and allows suppression of the dangerous
$d=5$ B violation operators. The conflation of the Planck scale
and the unification scale goes a long way towards alleviating a
perennial problem of  renormalizable SO(10) models
\cite{trmin,tas} namely divergent couplings in the ultraviolet.
 The unification scale or rather the Landau pole above it becomes a  physical cutoff beyond
which the theory enters a strongly coupled phase together with
gravity.

 Till a susy breaking soft mass is pinned by experiment  the development of the NMSGUT
will continue by   facing up to technical challenges that we have postponed
in the first phase of the definition of the model as reported in a
series of papers(\cite{ag1,ag2,abmsv,gmblm,blmdm,precthresh,core,msgreb} and
the papers of the current triplet: \cite{nmsgutI,nmsgutII} and
this paper). We conclude by itemizing the important issues which
may materially alter the numbers obtained by our fitting program
so far.

\begin{itemize}
\item We have, following\cite{piercebagger}, chosen $M_Z$ as the
scale at which we match the SM to the MSSM. On the one hand this
is well motivated since, as we have seen, the weak  gauginos, i.e
the Bino and Wino tend to emerge as light as we allow them : which
is as light as experiment permits i.e $\sim M_Z$ for the Chargino
but much lower for the Bino( the connection between the two masses
and the limit $M_{{\tilde W}^\pm}$ however does not let the Bino
mass descend below about 25 GeV). On the other hand we cannot deny
that the model has its own little hierarchy problem with the $\mu,
A_0,M^2_H$ parameters all lying in the range of 10-100 TeV.
Clearly the threshold effects due to the large differences of the
Susy particle spectra from $M_Z$ will cause appreciable threshold
corrections to the naive RG running which takes a single
undifferentiated Susy breaking scale and identifies it arbitrarily
with $M_Z$. Some development of the
  techniques for incorporating multiple thresholds between the SM and MSSM
   has already taken place\cite{box} and should prove useful.

\item We have calculated only tree  tree masses for the susy particles  in the
theory (the Electro weak symmetry breaking and thus the  Higgs masses were
 however calculated using  1-loop effective potential\cite{piercebagger,porod}.
  This is not a very good approximation in the MSSM with small $A_0$ and it may be
  worse in scenarios with large $A_0$ parameters.
The large $A_0$ parameters lead to large trilinear couplings
($A=A_0 Y_f$)   only for the third sgeneration since the Yukawas
$Y_f$ for the other generations are so small and may
significantly modify   the third generation sfermion masses. The
complete formulae for calculating these modifications-at least in
the flavour diagonal case - have long been
known\cite{piercebagger}. They will be incorporated in the next
version  of our search codes.

\item We have not yet incorporated the three thresholds associated
with the heavy right handed neutrino masses in the theory. Since
one progressively introduces the neutrino Dirac couplings as one
passes these thresholds going higher in energy, significant
effects on the yukawa unification may be anticipated. These
thresholds will also be significant when calculating the amount of
lepton flavour violation introduced when integrating down from the
high scale (driven by the non diagonality of the SO(10) yukawas
required to account for the observed  quark flavour mixing).
 It is for this reason, and because the mass insertion formalism
  is ill adapted to securely  evaluate novel scenarios,
  that we have not generated tables of
 Lepton flavour violating mass insertions for comparison with
 the existing analyses on lepton flavour constraints\cite{masvemp}.
 We note however that we  have calculated some of the common
 Susy sensitive quantities and found them to be
  compatible with existing limits. Nevertheless the incorporation
   of constraints on  our searches based upon electric dipole moments and  the strength of flavour
   violation operators involving the sfermions of the first two
   sgenerations are a priority issue for our program.

\item It will be clear to the reader familiar with the nuances of
the Susy Flavour problem that in terms of a Bottom-up approach our
results suggest that a radical extension of the consistent and
viable susy parameter space at large $\tan\beta,
A_0,|mu|,|M^2_{,\bar H}|$  and with $M_{\tilde f_3}>>M_{\tilde
f_3}$  may be possible and the characteristics of such an
extension as derived from the NMSO(10)GUT approach could be
diametrically opposed to those from all previous GUTs. However
that reader will also realize that the inversion of the standard
inverted hierarchy of sfermion masses to a normal hierarchy may
have drastic implications for the consistency of the theory with
flavour violation constraints and EDM constraints for the first
two generations (the heavy sthird is likely to pass such
constraints -as already seen in the case of the $b\rightarrow
s\gamma$ constraint : which is commonly a stringent one when the
third sgeneration is light but is here totally insensitive see
Table \ref{LWENRGYCNSTRTS}). Prominent among these well known
constraints are those on the electric dipole moments of the
electron, muon,  neutron etc, the severely constrained Branching
rations of the $\Delta F=1$ decays of Mesons such as
$B_d\rightarrow \mu\mu$, and the limits on the supersymmetric
contributions to $\Delta F=2$ quantities like $\epsilon_K ,\Delta
K$ etc. These contributions can now be calculated using
codes\cite{Susy-flavour} that do not make any simplifying
approximations as done in the mass insertion method\cite{gbgbrmas}
which allow only order of magnitude estimates of individual terms
but cannot be used to evaluate total contributions including the
effect of cancellations, in unconventional scenarios like ours. It
may well be that the parameter sets we have presented will fail
these tests when they put to them. However we emphasize that, from
the point of view of the this series of papers which seeks to pin
down the viable points   of the 45 dimensional  parameter space
SO(10) NMSGUT, such a challenge would be no different than the one
that was posed by the onus to show compatibility with B violation
rates:  which the theory overcame by novel use of expedients
available to it in a way  that pointed out fresh approaches to
hoary questions. In the same way it may be that when the
requirement of the consistency of the \emph{total} value of such
parameters predicted by the NMSGUT is taken into consideration
then again new viable parameter sets may emerge. In view of the
length of the present paper as well as the considerable further
effort required to answer definitively to these important
challenges the flavour violation constraints will be taken up in
the sequels. We urge the tolerant reader not to prejudge this
vital issue but join us in reflecting on the
 behaviour of the  novel type of Susy  parameter sets suggested by
 us.

\item In this calculation we  found parameter  sets leading to
quasi-stable Baryons by a shotgun carrying brute search. However
the characterization of the possible cancellations and the least
constraining  versions compatible with experimental limits remains
to be done. The search for fits with suppressed Baryon Number
violation was carried out by simply limiting the size of the
maximal element of the LLLL and RRRR operator coefficients with no
heed paid to the detailed  effects of that coefficient on the
decay rate in specific channels. A more sophisticated  (but hugely
more computer intensive) way of doing this would be to calculate
the Baryon violation rates in each channel at every iteration and
limit the total lifetime. In principle one could hope to implement
this given enough (super)computing power.

\item  Due to the large amount of running time required to find an
acceptable solution, we have only scratched the surface of the
enormous parameter space and can by no means pronounce on the
general structure of the solution space. It will take  long runs
on a super computing cluster to develop a statistical picture of
where the solutions tend to lie. We are now preparing for both the
improvements mentioned above and the harnessing of a cluster  for
the task.

\item We have run down only the diagonalized yukawa couplings from
$M_X^0$ to $M_Z$ along with the susy breaking parameters which are
optimized to fit the eigenvalues of the SM yukawa couplings to the
run down diagonal NMSGUT matter yukawas. A complete treatment
would run down the full set of coupling matrices  obtained from
the GUT and apply the large $\tan\beta$ driven Susy threshold
corrections to the off diagonal couplings before matching the two
sets : or at least the ``physical'' parameters (eigenvalues,
mixing angles and phases) coded in the two pairs. The formalism
for applying off diagonal corrections is still somewhat murky  so
some theoretical progress towards setting out clear algorithms for
including all off diagonal 1-loop effects and renormalizations is
called for.

\item We argued that the CCB and UFB minimae(that certainly exist
at the large values of $|\mu|,|A_0|$ central to the needs of the
NMSGUT) need not destabilize the metastable standard vacua we have
found and cited previous investigations of this
issue\cite{kuslangseg} which are specially encouraging in this
regard. The fact remains, however,  that these decay rates must actually be
estimated for each set of  parameters found. On the other hand
once the requisite subroutines for calculating the vacuum
tunneling rate are in hand one can add them to the search routines
to filter the parameter sets.

\item Threshold corrections play a central role  in our calculation and the
large wavefunction dressing of the Higgs doublet lines that we find at one
 loop demand an investigation of the two loop effects to determine whether
 they are yet larger still. If so our model, and by implication realistic
  GUTs generally,  may survive only after the Higgs wavefunction dressing
  has somehow been re-summed to all orders. However we note that this growth
   of wave function dressing leads to reduction in the actual size of the SO(10)
    yukawas actually needed to fit the low energy fermion data by a factor of
    10-100. Thus the contributions to the fermion lines are effectively degenerate
    and it is only the large number of heavy fields running in dressing of the
    light Higgs boson lines that leads to  such a large  wave function renormalization
    of the Higgs fields. It is possible that in case of further growth at the multiloop
    level the theory is being driven to a quantum fixed point dominated by the
     gauge coupling and therefore re-summable  using the the exact beta functions
     available for supersymmetric gauge theories\cite{vainshif}. In any event we
      consider that our calculations show that GUTs aiming to be realistic have
      been ``up-ended'':  in the sense that any Grand unified model  with
      pretensions to realism must define itself explicitly enough to permit calculation
      of   threshold corrections using calculated superheavy field spectra or else
       risk being under suspicion of being an qualitative scenario that may be
       destroyed as soon as quantum effects are included.

\item As Karl Popper emphasized  so insightfully, the virtue of a
comprehensive scientific theory that accounts for all known data
is that it accepts the challenge and charm of living dangerously
and in constant confrontation with experimental data that may
falsify it. It must face every new measurement that it claims to
be able to account for with its fate hanging  in the balance. This
indeed is what gives properly scientific models   their  peculiar
power and utility: that speculations living at a safe distance
from the cutting edge of Occam's razor rarely possess. The
(N)MSO(10)GUT has braved multiple iterations of challenges over
the three decades since it was proposed\cite{aulmoh,ckn} and, so
far,  has emerged stronger from every challenge,  defining the
possible in its realm ever more clearly and distinctly. If it dies
it will not have lived in vain.
\end{itemize}
\bea\nonumber\eea

 \section*{Acknowledgements}
 \vspace{.5 truecm}
 It is a pleasure to  acknowledge useful
correspondence  with  W.A. Porod in connection with the SPheno
Susy Spectrum calculator that we have used both whole, and in
  parts to build our code for the NMSGUT. I thank A.
Kusenko for permission to reproduce  a figure
from\cite{kuslangseg}.  I thank Ila Garg and Charanjit Kaur for
assistance in the preparation of the manuscript and specially the
tables. I thank the many  people who  have helped in the
completion of this paper through discussions, encouragement and
hospitality, notably,  B. Bajc, K.S. Babu, K.Huitu, S. Mohanty,
D.P.Roy, G. Senjanovic, and S. Vempati. I am also grateful to my
family, Satbir, Simran and Noorvir, for patience and support
during the   prolonged  struggle to complete these calculations.
\bea\nonumber\eea

 \section*{ Appendix A: Gut Scale threshold corrections to matter
 yukawas}

We give below our results for the threshold corrections to the
Yukawa couplings of the matter fields due to   heavy fields
running  in a self energy loop on a line leading into the Yukawa
vertex(here we assume the Higgs line is predominantly the $\textbf{10}$-plet
derived doublet).

 Let us denote by  $U^{A}(V^{A})$  the matrices that diagonalize the mass terms for   fields of
the alphabetized label type A : \bea {\overline{\Phi}}^T M \Phi
={\overline{\Phi'}}^T M_{Diag}
 \Phi'\quad\Leftarrow \quad    {\overline{\Phi }}= U^{\Phi}{\overline{\Phi' }} \quad ;
 \qquad {\Phi } =  V^{\Phi}\Phi'\nonumber \qquad(\mathbf{A1})\eea
generically by $W^{A}$ and the corresponding index ranging over
the multiplicity of that field type also by the corresponding
lower case roman letter (a). For example for the 6-fold set $[\bar
3,2,-{1\over 3}](\bar E_1, \bar E_2,\bar E_3,\bar E_4,\bar
E_5,\bar E_6) \oplus [3,2,{1\over 3}](E_1,E_2,E_3,E_4,E_5,E_6) $
the generic rotation matrix is denoted $W^{E}$( i.e $U^{E} or
V^{E}$ ) carrying indices $e,e'=1,2,3,4,5,6$. Then it is useful to
define

\bea {\cal F}_1 (W^A,a)&=&\sum_{ a' =1}^{ a'=dim( W^
A)}|W^A_{a,a'}|^2 F_{11}(m_{a'},Q) \nnu {\cal F}_1^{'}
(W^{H},h)&=&\sum_{ h' =2}^{ h' = 6}|W^{A}_{h,h'}|^2
F_{11}(m_{{h'}},Q)  \nnu {\cal F}_1^{u }
(W^{A},a,m^{(u)})&=&\sum_{ a' =1}^{ a' =dim(W^{A})} |W^{A}_{a, a'
}|^2 F_{12}( m^{(u)},m_{a'} ,Q) \nnu {\cal F}_1^{u '}
(W^{H},h,m^{(u)})&=&\sum_{ h' =2}^{ h' =dim(W^{H})} |W^{H}_{h, h'
}|^2 F_{12}( m^{(u)},m_{h'} ,Q)+|W^{H}_{h, 1 }|^2 F_{11}( m^{(u)},
Q) \nnu
 {\cal F}_2 (W^{A},W^{B},a, b )&=&\sum_{ b' =1}^{ b'=dim(W^B) }\quad
\sum_{a' =1}^{a' =dim( W^A)} |W^{A}_{a, a' }|^2 |W^{B}_{b,{b'}}|^2
 F_{12}( m_{{a'}},m_{{b'}},Q)   \nnu
{\cal F}_2^{'}(W^{A},W^{H},a, h )&=&\sum_{ h' =2}^{ h'=6 }\quad
\sum_{a' =1}^{a' =dim( W^A)} |W^{A}_{a, a' }|^2 |W^{H}_{h,{h'}}|^2
 F_{12}( m_{{a'}},m_{{h'}},Q) \nnu
 &+&  \sum_{a' =1}^{a' =dim( W^A)} |W^{A}_{a, a' }|^2
|W^{H}_{h,1}|^2 F_{11}( m_{{a'}} ,Q)    \nnu {\cal C}_1 (W^{A},a,
a' )&=&\sum_{ a'' =1}^{a''=dim(W^A)} (W^A_{a, a'' })^* W^A_{a'  ,
a'' }  F_{11}(m_{a''} ,Q)   \quad(\mathbf{A2})\nnu {\cal C}_1^{ '
}(W^{H},h, h' )&=&\sum_{ h'' =2}^{a''=dim(W^H)} (W^H_{h, h'' })^*
W^A_{h'  , h'' } F_{11}(m_{h''} ,Q) \nnu
 {\cal C}_2( W^{A},W^{B},a,
b,b' ) &=&\sum_{ b'' =1}^{ b''=dim(W^B) }\sum_{ a' =1}^{ a' =dim(
W^ A  )}|W^{A}_{a, a'}|^2 (W^{B}_{b,b''})^* W^{B }_{b',b'' }
 F_{12}( m_a ,m_{b''},Q)  \nnu
 {\cal C}_2^{ ' }(W^{H},W^{B},a,h,h' ) &=&  \sum_{ h'' =2}^{ h''=6 }\sum_{ a' =1}^{ a' =dim( W^ A
)}|W^{A}_{a, a'}|^2 (W^{H}_{h,h''})^* W^{H }_{h',h'' }
 F_{12}( m_a ,m_{h''},Q) \nnu &+& \sum_{ a' =1}^{ a' =dim(W^A)}|W^{A}_{a, a'}|^2
(W^{H}_{h,1})^* W^{H }_{h',1}
 F_{11}( m_a ,Q)     \nonumber
\eea

The primes on the function names and summations instruct an
omission of any  light fields (in practice the light Higgs
$[1,2,\pm 1]$ doublets only) from the sum over the heavy fields of
the given type. If the field is one of the unmixed types (i.e $u=$
A, B, I, M, N, O, S, T, U, V, W, Y, Z)   then the function carries
a superscript (u) thus e.g ${\cal F}_1^{(u)}(V_F,1,M_V) $ arises
from a coupling between $F_1[1,1,2]$ and $V[1,2,-3]$. Such
functions arise in the dressing of the Higgs lines. The
calculation is quite tedious but we applied various consistency
checks to ensure that we had included contributions from all
members of multiplets. Naturally we await the contributions of
those patient and interested enough to check our results.

\bea({ {32\pi^2}}) {\Delta}_{\bar{u}}&=&
2{\bar{h}}^*{\bar{h}}{\cal{F}}_1(U_T,1)
-4{\bar{g}}^*{\bar{g}}{\cal{F}}_1(U_T,7) -2i
{\sqrt{2}}{\bar{h}}^*{\bar{g}}{\cal{C}}_1(U_T,1,7)  -2i
{\sqrt{2}}{\bar{g}}^*{\bar{h}}{\cal{C}}_1(U_T,7,1) \nnu
&+&{\bar{h}}^*{\bar{h}}{\cal{F}}_1(V_T,1)
-2{\bar{g}}^*{\bar{g}}{\cal{F}}_1(V_T,7)
-2{\bar{g}}^*{\bar{g}}{\cal{F}}_1(V_T,6) \nnu
&-&i({\sqrt{2}}{\bar{h}}^*{\bar{g}}{\cal{C}}_1(V_T,1,7))
 -i({\sqrt{2}}{\bar{g}}^*{{\bar{h}}}{\cal{C}}_1(V_T,7,1))
-{\sqrt{2}}{\bar{h}}^*{\bar{g}}{\cal{C}}_1(V_T,1,6) \nnu
 &+& {\sqrt{2}}{\bar{g}}^*{{\bar{h}}}{\cal{C}}_1(V_T,6,1)
 -2i({\bar{g}}^*{\bar{g}}{\cal{C}}_1(V_T,6,7))
 +2i({\bar{g}}^*{\bar{g}}{\cal{C}}_1(V_T,7,6)) \nnu
&+& 2({{\bar{h}}}^*{{\bar{h}}}{\cal{F}}_1'(V_H,1)
-(1/3){\bar{g}}^*{\bar{g}}{\cal{F}}_1'(V_H,6) + ({i\over
\sqrt{3}}){\bar{h}}^*{\bar{g}}{\cal{C}}_1'(V_H,1,6)) \nnu
&+&(({i\over\sqrt{3}}){\bar{g}}^*{{\bar{h}}}{\cal{C}}_1'(V_H,6,1))
-{\bar{g}}^*{\bar{g}}{\cal{F}}_1'(V_H,5)
+{\bar{h}}^*{\bar{g}}{\cal{C}}_1'(V_H,1,5) \nnu
&-&{\bar{g}}^*{{\bar{h}}}{\cal{C}}_1'(V_H,5,1)
 - (({i\over\sqrt{3}}){\bar{g}}^*{\bar{g}}{\cal{C}}_1'(V_H,5,6))
+(({i\over\sqrt{3}}){\bar{g}}^*{\bar{g}}{\cal{C}}_1'(V_H,6,5)))
\nnu &-&{\bar{g}}^*{\bar{g}}((64/3){\cal{F}}_1(V_C,3)
+8{\cal{F}}_1(V_D,3) \nnu&+& 8{\cal{F}}_1(U_K,2)
 + 4{\cal{F}}_1(V_J,5)+8{\cal{F}}_1(V_L,2)))
  - (2 g^2)(25{\cal{F}}_1(V_G,6)\nnu &+&0.5{\cal{F}}_1(V_J,4)
+0.5{\cal{F}}_1(V_F,3) +2{\cal{F}}_1(V_X,3) +{\cal{F}}_1(V_E,5)
)\nonumber\quad\quad \quad\quad(\mathbf{A3}) \eea

\bea ({ {32\pi^2}}){\Delta}_{\bar{d}}&=&
(2{{\bar{h}}}^*{{\bar{h}}}{\cal{F}}_1(U_T,1)
-4{\bar{g}}^*{\bar{g}}{\cal{F}}_1(U_T,7) +2
i({\sqrt{2}}{\bar{h}}^*{\bar{g}}{\cal{C}}_1(U_T,1,7)) \nnu &+&  2
{\sqrt{2}}{\bar{g}}^*{{\bar{h}}}{\cal{C}}_1(U_T,7,1))
+{{\bar{h}}}^*{{\bar{h}}}{\cal{F}}_1(V_T,1)
-2{\bar{g}}^*{\bar{g}}{\cal{F}}_1(V_T,7)\nnu &-&
2{\bar{g}}^*{\bar{g}}{\cal{F}}_1(V_T,6)
+i({\sqrt{2}}{\bar{h}}^*{\bar{g}}{\cal{C}}_1(V_T,1,7))
+i({\sqrt{2}}{\bar{g}}^*{{\bar{h}}}{\cal{C}}_1(V_T,7,1))\nnu
&-&{\sqrt{2}}{\bar{h}}^*{\bar{g}}{\cal{C}}_1(V_T,1,6)
+{\sqrt{2}}{\bar{g}}^*{{\bar{h}}}{\cal{C}}_1(V_T,6,1) +2
i({\bar{g}}^*{\bar{g}}{\cal{C}}_1(V_T,6,7))\nnu & -&2
i({\bar{g}}^*{\bar{g}}{\cal{C}}_1(V_T,7,6))
+2({{\bar{h}}}^*{{\bar{h}}}{\cal{F}}_1'(U_H,1)
-(1/3){\bar{g}}^*{\bar{g}}{\cal{F}}_1'(U_H,6)\nnu
&-&{\bar{g}}^*{\bar{g}}{\cal{F}}_1'(U_H,5)
+i(({1\over\sqrt{3}}){\bar{h}}^*{\bar{g}}{\cal{C}}_1'(U_H,1,6))
+i(({1\over\sqrt{3}}){\bar{g}}^*{{\bar{h}}}{\cal{C}}_1'(U_H,6,1))\nnu
&+&{\bar{h}}^*{\bar{g}}{\cal{C}}_1'(U_H,1,5)
-{\bar{g}}^*{{\bar{h}}}{\cal{C}}_1'(U_H,5,1)
-i(({1\over\sqrt{3}}){\bar{g}}^*{\bar{g}}{\cal{C}}_1'(U_H,5,6))\nnu
&+&i(({1\over\sqrt{3}}){\bar{g}}^*{\bar{g}}{\cal{C}}_1'(U_H,6,5)))
-{\bar{g}}^*{\bar{g}}((64/3){\cal{F}}_1(U_C,3)
+8{\cal{F}}_1(V_E,6)+4{\cal{F}}_1(V_K,2)\nnu
&+&8{\cal{F}}_1(U_J,5)+8{\cal{F}}_1(V_L,2)) - (2
g^2)(0.225{\cal{F}}_1(V_G,6) \nnu &+& 0.5{\cal{F}}_1(V_J,4)
+0.5{\cal{F}}_1(V_F,3) +{\cal{F}}_1(V_X,3) +2{\cal{F}}_1(V_E,5) ))
\nonumber\quad\quad \quad\quad(\mathbf{A4}) \eea

\bea ({ {32\pi^2}}){\Delta}_{\bar{\nu}}&=&
(2{{\bar{h}}}^*{{\bar{h}}}{\cal{F}}_1'(V_H,1)
-2{\bar{g}}^*{\bar{g}}{\cal{F}}_1'(V_H,5)
-6{\bar{g}}^*{\bar{g}}{\cal{F}}_1'(V_H,6)\nnu &&
+2{\bar{h}}^*{\bar{g}}{\cal{C}}_1'(V_H,1,5)
-i2\sqrt{3}{\bar{h}}^*{\bar{g}}{\cal{C}}_1'(V_H,1,6)
+i2\sqrt{3}{\bar{g}}^*{\bar{g}}{\cal{C}}_1'(V_H,5,6)\nnu && +
(2{\bar{h}}^*{\bar{g}}{\cal{C}}_1'(V_H,1,5)
-i2\sqrt{3}{\bar{h}}^*{\bar{g}}{\cal{C}}_1'(V_H,1,6)
+i2\sqrt{3}{\bar{g}}^*{\bar{g}}{\cal{C}}_1'(V_H,5,6))^\dagger\nnu
&& +3{{\bar{h}}}^*{{\bar{h}}}{\cal{F}}_1(V_T,1)
-6{\bar{g}}^*{\bar{g}}{\cal{F}}_1(V_T,7)
-6{\bar{g}}^*{\bar{g}}{\cal{F}}_1(V_T,6)\nnu &&
+3{\sqrt{2}}{\bar{h}}^*{\bar{g}}{\cal{C}}_1(V_T,1,6)
-i3({\sqrt{2}}{\bar{h}}^*{\bar{g}}{\cal{C}}_1(V_T,1,7)) +6 i
({\bar{g}}^*{\bar{g}}{\cal{C}}_1(V_T,6,7)) \nnu &&+
(3{\sqrt{2}}{\bar{h}}^*{\bar{g}}{\cal{C}}_1(V_T,1,6)
-i3({\sqrt{2}}{\bar{h}}^*{\bar{g}}{\cal{C}}_1(V_T,1,7)) +6 i
({\bar{g}}^*{\bar{g}}{\cal{C}}_1(V_T,6,7)))^\dagger\nnu &&
-{\bar{g}}^*{\bar{g}}(4{\cal{F}}_1(U_F,4) +24{\cal{F}}_1(U_E,6)
+12{\cal{F}}_1(V_J,5))  \nnu && - (2 g^2)(0.625{\cal{F}}_1(V_G,6)
+1.5{\cal{F}}_1(V_J,4) +0.5{\cal{F}}_1(V_F,3) +3{\cal{F}}_1(V_E,5)
)) \nonumber\quad\quad \quad\quad(\mathbf{A5})\eea

\bea ({ {32\pi^2}}) {\Delta}_{\bar e}&=&
(2{{\bar{h}}}^*{{\bar{h}}}{\cal{F}}_1'(U_H,1)
-2{\bar{g}}^*{\bar{g}}{\cal{F}}_1'(U_H,5)
-6{\bar{g}}^*{\bar{g}}{\cal{F}}_1'(U_H,6)\nnu &&
+2{\bar{h}}^*{\bar{g}}{\cal{C}}_1'(U_H,1,5)
-i2\sqrt{3}{\bar{h}}^*{\bar{g}}{\cal{C}}_1'(U_H,1,6)
+i2\sqrt{3}{\bar{g}}^*{\bar{g}}{\cal{C}}_1'(U_H,5,6)\nnu && +
(2{\bar{h}}^*{\bar{g}}{\cal{C}}_1'(U_H,1,5)
-i2\sqrt{3}{\bar{h}}^*{\bar{g}}{\cal{C}}_1'(U_H,1,6)
+i2\sqrt{3}{\bar{g}}^*{\bar{g}}{\cal{C}}_1'(U_H,5,6))^\dagger\nnu
&& +3{{\bar{h}}}^*{{\bar{h}}}{\cal{F}}_1(V_T,1)
-6{\bar{g}}^*{\bar{g}}{\cal{F}}_1(V_T,7)
-6{\bar{g}}^*{\bar{g}}{\cal{F}}_1(V_T,6)\nnu &&
+3{\sqrt{2}}{\bar{h}}^*{\bar{g}}{\cal{C}}_1(V_T,1,6)
+i3({\sqrt{2}}{\bar{h}}^*{\bar{g}}{\cal{C}}_1(V_T,1,7)) -2
i3({\bar{g}}^*{\bar{g}}{\cal{C}}_1(V_T,6,7)) \nnu &&+
(3{\sqrt{2}}{\bar{h}}^*{\bar{g}}{\cal{C}}_1(V_T,1,6)
+i3({\sqrt{2}}{\bar{h}}^*{\bar{g}}{\cal{C}}_1(V_T,1,7)) -2
i3({\bar{g}}^*{\bar{g}}{\cal{C}}_1(V_T,6,7)))^\dagger\nnu &&
-{\bar{g}}^*{\bar{g}}(4{\cal{F}}_1(U_F,4) +24{\cal{F}}_1(U_D,3)
+12{\cal{F}}_1(V_K,2)) \nnu &&- (2 g^2)(025{\cal{F}}_1(V_G,6)
+1.5{\cal{F}}_1(V_J,4) +0.5{\cal{F}}_1(V_F,3) +3{\cal{F}}_1(V_X,3)
)) \nonumber\quad\quad \quad\quad(\mathbf{A6})\eea

\bea ({ {32\pi^2}}) {\Delta}_u &=&
({{\bar{h}}}^*{{\bar{h}}}{\cal{F}}_1(U_T,1)
-2{\bar{g}}^*{\bar{g}}{\cal{F}}_1(U_T,6)
-{\sqrt{2}}{\bar{h}}^*{\bar{g}}{\cal{C}}_1(U_T,1,6)\nnu &&
+{\sqrt{2}}{\bar{g}}^*{{\bar{h}}}{\cal{C}}_1(U_T,6,1)
+2{{\bar{h}}}^*{{\bar{h}}}{\cal{F}}_1(V_T,1)
+{{\bar{h}}}^*{{\bar{h}}}{\cal{F}}_1'(U_H,1)\nnu &&
-{\bar{g}}^*{\bar{g}}{\cal{F}}_1'(U_H,5)
-(1/3){\bar{g}}^*{\bar{g}}{\cal{F}}_1'(U_H,6)
-{\bar{h}}^*{\bar{g}}{\cal{C}}_1'(U_H,1,5)\nnu &&
-(i/{\sqrt{3}}){\bar{h}}^*{\bar{g}}{\cal{C}}_1'(U_H,1,6)
-(i/{\sqrt{3}}){\bar{g}}^*{\bar{g}}{\cal{C}}_1'(U_H,5,6) +
(-{\bar{h}}^*{\bar{g}}{\cal{C}}_1'(U_H,1,5)\nnu &&
-(i/{\sqrt{3}}){\bar{h}}^*{\bar{g}}{\cal{C}}_1'(U_H,1,6)
-(i/{\sqrt{3}}){\bar{g}}^*{\bar{g}}{\cal{C}}_1'(U_H,5,6))^\dagger
+{{\bar{h}}}^*{{\bar{h}}}{\cal{F}}_1'(V_H,1)\nnu &&
-{\bar{g}}^*{\bar{g}}{\cal{F}}_1'(V_H,5)
-(1/3){\bar{g}}^*{\bar{g}}{\cal{F}}_1'(V_H,6)
-{\bar{h}}^*{\bar{g}}{\cal{C}}_1'(V_H,1,5)\nnu &&
-(i/{\sqrt{3}}){\bar{h}}^*{\bar{g}}{\cal{C}}_1'(V_H,1,6)
-(i/{\sqrt{3}}){\bar{g}}^*{\bar{g}}{\cal{C}}_1'(V_H,5,6) +
(-{\bar{h}}^*{\bar{g}}{\cal{C}}_1'(V_H,1,5)\nnu &&
-(i/{\sqrt{3}}){\bar{h}}^*{\bar{g}}{\cal{C}}_1'(V_H,1,6)
-(i/{\sqrt{3}}){\bar{g}}^*{\bar{g}}{\cal{C}}_1'(V_H,5,6))^\dagger
+{\bar{g}}^*{\bar{g}}(-(32/3){\cal{F}}_1(U_C,3)\nnu &&
-(32/3){\cal{F}}_1(V_C,3) -4{\cal{F}}_1(U_D,3)
-4{\cal{F}}_1(U_E,6) \nnu && -4{\cal{F}}_1(V_P,2)
-8{\cal{F}}_1(V_P,2) -6{\cal{F}}_1(U_P,2) \nnu
&&-8{\cal{F}}_1(U_L,2)) - (2 g^2)(25{\cal{F}}_1(V_G,6)
+0.5{\cal{F}}_1(V_J,4) \nnu && +1.5{\cal{F}}_1(V_X,3)
+1.5{\cal{F}}_1(V_E,5) ))= ({ {32\pi^2}}) {\Delta}_d\nonumber
\quad\quad \quad\quad(\mathbf{A7})\eea

\bea ({ {32\pi^2}}) {\Delta}_e &=&
 (3{{\bar{h}}}^*{{\bar{h}}}{\cal{F}}_1(U_T,1)
-6{\bar{g}}^*{\bar{g}}{\cal{F}}_1(U_T,6)
+3{\sqrt{2}}{\bar{h}}^*{\bar{g}}{\cal{C}}_1(U_T,1,6)\nnu &&
-3{\sqrt{2}}{\bar{g}  }.{{\bar{h}}}{\cal{C}}_1(U_T,6,1)
+{{\bar{h}}}^*{{\bar{h}}}{\cal{F}}_1'(V_H,1)
-{\bar{g}}^*{\bar{g}}{\cal{F}}_1'(V_H,5)\nnu &&
-3{\bar{g}}^*{\bar{g}}{\cal{F}}_1'(V_H,6)
+{\bar{h}}^*{\bar{g}}{\cal{C}}_1'(V_H,1,5)
-(i{\sqrt{3}}){\bar{h}}^*{\bar{g}}{\cal{C}}_1'(V_H,1,6)\nnu &&
+(i{\sqrt{3}}){\bar{g}}^*{\bar{g}}{\cal{C}}_1'(V_H,5,6) +
({\bar{h}}^*{\bar{g}}{\cal{C}}_1'(V_H,1,5)
-(i{\sqrt{3}}){\bar{h}}^*{\bar{g}}{\cal{C}}_1'(V_H,1,6)\nnu &&
+(i{\sqrt{3}}){\bar{g}}^*{\bar{g}}{\cal{C}}_1'(V_H,5,6))^\dagger
+{{\bar{h}}}^*{{\bar{h}}}{\cal{F}}_1'(U_H,1)
-{\bar{g}}^*{\bar{g}}{\cal{F}}_1'(U_H,5)\nnu &&
-3{\bar{g}}^*{\bar{g}}{\cal{F}}_1'(U_H,6)
-{\bar{h}}^*{\bar{g}}{\cal{C}}_1'(U_H,1,5)
+(i{\sqrt{3}}){\bar{h}}^*{\bar{g}}{\cal{C}}_1'(U_H,1,6)\nnu &&
+(i{\sqrt{3}}){\bar{g}}^*{\bar{g}}{\cal{C}}_1'(U_H,5,6) +
(-{\bar{h}}^*{\bar{g}}{\cal{C}}_1'(U_H,1,5)
+(i{\sqrt{3}}){\bar{h}}^*{\bar{g}}{\cal{C}}_1'(U_H,1,6)\nnu &&
+(i{\sqrt{3}}){\bar{g}}^*{\bar{g}}{\cal{C}}_1'(U_H,5,6))^\dagger
+{\bar{g}}^*{\bar{g}}( -4{\cal{F}}_1(V_F,4)
-12{\cal{F}}_1(V_E,6)\nnu && -12{\cal{F}}_1(V_D,3)
-18{\cal{F}}_1(U_P,2)) - (2 g^2)((9/40){\cal{F}}_1(V_G,6)\nnu &&
+1.5{\cal{F}}_1(V_J,4) +1.5{\cal{F}}_1(U_X,3)
+1.5{\cal{F}}_1(U_E,5 ))=({ {32\pi^2}}) {\Delta}_\nu \nonumber
\quad\quad \quad\quad(\mathbf{A8})\eea

\bea {{( 32\pi^2 ) \Delta_{H^0}}\over {|V^H_{11}|^2}} &=&  {\bar
\gamma}^2{\cal{F}}_2'(V_G,U_H,2,2) + {\gamma}
^2{\cal{F}}_2'(V_G,U_H,2,3)
+{\bar\gamma}{\gamma}{\cal{C}}_2'(V_G,U_H,2,2,3)\nnu && +{\bar
\gamma}{\gamma} ({\cal{C}}_2'(V_G,U_H,2,2,3))^8 +8( {\bar
 \gamma}^2{\cal{F}}_2(V_R,U_C,1,1)
+ \gamma^2{\cal{F}}_2(V_R,U_C,1,2)\nnu && +{\bar \gamma}
{\gamma}{\cal{C}}_2(V_R,U_C,1,1,2) +{\bar \gamma} {\gamma}
({\cal{C}}_2(V_R,U_C,1,1,2))^*) +3( {\bar \gamma }
^2{\cal{F}}_2(V_J,U_D,2,2)  \nnu &&
+({\gamma})^2{\cal{F}}_2(V_J,U_D,2,1) +{\bar \gamma
}{\gamma}{\cal{C}}_2(V_J,U_D,2,2,1) +{\bar \gamma} {\gamma}
({\cal{C}}_2(V_J,U_D,2,2,1))^*)\nnu && +3(({\bar \gamma}
)^2{\cal{F}}_2(U_J,V_E,2,1) +({\gamma})^2{\cal{F}}_2(U_J,V_E,2,2)
+{\bar \gamma} {\gamma}{\cal{C}}_2(U_J,V_E,2,1,2)\nnu && +{\bar
\gamma} {\gamma} ({\cal{C}}_2(U_J,V_E,2,1,2))^*) +3( {\bar \gamma}
^2{\cal{F}}_2(U_E,V_T,4,2) + {\gamma}
^2{\cal{F}}_2(U_E,V_T,4,3)\nnu && +{\bar \gamma}
{\gamma}{\cal{C}}_2(U_E,V_T,4,3,2) +{\bar \gamma} {\gamma}
({\cal{C}}_2(U_E,V_T,4,3,2))^*) +3( {\bar \gamma}
^2{\cal{F}}_2(V_X,U_T,2,2)\nnu && + {\gamma}
^2{\cal{F}}_2(V_X,U_T,2,3) +{\bar \gamma}
{\gamma}{\cal{C}}_2(V_X,U_T,2,3,2) +{\bar \gamma} {\gamma}
({\cal{C}}_2(V_X,U_T,2,3,2))^*)\nnu &&
+6{\gamma}^2{\cal{F}}_1^u(V_L,1,M_Y) +12{\gamma}^2
{\cal{F}}_1(M_B,M_M) +3{\gamma}^2{\cal{F}}_2(V_X,U_T,1,4)\nnu &&
+6{\gamma}^2{\cal{F}}_2(V_E,U_J,3,1) +18{\bar \gamma }^2
F_1(M_Y,M_W) +9{\bar \gamma}^2{\cal{F}}_2(V_X,U_P,1,1)\nnu &&
+{\gamma}^2{\cal{F}}_1^u(V_F,1,M_V)
+3{\gamma}^2{\cal{F}'}_1^{u}(V_H,4,M_O)
+2{\gamma}^2{\cal{F}}_2'(V_G,U_H,4,4)\nnu &&
+18{\gamma}^2F_1(M_B,M_W) +9{\gamma}^2{\cal{F}}_2(V_P,U_E,1,3)
+6{\bar \gamma} ^2{\cal{F}}_1^u(V_L,1,M_B)\nnu && +3{\bar \gamma}
^2{\cal{F}}_2(V_T,U_E,4,3) +12{\bar\gamma}^2 F_1(M_Y,M_N)
+3{\bar \gamma} ^2F_1(M_V,M_O)\nnu && +2{\bar \gamma}^2
F_1(M_V,M_A) +{\bar \gamma}^2{\cal{F}}_2'(U_F,V_H,1,4)  +6{\bar
\gamma}^2{\cal{F}}_2(V_K,U_X,1,1)\nnu && +4{\bar
\gamma}^2{\cal{F}}_2(V_R,U_C,2,1)
+4{\gamma}^2{\cal{F}}_2(V_R,U_C,2,2) +0.5{\bar \gamma}
^2{\cal{F}}_2'(V_G,U_H,3,2)\nnu &&
+0.5{\gamma}^2{\cal{F}}_2'(V_G,U_H,3,3)
+1.5{\gamma}^2{\cal{F}}_2(V_J,U_D,3,1) +1.5{\bar
\gamma}^2{\cal{F}}_2(V_J,U_D,3,2)\nnu &&
+1.5{\gamma}^2{\cal{F}}_2(V_J,U_E,3,2) +1.5{\bar
\gamma}^2{\cal{F}}_2(V_J,U_E,3,1)
+8{\gamma}^2{\cal{F}}_1^u(V_C,1,M_Z)\nnu && +8{\bar
\gamma}^2{\cal{F}}_1^u(V_C,2,M_Z)
+{\gamma}^2{\cal{F}}_2'(U_F,V_H,2,3) +{\bar
\gamma}^2{\cal{F}}_2'(U_F,V_H,2,2)\nnu &&
+3{\gamma}^2{\cal{F}}_2(V_T,U_E,5,1) +3{\bar
\gamma}^2{\cal{F}}_2(V_T,U_E,5,2) +3{\bar
\gamma}^2{\cal{F}}_1^u(V_D,1,M_I)\nnu &&
+3{\gamma}^2{\cal{F}}_1^u(V_D,2,M_I)
+12{\gamma}^2{\cal{F}}_1^u(U_C,2,M_Q) +12{\bar
\gamma}^2{\cal{F}}_1^u(U_C,1,M_Q)\nnu &&
+1.5{\gamma}^2{\cal{F}'}_1^{u}(U_H,3,M_S) +1.5{\bar
\gamma}^2{\cal{F}'}_1^{u}(U_H,2,M_S) +4.5{\bar
\gamma}^2{\cal{F}}_1^u(U_D,1,M_U)\nnu &&
+4.5{\gamma}^2{\cal{F}}_1^u(U_D,2,M_U) +4.5{\bar
\gamma}^2{\cal{F}}_1^u(V_E,1,M_U)
+4.5{\gamma}^2{\cal{F}}_1^u(V_E,2,M_U)\nnu &&
+|k|^2({\cal{F}}_2'(V_G,U_H,1,5) +6{\cal{F}}_1^u(V_L,2,M_B)
+3{\cal{F}}_2(V_T,U_E,6,3)\nnu && +6{\cal{F}}_1^u(U_L,2,M_Y)
+3{\cal{F}}_2(V_X,U_T,1,6) +{\cal{F}}_2'(U_F,V_H,4,4)\nnu &&
+{\cal{F}}_1^u(V_F,4,M_V) +4{\cal{F}}_2(V_R,U_C,2,3)
+8{\cal{F}}_1^u(V_C,3,M_Z) \nnu && +0.5{\cal{F}}_2'(V_G,U_H,2,6)
+{\cal{F}}_2'(U_F,V_H,2,6) +1.5{\cal{F}}_2(V_E,U_J,6,3)\nnu &&
+3{\cal{F}}_1^u(V_D,3,M_I) +1.5{\cal{F}}_2(V_J,U_D,3,3)
+3{\cal{F}}_2(V_T,U_E,5,6)\nnu && +1.5{\cal{F}'}_1^{u}(V_H,6,M_S)
+12{\cal{F}}_1^u(V_C,3,M_Q) +4.5{\cal{F}}_1^u(V_D,3,M_U)\nnu &&
+4.5{\cal{F}}_1^u(U_E,6,M_U)+1.5{\cal{F}}_2(V_T,U_E,7,4)
+3{\cal{F}}_2(V_K,U_X,2,2)\nnu && +1.5{\cal{F}}_2(V_X,U_T,2,7)
+3{\cal{F}}_2(V_E,U_J,4,5) +4.5{\cal{F}}_2(V_P,U_E,2,4)\nnu &&
+4.5{\cal{F}}_2(V_X,U_P,2,2) ) -(2 g^2)(
+1.5{\cal{F}}_2(V_T,U_X,1,3)\nnu && +1.5{\cal{F}}_2(V_E,U_T,5,1)
+0.1{\cal{F}}_2'(V_G,V_H,6,1) +0.5{\cal{F}}_2'(V_F,U_H,3,1) ))
\nonumber \quad\quad \quad\quad(\mathbf{A9})\eea

\bea{{ (32\pi^2) \Delta_{\bar{H}^0}}\over{|U^H_{11}|^2}} &=& ( 8(
{\bar \gamma}^2{\cal{F}}_2(U_R,V_C,1,2)
+({\gamma})^2{\cal{F}}_2(U_R,V_C,1,1) +{\bar \gamma}
{\gamma}{\cal{C}}_2(U_R,V_C,1,2,1) \nnu &&+{\bar \gamma} {\gamma}
({\cal{C}}_2(U_R,V_C,1,2,1))^*) +{\bar\gamma}^2
{\cal{F}}_2'(U_G,V_H,2,2)
+({\gamma})^2{\cal{F}}_2'(U_G,V_H,2,3)\nnu && +{\bar \gamma}
{\gamma}{\cal{C}}_2'(U_G,V_H,2,2,3) +{\bar \gamma} {\gamma}
({\cal{C}}_2'(U_G,V_H,2,2,3))^*
+3({\bar\gamma}^2{\cal{F}}_2(U_J,V_D,2,1)\nnu &&
+({\gamma})^2{\cal{F}}_2(U_J,V_D,2,2) +{\bar \gamma}
{\gamma}{\cal{C}}_2(U_J,V_D,2,1,2) +{\bar \gamma }{\gamma}
({\cal{C}}_2(U_J,V_D,2,1,2))^*)\nnu && +3({\bar\gamma}
^2{\cal{F}}_2(V_J,U_E,2,2) +({\gamma})^2{\cal{F}}_2(V_J,U_E,2,1)
+{\bar \gamma}{\gamma}{\cal{C}}_2(V_J,U_E,2,2,1)\nnu && +{\bar
\gamma}{\gamma} ({\cal{C}}_2(V_J,U_E,2,2,1))^*)
+3({\bar\gamma}^2{\cal{F}}_2(U_X,V_T,2,2)\nnu &&
+({\gamma})^2{\cal{F}}_2(U_X,V_T,2,3) +{\bar
\gamma}{\gamma}{\cal{C}}_2(U_X,V_T,2,2,3) +{\bar\gamma}{\gamma}
({\cal{C}}_2(U_X,V_T,2,2,3))^*)\nnu &&
+3({\bar\gamma}^2{\cal{F}}_2(V_E,U_T,4,2) +(
{\gamma}^2{\cal{F}}_2(V_E,U_T,4,3) +{\bar
\gamma}{\gamma}{\cal{C}}_2(V_E,U_T,4,3,2)\nnu && +{\bar
\gamma}{\gamma} ({\cal{C}}_2(V_E,U_T,4,3,2))^*) +12{\gamma}^2
F_1(M_Y,M_N) +6{\gamma}^2{\cal{F}}_1^u(U_L,1,M_B)\nnu &&
+3{\gamma}^2{\cal{F}}_2(U_T,V_E,4,3)
+6{\gamma}^2{\cal{F}}_2(U_K,V_X,1,1) +18{\bar
\gamma}^2F_1(M_B,M_W)\nnu && +9{\bar
\gamma}^2{\cal{F}}_2(V_E,U_P,3,1) + 3{\gamma}^2F_1(M_V,M_O) +
2{\gamma}^2F_1(M_V,M_A) +{\gamma}^2{\cal{F}}_2'(V_F,U_H,1,4)\nnu
&& +18{\gamma}^2F_1(M_Y,M_W) +{\bar
\gamma}^2{\cal{F}}_1^u(U_F,1,M_V) +3{\bar
\gamma}^2F_1(M_V,M_O)\nnu && +2{\bar
\gamma}^2{\cal{F}}_2'(V_G,V_H,4,4)
+9{\gamma}^2{\cal{F}}_2(V_P,U_X,1,1) +3{\bar
\gamma}^2{\cal{F}}_2(V_T,U_X,4,1) \nnu && +6{\bar
\gamma}^2{\cal{F}}_2(V_J,U_E,1,4) +6{\bar
\gamma}^2{\cal{F}}_1^u(V_L,1,M_Y) +12{\bar
\gamma}^2F_1(M_B,M_M)\nnu && +4{\bar
\gamma}^2{\cal{F}}_2(V_R,V_C,2,2)
+4{\gamma}^2{\cal{F}}_2(V_R,V_C,2,1) +0.5{\bar
\gamma}^2{\cal{F}}_2'(V_G,U_H,3,2)\nnu &&
+0.5{\gamma}^2{\cal{F}}_2'(V_H,U_G,3,3)
+1.5{\gamma}^2{\cal{F}}_2(V_D,U_J,2,3) +1.5{\bar
\gamma}^2{\cal{F}}_2(V_D,U_J,1,3)\nnu &&
+1.5{\gamma}^2{\cal{F}}_2(V_J,U_E,3,1) +1.5{\bar
\gamma}^2{\cal{F}}_2(V_J,U_E,3,2)
+8{\gamma}^2{\cal{F}}_1^u(U_C,2,M_Z) \nnu &&+8{\bar
\gamma}^2{\cal{F}}_1^u(U_C,1,M_Z)
+{\gamma}^2{\cal{F}}_2'(V_F,U_H,2,3)\nnu && +{\bar
\gamma}^2{\cal{F}}_2'(V_F,U_H,2,2)
+3{\gamma}^2{\cal{F}}_2(U_T,V_E,5,2) +3{\bar
\gamma}^2{\cal{F}}_2(U_T,V_E,5,1)\nnu && +3{\bar
\gamma}^2{\cal{F}}_1^u(U_D,2,M_I)
+3{\gamma}^2{\cal{F}}_1^u(U_D,1,M_I)
+12{\gamma}^2{\cal{F}}_1^u(V_C,1,M_Q) \nnu &&+12{\bar
\gamma}^2{\cal{F}}_1^u(V_C,2,M_Q)
+1.5{\gamma}^2{\cal{F}'}_1^{u}(V_H,3,M_S) +1.5{\bar
\gamma}^2{\cal{F}'}_1^{u}(V_H,2,M_S,M_H,6) \nnu &&+4.5{\bar
\gamma}^2{\cal{F}}_1^u(U_E,2,M_U
+4.5{\gamma}^2{\cal{F}}_1^u(U_E,1,M_U +4.5{\bar
\gamma}^2{\cal{F}}_1^u(V_D,1,M_U)\nnu &&
+4.5{\gamma}^2{\cal{F}}_1^u(V_D,2,M_U)
+|k|^2({\cal{F}}_2'(V_G,V_H,1,5) +6{\cal{F}}_1^u(V_L,2,M_Y)\nnu &&
+3{\cal{F}}_2(U_X,V_T,1,6) +{\cal{F}}_2'(V_F,U_H,4,4)
+{\cal{F}}_1^u(U_F,4,M_V)\nnu && +6{\cal{F}}_1^u(U_L,2,M_B)
+3{\cal{F}}_2(U_T,V_E,6,3) +8{\cal{F}}_1^u(U_C,3,M_Z)\nnu &&
+{\cal{F}}_2'(V_F,U_H,2,6) +4{\cal{F}}_2(V_R,V_C,2,3)
+0.5{\cal{F}}_2'(V_G,V_H,3,6) \nnu &&+1.5{\cal{F}}_2(U_E,V_J,6,3)
+3{\cal{F}}_1^u(U_D,3,M_I) +1.5{\cal{F}}_2(V_D,U_J,3,3)\nnu &&
+3{\cal{F}}_2(V_E,U_T,6,5) +1.5{\cal{F}'}_1^{u}(U_H,6,M_S)
+12{\cal{F}}_1^u(U_C,3,M_Q) \nnu &&+4.5{\cal{F}}_1^u(U_D,3,M_U)
+4.5{\cal{F}}_1^u(V_E,6,M_U) +1.5{\cal{F}}_2(V_E,U_T,4,7)\nnu &&
+3{\cal{F}}_2(U_K,V_X,2,2) +1.5{\cal{F}}_2(U_X,V_T,2,7)
+3{\cal{F}}_2(U_E,V_J,4,5) \nnu &&+4.5{\cal{F}}_2(U_P,V_E,2,4)
+4.5{\cal{F}}_2(U_X,V_P,2,2) ) -(2 g^2)(
+1.5{\cal{F}}_2(U_T,V_X,1,3) \\ &&+1.5{\cal{F}}_2(U_E,T,5,1)
+0.1{\cal{F}}_2'(G,U_H,6,1) +0.5{\cal{F}}_2'(U_F,H,3,1) ))
\nonumber \quad\quad \quad\quad(\mathbf{A10}) \eea \bea
\nonumber\eea

 \section*{Appendix B : Discussion of RGE features}

We  used the two loop Renormalization group evolution equations
for the softly broken MSSM given in \cite{martinRG} to evolve
randomly chosen SUGRY-NUHM parameters
 $\{m_{\tilde f},m_{1/2},A_0,B, \hfil\break M^2_{H,\bar H}\} $  together with the  $ \mu
 $ parameter and the rest of the superpotential parameters down
 from $M_X^0=10^{16.25}$ GeV to $M_Z$. It is notable that the
universal  gaugino mass $M_{1/2}$  at $M_X$ is found to be
negative in all our fits. Furthermore the Higgs mass parameters
$M^2_{H,{\bar H}}$   are very large and negative being typically
$\sim - 10^4$ TeV${}^2$.  The sfermion beta functions at one loop
contain terms  proportional to the Higgs mass parameters squared
\cite{ramondSMMSSM,martinRG}. For example :

\bea \beta^{(1)}_{{\bf{m}}_Q^2} \> = \> & ({{\bf{m}}_Q^2} + 2
M^2_{H}) {\bf Y}_u^\dagger \mathbf{Y}_u + ({{\bf{m}}_Q^2} + 2
M^2_{{\bar H}})\mathbf{Y}_d^\dagger \mathbf{Y}_d  +........\eea

which dominate the RGE for the third generation sfermions and
drive their masses far above the those of the first two
generations as one flows from $M_X$ to $M_Z$. This behaviour is
only slightly modulated by the contributions of $A_0$ and is a
  one-loop feature immune to significant modification by the two loop contributions.
 The presence of terms $A^\dagger A$ added to
the Higgs contributions has a countervailing effect on the scalar
mass evolution since it tends to decrease the mass squared in the
infrared. However the huge Higgs masses always prevail resulting
in the third sgeneration always being heavier than the first two.
This is an invariant feature of our spectra.  Heavy third
sgeneration is a distinctive feature of our fits and counterposes
them  to all(to our knowledge)  previous GUT based predictions
which have a third sgeneration \emph{lighter} than the first
two.

These tendencies can be clearly seen in the plots of the RG
evolution given in Figures 2-5 which refer to the actual two loop
RG evolution (Fig. 2) of $M^2_{\tilde d}$ , and the next three
figures refer to hypothetical cases with $\{A_0(M_X)=0$\}(Fig.3),
$\{M^2_H(M_X) =0=M^2_{\bar H}(M_X)\}$(Fig.4),$\{M^2_H(M_X)
=0=M^2_{\bar H}(M_X)=A_0(M_X)\}$(Fig.5).

\begin{figure}
 \epsfxsize15cm\epsffile{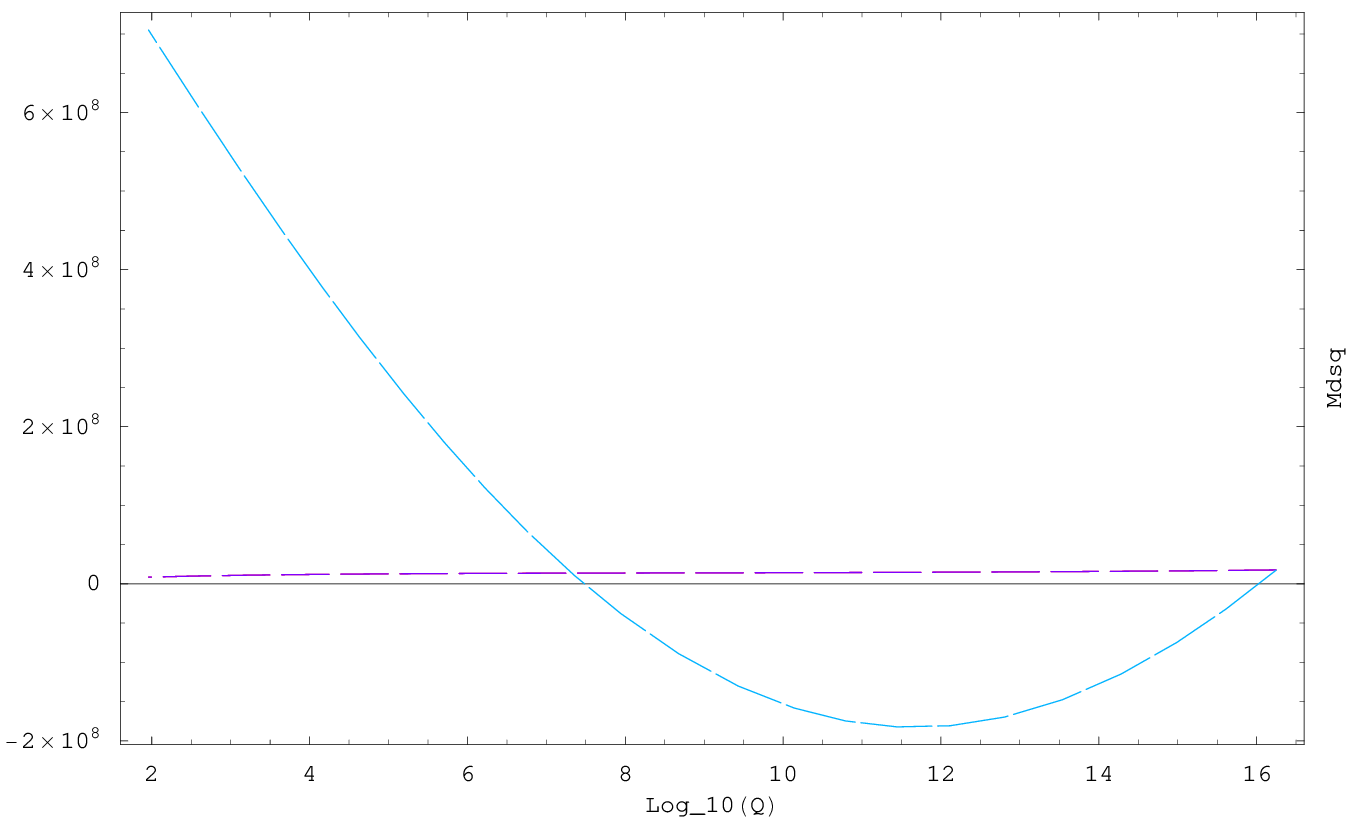}
\caption{Two loop RG evolution of $M_{\tilde {\bar
d}}^2 $ from $M_X^0$ to $M_Z$  for Case I-1. Red: $M_{\tilde {\bar
d}}^2 $, Blue: $M_{\tilde {\bar s}}^2 $, Green: $M_{\tilde {\bar
t}}^2 $. Note the strong growth in the the third sgeneration mass
at low energies. The same behaviour is exhibited by all
sfermions.}
\label{mdsq1}\end{figure}

\begin{figure}
 \epsfxsize15cm\epsffile{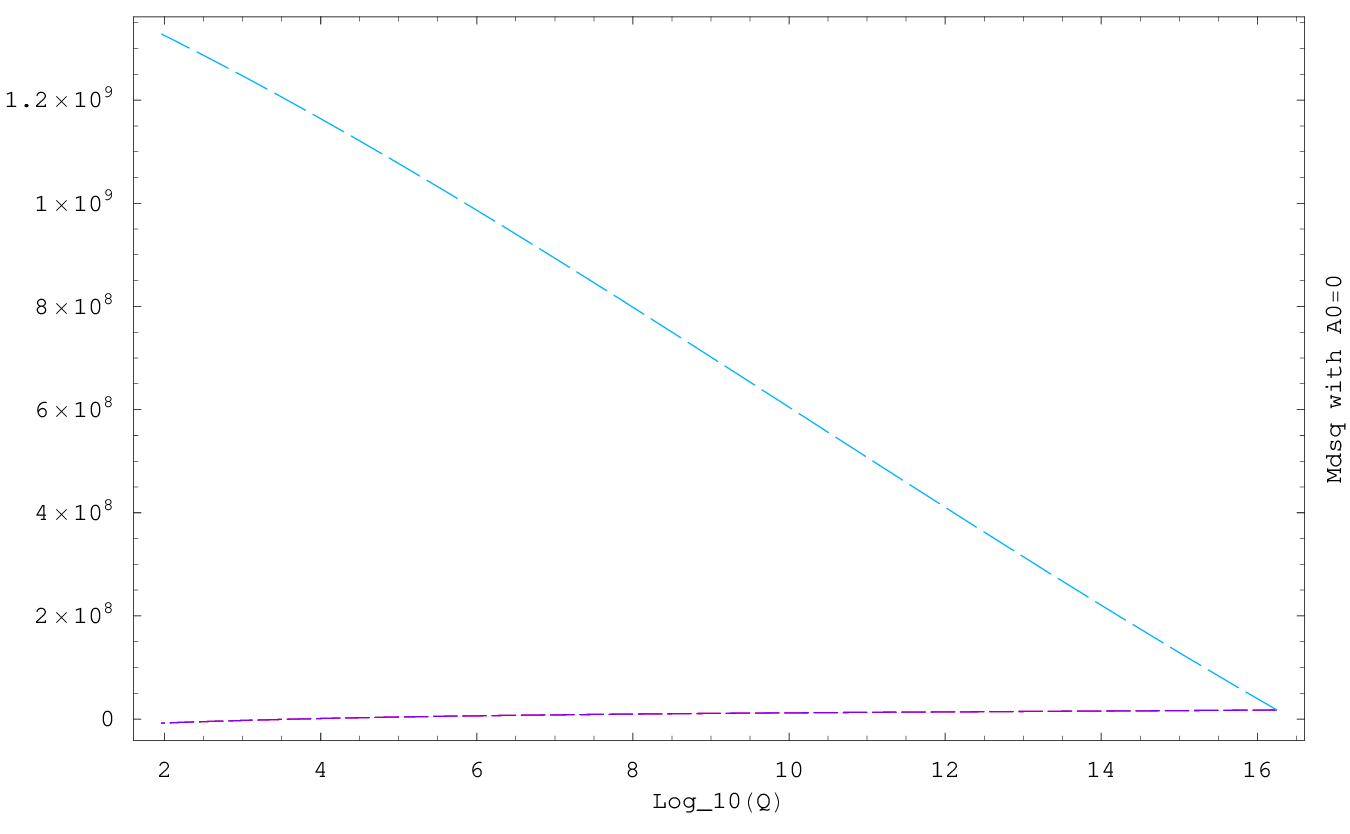}
\caption{Little  effect of $A_0$ : Hypothetical Two
loop RG evolution of $M_{\tilde d}^2 $ from $M_X^0$ to $M_Z$  with
$A_0(M_X)=0$  for Case I-1. Red: $M_{\tilde {\bar d}}^2 $, Blue:
$M_{\tilde {\bar s}}^2 $, Green: $M_{\tilde {\bar t}}^2 $. Note
the strong growth in the the third sgeneration mass at low
energies. The same behaviour is exhibited by all sfermions.  }
\label{mdsq2}\end{figure}

\begin{figure}
 \epsfxsize15cm\epsffile{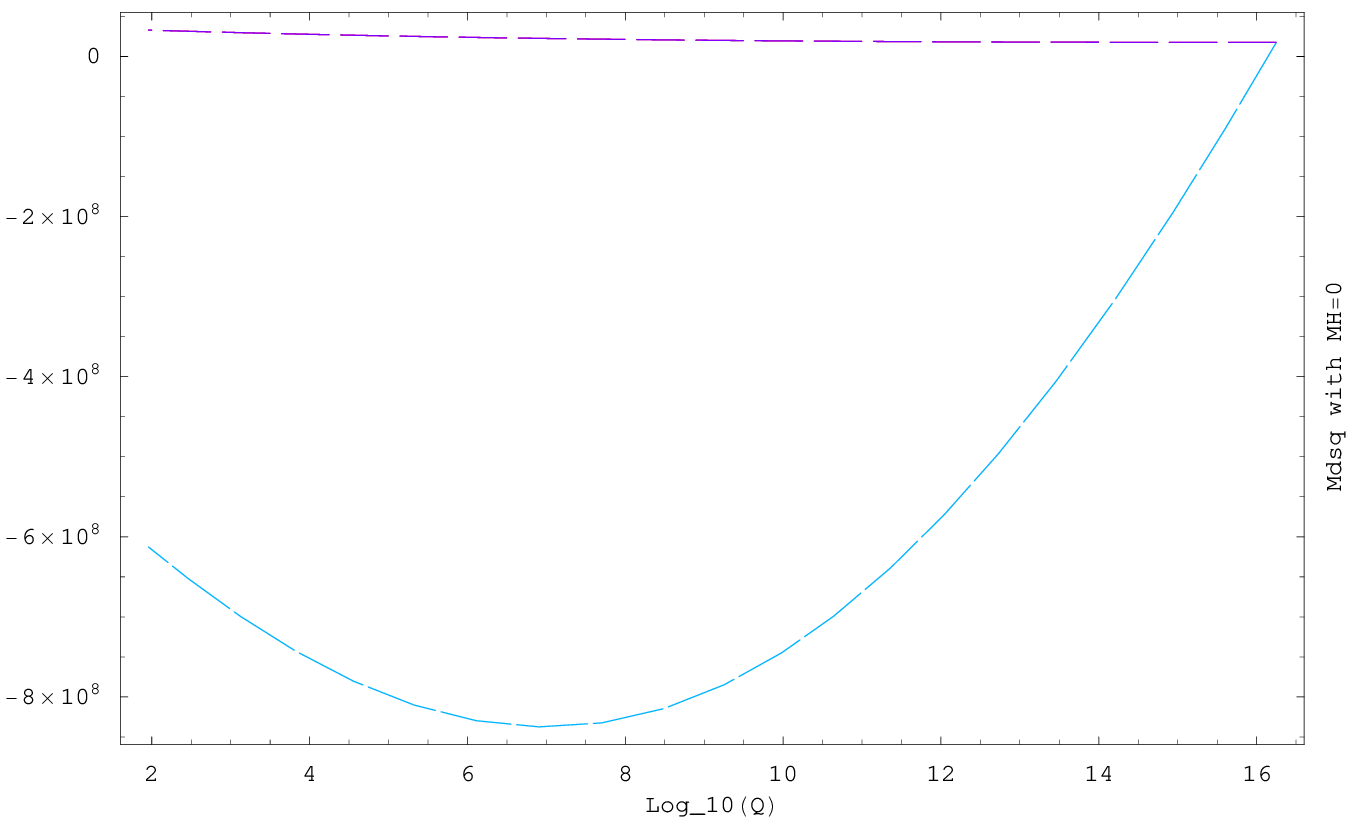}
\caption{Effect of large  $M^2_{H,{\bar H}}$ :
Hypothetical Two loop RG evolution of $M_{\tilde d}^2 $ from
$M_X^0$ to $M_Z$ with $M^2_H(M_X)=M^2_{\bar H}(M_X)=0$ for Case
I-1. Red: $M_{\tilde {\bar d}}^2 $, Blue: $M_{\tilde {\bar s}}^2
$, Green: $M_{\tilde {\bar t}}^2 $. Note the strong
\emph{decrease}  in the the third sgeneration mass at low energies
while the first two generations are unaffected. Putting $A_0=0$
has essentially no effect except that the increase becomes linear.
The same behaviour is exhibited by all sfermions.}
\label{mdsq3}\end{figure}

\begin{figure}
 \epsfxsize15cm\epsffile{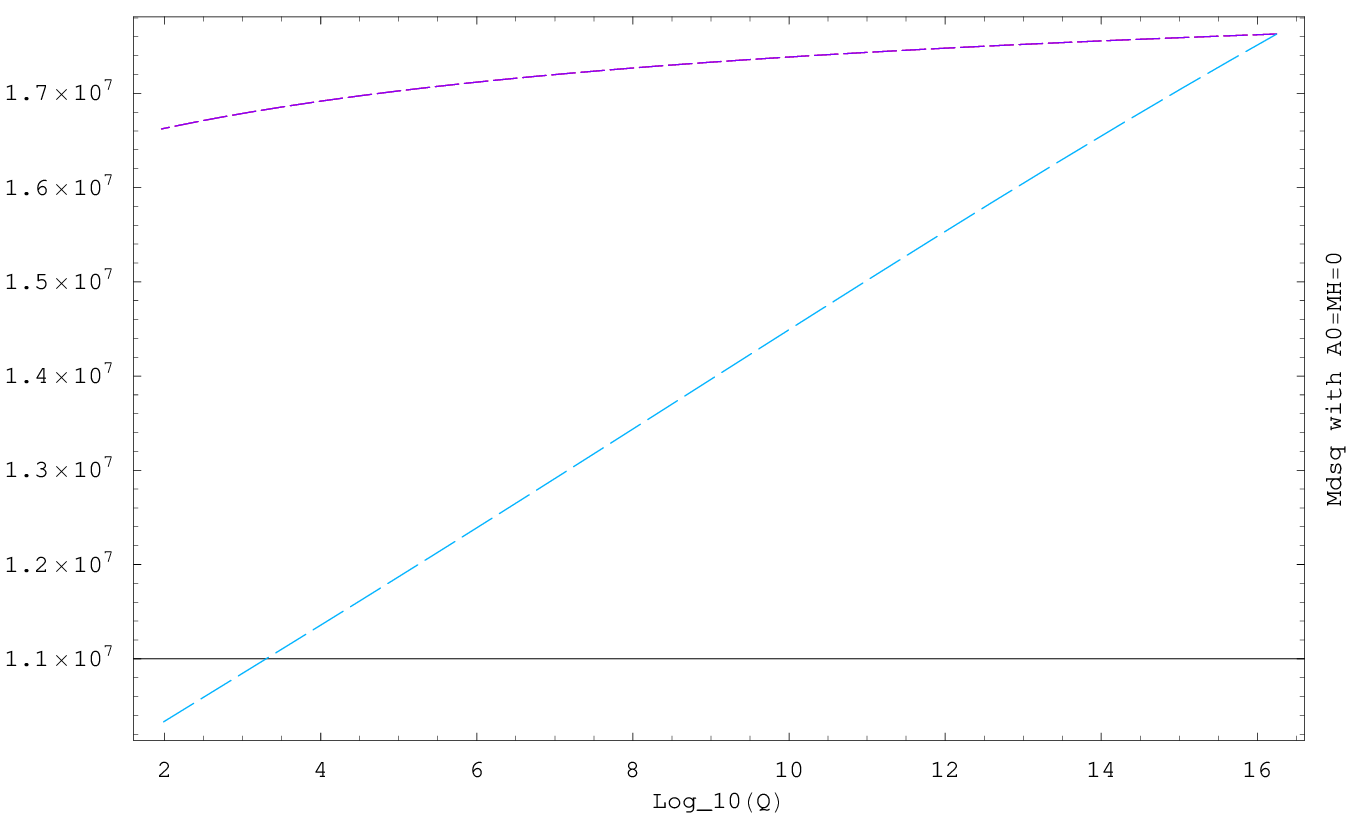}
\caption{Little  effect of $A_0$ : Hypothetical Two
loop RG evolution of $M_{\tilde d}^2 $ from $M_X^0$ to $M_Z$  with
$M^2_H(M_X)=M^2_{\bar H}(M_X)=0=A_0(M_X)$ for Case I-1. Note the
strong \emph{decrease}  in the the third sgeneration mass at low
energies while the first two generations are unaffected. The
removal of the curvilinear decrease in favour of a linear one is
the only effect of putting $A_0=0$ in addition. The same behaviour
is exhibited by all sfermions. }
\label{mdsq4}\end{figure}

Similarly we can understand why the one loop  prediction ($M_1
:M_2:M_3$  as $g_1^2:g_2^2:g_3^2$  which is  $\simeq 1:2:7$  at
$M_Z$.) for the ratio of gaugino masses $M_i$  which   follows
from the 1-loop RG invariance of $M_i/g_i^2, i=1,2,3$  is badly
violated at two loops when $A_0$ is large. Although the $1:2:7$
seems set in stone by the known gauge couplings at $M_Z$ if GUT
mandated equality of gaugino masses at $M_X$ is accepted, the
influence of the additional terms\cite{ramondSMMSSM,martinRG} at
two loop in the gaugino mass RGE :

\bea  { {d\over dt} M_a } &=&   {2 g_a^2\over 16\pi^2} B^{(1)}_a
M_a + {2 g_a^2\over (16 \pi^2)^2} \biggl [ \sum_{b=1}^3
B^{(2)}_{ab} g_b^2 (M_a + M_b) \\ &&  \qquad\qquad\qquad +
\sum_{x=u,d,e} C_a^x \left ( {\rm Tr} [Y_x^\dagger A_x] - M_a {\rm
Tr } [Y_x^\dagger Y_x] \right ) \biggr ]      \eea

The terms containing the product of the  Yukawa gauge couplings
and the corresponding soft trilinear couplings that are generated
from $A_0$, imply that in practice the ratios can vary widely if
$A_0$ is large and gluinos can even be lighter than winos : This
is seen clearly from the graphs of the RG flow of the gaugino
masses with and without $A_0$ (Fig. 6) and the graph of the ratios
of the gaugino masses with and without $A_0$(Fig. 7).

\begin{figure}
 \epsfxsize15cm\epsffile{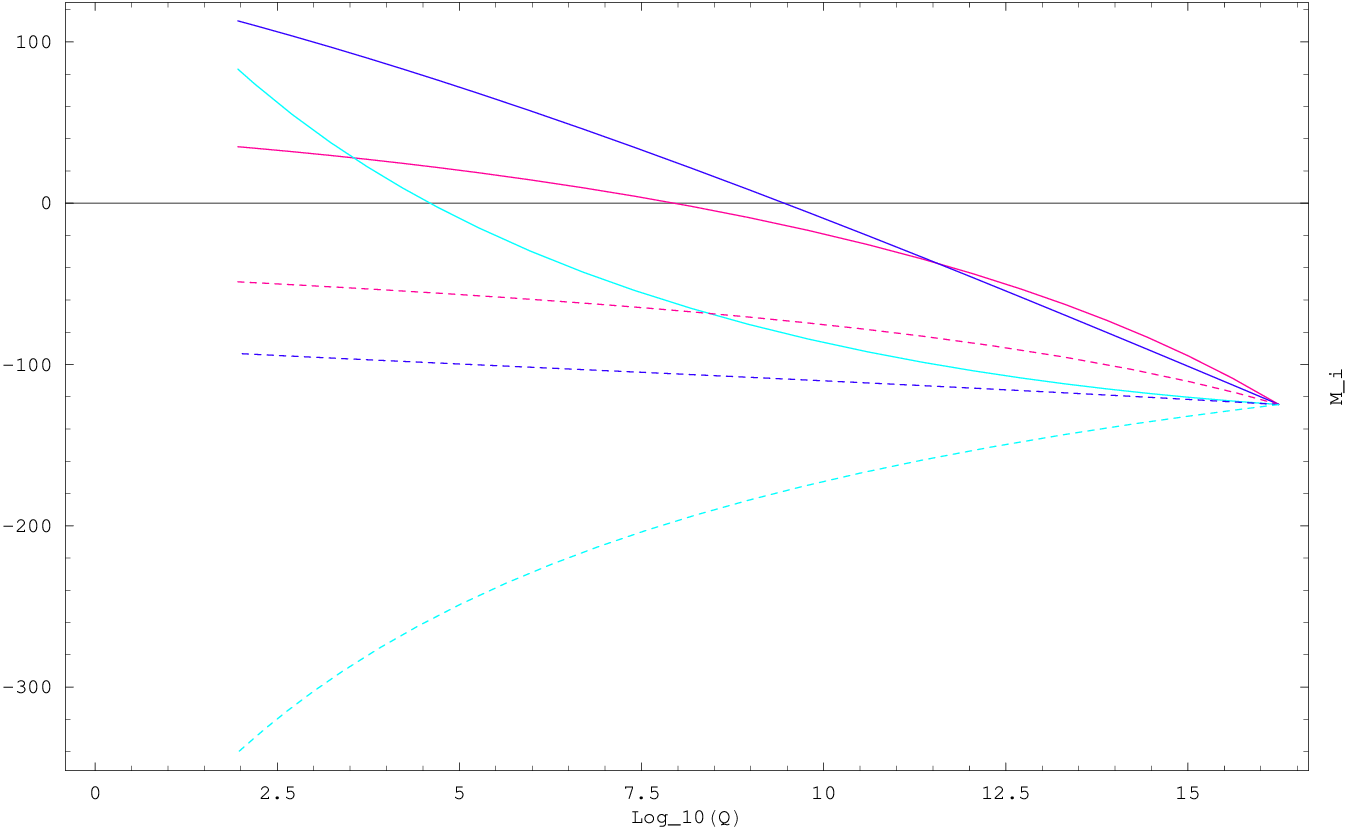}
\caption{ Hypothetical  2-loop RG evolution of
 gaugino masses with $A_0\neq 0$ (full lines) and with  $A_0= 0$ (dashed  lines)
   for Case II-1. Red:$M_1$, Blue $M_2$, Green $M_3$. Notice how
   the Gluino mass can even fall below the Wino mass when $A_0\neq
   0$.}
  \label{gino1}\end{figure}
\clearpage
\begin{figure}
 \epsfxsize15cm\epsffile{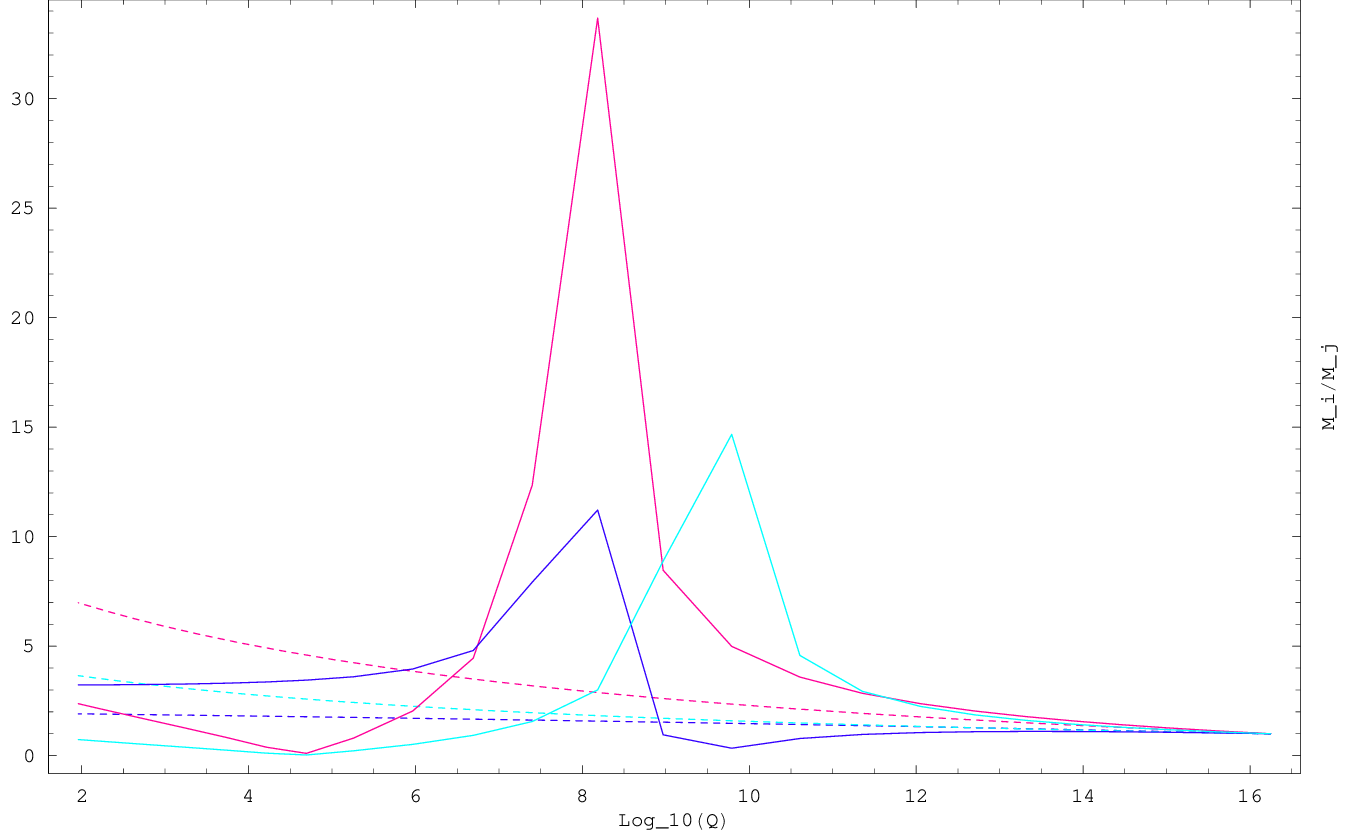}
\caption{ Hypothetical Two loop RG evolution of ratios of gaugino
mass ratos with $A_0\neq 0$ (full lines) and with $A_0= 0$ (dashed
lines)  for Case II-1.  Red : $M_3/M_1$, Blue: $M_2/M_1$ , Green
$M_3/M_2$. In the case $A_0=0$,  the gaugino masses follow the
standard evolution to the $1:2:7$ ratio at low energies.  }
\label{gino2}\end{figure}

\section*{Appendix C: Additional  Tables of Parameter values }

    \begin{table}
 $$
\begin{array}{|c|c|c|c|}
 \hline
 {\rm Parameter }&Value &{\rm  Field }&\hspace{10mm} Masses\\
 &&{\rm}[SU(3),SU(2),Y]&\hspace{10mm}( Units\,\,of 10^{16} GeV)\\ \hline
       \chi_{X}&  0.0232           &A[1,1,4]&  17462.07 \\ \chi_{Z}&
    0.0712
                &B[6,2,{5/3}]&            0.4028\\
           h_{11}/10^{-6}& -0.4808         &C[8,2,1]&{    733.73,   2706.98,   4133.40 }\\
           h_{22}/10^{-4}&  0.3364    &D[3,2,{7/ 3}]&{   1514.69,   9791.18,  11582.78 }\\
                   h_{33}&  0.0141     &E[3,2,{1/3}]&{      0.91,    529.34,    529.34 }\\
 f_{11}/10^{-6}&
 -0.0989-  0.0244i
                      &&{  1724.816,   5358.19,   7929.76 }\\
\hline\end{array}
 $$
  \caption{\small{I-2-a:   Column 1 contains values   of the NMSGUT-SUGRY-NUHM  parameters at $M_X$
  derived from an  accurate fit to all 18 fermion data and compatible with RG constraints.
  See caption to Table 2 for explanation.}}\label{I-2-a}\end{table}
 \begin{table}
 $$
 \begin{array}{|c|c|c|c|c|}
 \hline
 &&&&\\
 {\rm  Parameter }&Target =\bar O_i &Uncert.= \delta_i    &Achieved= O_i &Pull =(O_i-\bar O_i)/\delta_i\\
 \hline
    y_u/10^{-6}&  2.106596&  0.804720&  2.107032&  0.000542\\
    y_c/10^{-3}&  1.026873&  0.169434&  1.027891&  0.006009\\
            y_t&  0.388445&  0.015538&  0.388409& -0.002299\\
    y_d/10^{-5}&  5.979116&  3.485824&  6.027227&  0.013802\\
    y_s/10^{-3}&  1.131477&  0.534057&  1.125268& -0.011625\\
            y_b&  0.544641&  0.282669&  0.545745&  0.003905\\
\hline
 \end{array}
 $$
  \caption{\small{I-2-bFit   with $\chi_X=\sqrt{ \sum_{i=1}^{17}
 (O_i-\bar O_i)^2/\delta_i^2}=
    0.0232
 $. Target values,  at $M_X$ of the fermion yukawa
 couplings and mixing parameters, together with the estimated uncertainties, achieved values and pulls.
  See caption to Table 3 for explanation.}}
 \label{ I-2-b}\end{table}
 \begin{table}
 $$
 \begin{array}{|c|c|c|c|}
 \hline &&&\\ {\rm  Parameter }&SM(M_Z) & m^{GUT}(M_Z) & m^{MSSM}=(m+\Delta m)^{GUT}(M_Z) \\
 \hline
    m_d/10^{-3}&   2.90000&   0.52473&   3.07224\\
    m_s/10^{-3}&  55.00000&   9.79652&  57.51101\\
            m_b&   2.90000&   3.24592&   2.90989\\
    m_e/10^{-3}&   0.48657&   0.47320&   0.48644\\
         m_\mu &   0.10272&   0.09578&   0.10240\\
         m_\tau&   1.74624&   1.74402&   1.74715\\
    m_u/10^{-3}&   1.27000&   1.10467&   1.27126\\
            m_c&   0.61900&   0.53890&   0.62018\\
            m_t& 172.50000& 148.26724& 172.47971\\
 \hline
 \end{array}
 $$

 \caption{\small{I-2-c: Values of standard model
 fermion masses in GeV at $M_Z$ compared with the masses obtained from
 values of GUT derived  yukawa couplings  run down from $M_X^0$ to
 $M_Z$  both before and after threshold corrections.
  Fit with $\chi_Z=\sqrt{ \sum_{i=1}^{9} (m_i^{MSSM}- m_i^{SM})^2/ (m_i^{MSSM})^2} =
0.0712$.}}
  \label{ I-2-c}\end{table}

  \begin{table}
 $$
 \begin{array}{|c|c|c|c|}
 \hline
 {\rm  Parameter}  &Value&  {\rm  Parameter}&Value \\
 \hline
                       M_{1}&            104.66&   M_{{\tilde {\bar {u}}_1}}&           3630.75\\
                       M_{2}&            353.60&   M_{{\tilde {\bar {u}}_2}}&           3627.51\\
                       M_{3}&            268.96&   M_{{\tilde {\bar {u}}_3}}&          20705.08\\
     M_{{\tilde {\bar l}_1}}&           1887.68&               A^{0(l)}_{11}&        -110809.91\\
     M_{{\tilde {\bar l}_2}}&            265.45&               A^{0(l)}_{22}&        -110677.57\\
     M_{{\tilde {\bar l}_3}}&          11488.56&               A^{0(l)}_{33}&         -67135.87\\
        M_{{\tilde {L}_{1}}}&           6440.25&               A^{0(u)}_{11}&        -133970.62\\
        M_{{\tilde {L}_{2}}}&           6303.95&               A^{0(u)}_{22}&        -133969.80\\
        M_{{\tilde {L}_{3}}}&          10623.40&               A^{0(u)}_{33}&         -66830.19\\
     M_{{\tilde {\bar d}_1}}&            825.29&               A^{0(d)}_{11}&        -110623.57\\
     M_{{\tilde {\bar d}_2}}&            819.67&               A^{0(d)}_{22}&        -110622.83\\
     M_{{\tilde {\bar d}_3}}&          42238.45&               A^{0(d)}_{33}&         -37954.46\\
          M_{{\tilde {Q}_1}}&           5443.89&                   \tan\beta&             50.00\\
          M_{{\tilde {Q}_2}}&           5442.40&                    \mu(M_Z)&         112330.33\\
          M_{{\tilde {Q}_3}}&          33482.05&                      B(M_Z)&
          2.1590 \times 10^{   9}
 \\
 M_{\bar {H}}^2&
         -1.2007 \times 10^{  10}
 &M_{H}^2&
         -1.3857 \times 10^{  10}
 \\
 \hline
 \end{array}
 $$
  \caption{ \small {I-2-d: Values (GeV) in  of the soft Susy parameters  at $M_Z$
 (evolved from the soft SUGRY-NUHM parameters at $M_X$).
   See caption to Table 5 for explanation.}}
\label{ I-2-d} \end{table}

  \begin{table}
 $$
 \begin{array}{|c|c|}
 \hline {\mbox {Field } }&Mass(GeV)\\
 \hline
                M_{\tilde{G}}&            268.96\\
               M_{\chi^{\pm}}&            353.59,         112330.39\\
       M_{\chi^{0}}&            104.66,            353.59,         112330.36    ,         112330.37\\
              M_{\tilde{\nu}}&          6439.906,          6303.596,         10623.190\\
                M_{\tilde{e}}&           1888.22,           6440.43,            254.80   ,           6304.74,          10423.50,          11670.42  \\
                M_{\tilde{u}}&           3630.56,           5443.61,           3627.28   ,           5442.15,          20701.92,          33484.59  \\
                M_{\tilde{d}}&            825.70,           5444.24,            820.02   ,           5442.76,          33474.52,          42244.47  \\
                        M_{A}&         328626.38\\
                  M_{H^{\pm}}&         328626.39\\
                    M_{H^{0}}&         328626.36\\
                    M_{h^{0}}&            112.60\\
 \hline
 \end{array}
 $$
\caption{\small{I-2-e: Spectra of supersymmetric partners calculated ignoring generation mixing effects.
    See caption to Table 6 for explanation.}} \label{ I-2-e}\end{table}

\begin{table}
 $$
 \begin{array}{|c|c|}
 \hline {\mbox {Field } }&Mass(GeV)\\
 \hline
                M_{\tilde{G}}&            269.14\\
               M_{\chi^{\pm}}&            353.67,         112325.66\\
       M_{\chi^{0}}&            104.69,            353.67,         112325.64    ,         112325.65\\
              M_{\tilde{\nu}}&           6303.72,           6440.01,         10625.538\\
                M_{\tilde{e}}&            296.70,           1887.75,           6294.24   ,           6440.47,          10487.30,          11631.85  \\
                M_{\tilde{u}}&           3627.98,           3631.26,           5423.60   ,           5444.05,          20695.60,          33484.45  \\
                M_{\tilde{d}}&            822.45,            828.15,           5424.23   ,           5444.67,          33474.38,          42241.87  \\
                        M_{A}&         328751.52\\
                  M_{H^{\pm}}&         328751.53\\
                    M_{H^{0}}&         328751.50\\
                    M_{h^{0}}&            112.62\\
 \hline
 \end{array}
 $$
\caption{\small{I-2-f: Spectra of supersymmetric partners calculated including  generation mixing effects.
  See caption to Table 7 for explanation.}} \label{ I-2-f}\end{table}

 \begin{table}
 $$
 \begin{array}{|c|c|c|c|}
 \hline
 {\rm Parameter }&Value &{\rm  Field }&\hspace{10mm} Masses\\
 &&{\rm}[SU(3),SU(2),Y]&\hspace{10mm}( Units\,\,of 10^{16} GeV)\\ \hline
       \chi_{X}&  0.0483           &A[1,1,4]&    809.66 \\ \chi_{Z}&
    0.0065
                &B[6,2,{5/3}]&            0.2796\\
           h_{11}/10^{-6}&  0.0210         &C[8,2,1]&{     43.65,    387.28,    404.49 }\\
           h_{22}/10^{-4}&  0.0353    &D[3,2,{7/ 3}]&{     42.78,    445.41,    456.44 }\\
                   h_{33}&  0.0026     &E[3,2,{1/3}]&{      0.45,     32.26,     32.26 }\\
 f_{11}/10^{-6}&
  0.0781-  0.1376i
                      &&{    34.752,    488.08,    543.11 }\\
\hline\end{array}
 $$
\caption{\small{I-3-a Fit : Column 1 contains values   of the NMSGUT-SUGRY-NUHM  parameters at $M_X$
  derived from an  accurate fit to all 18 fermion data and compatible with RG constraints.
 See caption to Table 2 for explanation. }} \label{ I-3-a} \end{table}
 \begin{table}
 $$
 \begin{array}{|c|c|c|c|c|}
 \hline
 &&&&\\
 {\rm  Parameter }&Target =\bar O_i &Uncert.= \delta_i    &Achieved= O_i &Pull =(O_i-\bar O_i)/\delta_i\\
 \hline
    y_u/10^{-6}&  2.084307&  0.796205&  2.084498&  0.000240\\
    y_c/10^{-3}&  1.016016&  0.167643&  1.017365&  0.008048\\
            y_t&  0.383569&  0.015343&  0.383558& -0.000709\\
    y_d/10^{-5}&  7.238813&  4.220228&  7.247117&  0.001968\\
    y_s/10^{-3}&  1.372490&  0.647815&  1.376178&  0.005694\\
            y_b&  0.505108&  0.262151&  0.505457&  0.001331\\
\hline
 \end{array}
 $$
  \caption{\small{I-3-b: Fit   with $\chi_X=\sqrt{ \sum_{i=1}^{17}
 (O_i-\bar O_i)^2/\delta_i^2}=
    0.0483
 $. Target values,  at $M_X$ of the fermion yukawa
 couplings and mixing parameters, together with the estimated uncertainties, achieved values and pulls.
 See caption to Table 3 for explanation. }}
 \label{ I-3-b}\end{table}
 \begin{table}
 $$
 \begin{array}{|c|c|c|c|}
 \hline &&&\\ {\rm  Parameter }&SM(M_Z) & m^{GUT}(M_Z) & m^{MSSM}=(m+\Delta m)^{GUT}(M_Z) \\
 \hline
    m_d/10^{-3}&   2.90000&   0.62818&   2.90275\\
    m_s/10^{-3}&  55.00000&  11.88721&  54.95005\\
            m_b&   2.90000&   3.07412&   2.89882\\
    m_e/10^{-3}&   0.48657&   0.47199&   0.48604\\
         m_\mu &   0.10272&   0.09706&   0.10272\\
         m_\tau&   1.74624&   1.73492&   1.73563\\
    m_u/10^{-3}&   1.27000&   1.09437&   1.26906\\
            m_c&   0.61900&   0.53416&   0.61943\\
            m_t& 172.50000& 148.47236& 172.66514\\
 \hline
 \end{array}
 $$

 \caption{\small{I-3-c: Values of standard model
 fermion masses in GeV at $M_Z$ compared with the masses obtained from
 values of GUT derived  yukawa couplings  run down from $M_X^0$ to
 $M_Z$  both before and after threshold corrections.
  Fit with $\chi_Z=\sqrt{ \sum_{i=1}^{9} (m_i^{MSSM}- m_i^{SM})^2/ (m_i^{MSSM})^2} =
0.0065$.}}
  \label{ I-3-c }\end{table}

\begin{table}
 $$
 \begin{array}{|c|c|c|c|}
 \hline
 {\rm  Parameter}  &Value&  {\rm  Parameter}&Value \\
 \hline
                       M_{1}&            147.27&   M_{{\tilde {\bar {u}}_1}}&           6318.67\\
                       M_{2}&            406.23&   M_{{\tilde {\bar {u}}_2}}&           6317.40\\
                       M_{3}&            598.97&   M_{{\tilde {\bar {u}}_3}}&          21142.35\\
     M_{{\tilde {\bar l}_1}}&           1614.43&               A^{0(l)}_{11}&         -98630.83\\
     M_{{\tilde {\bar l}_2}}&            320.14&               A^{0(l)}_{22}&         -98506.40\\
     M_{{\tilde {\bar l}_3}}&          16050.48&               A^{0(l)}_{33}&         -59069.03\\
        M_{{\tilde {L}_{1}}}&           7827.10&               A^{0(u)}_{11}&        -117145.75\\
        M_{{\tilde {L}_{2}}}&           7747.15&               A^{0(u)}_{22}&        -117144.96\\
        M_{{\tilde {L}_{3}}}&          13923.16&               A^{0(u)}_{33}&         -59199.13\\
     M_{{\tilde {\bar d}_1}}&           3334.46&               A^{0(d)}_{11}&         -98885.51\\
     M_{{\tilde {\bar d}_2}}&           3333.11&               A^{0(d)}_{22}&         -98884.62\\
     M_{{\tilde {\bar d}_3}}&          36975.96&               A^{0(d)}_{33}&         -37795.22\\
          M_{{\tilde {Q}_1}}&           6120.46&                   \tan\beta&             51.50\\
          M_{{\tilde {Q}_2}}&           6119.44&                    \mu(M_Z)&         101035.83\\
          M_{{\tilde {Q}_3}}&          30209.06&                      B(M_Z)&
          1.7444 \times 10^{   9}
 \\
 M_{\bar {H}}^2&
         -1.0097 \times 10^{  10}
 &M_{H}^2&
         -1.1107 \times 10^{  10}
 \\
 \hline
 \end{array}
 $$
 \caption{ \small {I-3-d: Values (GeV) in  of the soft Susy parameters  at $M_Z$
 (evolved from the soft SUGRY-NUHM parameters at $M_X$).
  See caption to Table 5 for explanation. }}
 \label{ I-3-d } \end{table}

 \begin{table}
 $$
 \begin{array}{|c|c|}
 \hline {\mbox {Field } }&Mass(GeV)\\
 \hline
                M_{\tilde{G}}&            598.97\\
               M_{\chi^{\pm}}&            406.23,         101035.90\\
       M_{\chi^{0}}&            147.27,            406.23,         101035.87    ,         101035.87\\
              M_{\tilde{\nu}}&          7826.822,          7746.869,         13922.999\\
                M_{\tilde{e}}&           1615.07,           7827.26,            316.41   ,           7747.59,          13877.17,          16090.39  \\
                M_{\tilde{u}}&           6120.21,           6318.57,           6119.06   ,           6317.42,          21138.66,          30212.30  \\
                M_{\tilde{d}}&           3334.56,           6120.76,           3333.19   ,           6119.76,          30199.68,          36983.67  \\
                        M_{A}&         299785.40\\
                  M_{H^{\pm}}&         299785.42\\
                    M_{H^{0}}&         299785.39\\
                    M_{h^{0}}&            114.58\\
 \hline
 \end{array}
 $$
 \caption{\small{I-3-e: Spectra of supersymmetric partners calculated ignoring generation mixing effects.
  See caption to Table 6 for explanation.
  }}\label{ I-3-e }\end{table}

 \begin{table}
 $$
 \begin{array}{|c|c|}
 \hline {\mbox {Field } }&Mass(GeV)\\
 \hline
                M_{\tilde{G}}&            599.14\\
               M_{\chi^{\pm}}&            406.47,         100968.77\\
       M_{\chi^{0}}&            147.34,            406.47,         100968.75    ,         100968.75\\
              M_{\tilde{\nu}}&           7744.59,           7824.46,         14034.822\\
                M_{\tilde{e}}&            393.14,           1632.84,           7745.08   ,           7824.89,          13992.95,          16284.21  \\
                M_{\tilde{u}}&           6108.76,           6117.73,           6311.63   ,           6312.80,          21005.60,          30236.11  \\
                M_{\tilde{d}}&           3331.17,           3332.57,           6109.44   ,           6118.29,          30223.69,          37095.48  \\
                        M_{A}&         300060.73\\
                  M_{H^{\pm}}&         300060.74\\
                    M_{H^{0}}&         300060.72\\
                    M_{h^{0}}&            114.85\\
 \hline
 \end{array}
 $$
\caption{\small{I-3-f: Spectra of supersymmetric partners calculated including  generation mixing effects.
  See caption to Table 7 for explanation.}} \label{ I-3-f}\end{table}

\clearpage

 \begin{table}
 $$
 \begin{array}{|c|c|c|c|}
 \hline
 {\rm Parameter }&Value &{\rm  Field }&\hspace{10mm} Masses\\
 &&{\rm}[SU(3),SU(2),Y]&\hspace{10mm}( Units\,\,of 10^{16} GeV)\\ \hline
       \chi_{X}&  0.0771           &A[1,1,4]&    168.86 \\ \chi_{Z}&
    0.0225
                &B[6,2,{5/3}]&            0.3197\\
           h_{11}/10^{-6}&  0.0156         &C[8,2,1]&{      5.54,    205.24,    206.53 }\\
           h_{22}/10^{-4}& -0.0131    &D[3,2,{7/ 3}]&{      4.90,    110.54,    138.09 }\\
                   h_{33}&  0.0033     &E[3,2,{1/3}]&{      0.38,      3.89,      6.18 }\\
 f_{11}/10^{-6}&
  0.0198+  0.1542i
                      &&{     6.184,    165.12,    192.29 }\\
\hline\end{array}
 $$
  \caption{\small{I-4-a Fit : Column 1 contains values   of the NMSGUT-SUGRY-NUHM  parameters at $M_X$
  derived from an  accurate fit to all 18 fermion data and compatible with RG constraints.
 See caption to Table 2 for explanation. }}\label{ I-4-a}\end{table}

\begin{table}
 $$
 \begin{array}{|c|c|c|c|c|}
 \hline
 &&&&\\
 {\rm  Parameter }&Target =\bar O_i &Uncert.= \delta_i    &Achieved= O_i &Pull =(O_i-\bar O_i)/\delta_i\\
 \hline
    y_u/10^{-6}&  2.099660&  0.802070&  2.099948&  0.000359\\
    y_c/10^{-3}&  1.023488&  0.168875&  1.022975& -0.003035\\
            y_t&  0.368075&  0.014723&  0.368096&  0.001417\\
    y_d/10^{-5}&  6.248810&  3.643056&  6.265778&  0.004658\\
    y_s/10^{-3}&  1.185571&  0.559590&  1.201681&  0.028788\\
            y_b&  0.424469&  0.220299&  0.436945&  0.056636\\
\hline
 \end{array}
 $$
 \caption{\small{I-4-b: Fit   with $\chi_X=\sqrt{ \sum_{i=1}^{17}
 (O_i-\bar O_i)^2/\delta_i^2}=
    0.0771
 $. Target values,  at $M_X$ of the fermion yukawa
 couplings and mixing parameters, together with the estimated uncertainties, achieved values and pulls.
 See caption to Table 3 for explanation. }}
  \label{ I-4-b}\end{table}
 \begin{table}
 $$
 \begin{array}{|c|c|c|c|}
 \hline &&&\\ {\rm  Parameter }&SM(M_Z) & m^{GUT}(M_Z) & m^{MSSM}=(m+\Delta m)^{GUT}(M_Z) \\
 \hline
    m_d/10^{-3}&   2.90000&   0.57869&   2.88556\\
    m_s/10^{-3}&  55.00000&  11.09336&  55.29916\\
            m_b&   2.90000&   2.96813&   2.93512\\
    m_e/10^{-3}&   0.48657&   0.45919&   0.48201\\
         m_\mu &   0.10272&   0.09719&   0.10198\\
         m_\tau&   1.74624&   1.72539&   1.72437\\
    m_u/10^{-3}&   1.27000&   1.11423&   1.27070\\
            m_c&   0.61900&   0.54254&   0.61873\\
            m_t& 172.50000& 147.09707& 172.15913\\
 \hline
 \end{array}
 $$

 \caption{\small{I-4-c: Values of standard model
 fermion masses in GeV at $M_Z$ compared with the masses obtained from
 values of GUT derived  yukawa couplings  run down from $M_X^0$ to
 $M_Z$  both before and after threshold corrections.
  Fit with $\chi_Z=\sqrt{ \sum_{i=1}^{9} (m_i^{MSSM}- m_i^{SM})^2/ (m_i^{MSSM})^2} =
0.0225$.}}
 \label{  I-4-c}\end{table}

  \begin{table}
 $$
 \begin{array}{|c|c|c|c|}
 \hline
 {\rm  Parameter}  &Value&  {\rm  Parameter}&Value \\
 \hline
                       M_{1}&            104.16&   M_{{\tilde {\bar {u}}_1}}&           6018.96\\
                       M_{2}&            302.05&   M_{{\tilde {\bar {u}}_2}}&           6018.69\\
                       M_{3}&            350.55&   M_{{\tilde {\bar {u}}_3}}&          32268.70\\
     M_{{\tilde {\bar l}_1}}&            115.35&               A^{0(l)}_{11}&         -92737.91\\
     M_{{\tilde {\bar l}_2}}&            123.64&               A^{0(l)}_{22}&         -92630.01\\
     M_{{\tilde {\bar l}_3}}&          27768.92&               A^{0(l)}_{33}&         -58880.24\\
        M_{{\tilde {L}_{1}}}&           8173.98&               A^{0(u)}_{11}&        -105676.88\\
        M_{{\tilde {L}_{2}}}&           8174.04&               A^{0(u)}_{22}&        -105676.19\\
        M_{{\tilde {L}_{3}}}&          21289.08&               A^{0(u)}_{33}&         -55613.06\\
     M_{{\tilde {\bar d}_1}}&           1701.52&               A^{0(d)}_{11}&         -92749.44\\
     M_{{\tilde {\bar d}_2}}&           1702.61&               A^{0(d)}_{22}&         -92748.73\\
     M_{{\tilde {\bar d}_3}}&          40148.17&               A^{0(d)}_{33}&         -41722.79\\
          M_{{\tilde {Q}_1}}&           5628.19&                   \tan\beta&             51.00\\
          M_{{\tilde {Q}_2}}&           5628.21&                    \mu(M_Z)&         106300.71\\
          M_{{\tilde {Q}_3}}&          36452.35&                      B(M_Z)&
          1.8340 \times 10^{   9}
 \\
 M_{\bar {H}}^2&
         -1.1556 \times 10^{  10}
 &M_{H}^2&
         -1.2234 \times 10^{  10}
 \\
 \hline
 \end{array}
 $$
  \caption{ \small {I-4-d: Values (GeV) in  of the soft Susy parameters  at $M_Z$
 (evolved from the soft SUGRY-NUHM parameters at $M_X$).
  See caption to Table 5 for explanation. }}
 \label{ I-4-d }\end{table}

\begin{table}
 $$
 \begin{array}{|c|c|}
 \hline {\mbox {Field } }&Mass(GeV)\\
 \hline
                M_{\tilde{G}}&            350.55\\
               M_{\chi^{\pm}}&            302.05,         106300.78\\
       M_{\chi^{0}}&            104.16,            302.05,         106300.75    ,         106300.76\\
              M_{\tilde{\nu}}&          8173.708,          8173.768,         21288.979\\
                M_{\tilde{e}}&            123.89,           8174.12,            114.16   ,           8174.45,          21282.54,          27774.02  \\
                M_{\tilde{u}}&           5627.92,           6018.85,           5627.87   ,           6018.64,          32265.13,          36456.04  \\
                M_{\tilde{d}}&           1701.72,           5628.53,           1702.77   ,           5628.56,          36439.70,          40159.71  \\
                        M_{A}&         305894.52\\
                  M_{H^{\pm}}&         305894.54\\
                    M_{H^{0}}&         305894.52\\
                    M_{h^{0}}&            128.29\\
 \hline
 \end{array}
 $$
\caption{\small{I-4-e: Spectra of supersymmetric partners calculated ignoring generation mixing effects.
  See caption to Table 6 for explanation.
  }} \label{ I-4-e }\end{table}

 \begin{table}
 $$
 \begin{array}{|c|c|}
 \hline {\mbox {Field } }&Mass(GeV)\\
 \hline
                M_{\tilde{G}}&            350.89\\
               M_{\chi^{\pm}}&            302.21,         106276.46\\
       M_{\chi^{0}}&            104.22,            302.21,         106276.43    ,         106276.44\\
              M_{\tilde{\nu}}&           8173.67,           8173.76,         21295.664\\
                M_{\tilde{e}}&             89.25,            117.38,           8174.09   ,           8174.44,          21289.49,          27784.15  \\
                M_{\tilde{u}}&           5627.05,           5648.66,           6018.00   ,           6018.22,          32278.73,          36462.50  \\
                M_{\tilde{d}}&           1699.10,           1700.14,           5627.66   ,           5649.32,          36446.20,          40166.59  \\
                        M_{A}&         306124.85\\
                  M_{H^{\pm}}&         306124.86\\
                    M_{H^{0}}&         306124.84\\
                    M_{h^{0}}&            128.38\\
 \hline
 \end{array}
 $$
 \caption{\small{I-4-f: Spectra of supersymmetric partners calculated including  generation mixing effects.
  See caption to Table 7 for explanation.}}\label{ I-4-f}\end{table}

\clearpage

  \begin{table}
 $$
 \begin{array}{|c|c|c|c|}
 \hline
 {\rm Parameter }&Value &{\rm  Field }&\hspace{10mm} Masses\\
 &&{\rm}[SU(3),SU(2),Y]&\hspace{10mm}( Units\,\,of 10^{16} GeV)\\ \hline
       \chi_{X}&  0.0050           &A[1,1,4]&    948.38 \\ \chi_{Z}&
    0.0038
                &B[6,2,{5/3}]&            0.2867\\
           h_{11}/10^{-6}&  0.0200         &C[8,2,1]&{     54.37,    436.41,    458.35 }\\
           h_{22}/10^{-4}&  0.0348    &D[3,2,{7/ 3}]&{     52.04,    521.13,    532.76 }\\
                   h_{33}&  0.0027     &E[3,2,{1/3}]&{      0.47,     38.02,     38.02 }\\
 f_{11}/10^{-6}&
  0.0796-  0.1364i
                      &&{    43.611,    563.34,    624.40 }\\
\hline\end{array}
 $$
 \caption{\small{II-3-a : Column 1 contains values   of the NMSGUT-SUGRY-NUHM  parameters at $M_X$
  derived from an  accurate fit to all 18 fermion data and compatible with RG constraints.
 See caption to Table 2 for explanation. }} \label{ II-3-a}\end{table}
 \begin{table}
 $$
 \begin{array}{|c|c|c|c|c|}
 \hline
 &&&&\\
 {\rm  Parameter }&Target =\bar O_i &Uncert.= \delta_i    &Achieved= O_i &Pull =(O_i-\bar O_i)/\delta_i\\
 \hline
    y_u/10^{-6}&  2.100774&  0.802496&  2.100767& -0.000009\\
    y_c/10^{-3}&  1.023520&  0.168881&  1.023538&  0.000104\\
            y_t&  0.393877&  0.015755&  0.393880&  0.000182\\
    y_d/10^{-5}&  7.305382&  4.259037&  7.317435&  0.002830\\
    y_s/10^{-3}&  1.384369&  0.653422&  1.382982& -0.002123\\
            y_b&  0.516686&  0.268160&  0.516774&  0.000329\\
\hline
 \end{array}
 $$
  \caption{\small{II-3-b: Fit   with $\chi_X=\sqrt{ \sum_{i=1}^{17}
 (O_i-\bar O_i)^2/\delta_i^2}=
    0.0050
 $. Target values,  at $M_X$ of the fermion yukawa
 couplings and mixing parameters, together with the estimated uncertainties, achieved values and pulls.
  See caption to Table 3 for explanation.}}
\label{ II-3-b} \end{table}
 \begin{table}
 $$
 \begin{array}{|c|c|c|c|}
 \hline &&&\\ {\rm  Parameter }&SM(M_Z) & m^{GUT}(M_Z) & m^{MSSM}=(m+\Delta m)^{GUT}(M_Z) \\
 \hline
    m_d/10^{-3}&   2.90000&   0.62793&   2.90404\\
    m_s/10^{-3}&  55.00000&  11.86773&  54.92429\\
            m_b&   2.90000&   3.08544&   2.90008\\
    m_e/10^{-3}&   0.48657&   0.48023&   0.48631\\
         m_\mu &   0.10272&   0.10130&   0.10265\\
         m_\tau&   1.74624&   1.74579&   1.74474\\
    m_u/10^{-3}&   1.27000&   1.09336&   1.26989\\
            m_c&   0.61900&   0.53271&   0.61905\\
            m_t& 172.50000& 149.44924& 171.97519\\
 \hline
 \end{array}
 $$

 \caption{\small{II-3-c: Values of standard model
 fermion masses in GeV at $M_Z$ compared with the masses obtained from
 values of GUT derived  yukawa couplings  run down from $M_X^0$ to
 $M_Z$  both before and after threshold corrections.
  Fit with $\chi_Z=\sqrt{ \sum_{i=1}^{9} (m_i^{MSSM}- m_i^{SM})^2/ (m_i^{MSSM})^2} =
0.0038$.}}
\label{ II-3-c } \end{table}

 \begin{table}
 $$
 \begin{array}{|c|c|c|c|}
 \hline
 {\rm  Parameter}  &Value&  {\rm  Parameter}&Value \\
 \hline
                       M_{1}&             43.51&   M_{{\tilde {\bar {u}}_1}}&            205.86\\
                       M_{2}&            144.40&   M_{{\tilde {\bar {u}}_2}}&            200.37\\
                       M_{3}&             89.92&   M_{{\tilde {\bar {u}}_3}}&          15100.44\\
     M_{{\tilde {\bar l}_1}}&           2153.08&               A^{0(l)}_{11}&         -47638.60\\
     M_{{\tilde {\bar l}_2}}&           2072.60&               A^{0(l)}_{22}&         -47572.19\\
     M_{{\tilde {\bar l}_3}}&          12763.49&               A^{0(l)}_{33}&         -28011.67\\
        M_{{\tilde {L}_{1}}}&           2923.85&               A^{0(u)}_{11}&         -56421.60\\
        M_{{\tilde {L}_{2}}}&           2894.78&               A^{0(u)}_{22}&         -56421.22\\
        M_{{\tilde {L}_{3}}}&           9407.34&               A^{0(u)}_{33}&         -27834.22\\
     M_{{\tilde {\bar d}_1}}&            455.68&               A^{0(d)}_{11}&         -47530.07\\
     M_{{\tilde {\bar d}_2}}&            455.15&               A^{0(d)}_{22}&         -47529.64\\
     M_{{\tilde {\bar d}_3}}&          20922.14&               A^{0(d)}_{33}&         -17493.87\\
          M_{{\tilde {Q}_1}}&           2445.28&                   \tan\beta&             51.50\\
          M_{{\tilde {Q}_2}}&           2445.00&                    \mu(M_Z)&          52905.86\\
          M_{{\tilde {Q}_3}}&          18336.34&                      B(M_Z)&
          4.0145 \times 10^{   8}
 \\
 M_{\bar {H}}^2&
         -2.6746 \times 10^{   9}
 &M_{H}^2&
         -3.0610 \times 10^{   9}
 \\
 \hline
 \end{array}
 $$
  \caption{ \small {II-3-d: Values (GeV) in  of the soft Susy parameters  at $M_Z$
 (evolved from the soft SUGRY-NUHM parameters at $M_X$).
  See caption to Table 5 for explanation.}}
 \label{ II-3-d}\end{table}

\begin{table}
 $$
 \begin{array}{|c|c|}
 \hline {\mbox {Field } }&Mass(GeV)\\
 \hline
                M_{\tilde{G}}&             89.92\\
               M_{\chi^{\pm}}&            144.39,          52905.99\\
       M_{\chi^{0}}&             43.51,            144.39,          52905.94    ,          52905.95\\
              M_{\tilde{\nu}}&          2923.095,          2894.022,          9407.108\\
                M_{\tilde{e}}&           2153.55,           2924.25,           2068.45   ,           2898.51,           9391.01,          12775.68  \\
                M_{\tilde{u}}&            202.54,           2444.65,            196.56   ,           2444.41,          15095.44,          18341.55  \\
                M_{\tilde{d}}&            456.42,           2446.04,            455.69   ,           2445.80,          18317.31,          20938.90  \\
                        M_{A}&         143814.30\\
                  M_{H^{\pm}}&         143814.33\\
                    M_{H^{0}}&         143814.29\\
                    M_{h^{0}}&            123.80\\
 \hline
 \end{array}
 $$
\caption{\small{II-3-e: Spectra of supersymmetric partners calculated ignoring generation mixing effects.
  See caption to Table 6 for explanation.
  }} \label{ II-3-e }\end{table}
\begin{table}
 $$
 \begin{array}{|c|c|}
 \hline {\mbox {Field } }&Mass(GeV)\\
 \hline
                M_{\tilde{G}}&             90.00\\
               M_{\chi^{\pm}}&            144.43,          52903.84\\
       M_{\chi^{0}}&             43.52,            144.43,          52903.79    ,          52903.80\\
              M_{\tilde{\nu}}&           2894.06,           2923.13,          9407.538\\
                M_{\tilde{e}}&           2065.45,           2153.46,           2899.51   ,           2924.33,           9392.27,          12775.95  \\
                M_{\tilde{u}}&            198.31,            204.25,           2444.78   ,           2451.67,          15094.69,          18340.05  \\
                M_{\tilde{d}}&            456.03,            456.94,           2446.17   ,           2453.03,          18315.87,          20938.32  \\
                        M_{A}&         143872.09\\
                  M_{H^{\pm}}&         143872.11\\
                    M_{H^{0}}&         143872.08\\
                    M_{h^{0}}&            123.91\\
 \hline
 \end{array}
 $$
 \caption{\small{II-3-f: Spectra of supersymmetric partners calculated including  generation mixing effects.
 See caption to Table 6 for explanation. }}\label{ II-3-f}\end{table}

\clearpage

  \begin{table}
 $$
 \begin{array}{|c|c|c|c|}
 \hline
 {\rm Parameter }&Value &{\rm  Field }&\hspace{10mm} Masses\\
 &&{\rm}[SU(3),SU(2),Y]&\hspace{10mm}( Units\,\,of 10^{16} GeV)\\ \hline
       \chi_{X}&  0.0061           &A[1,1,4]&    137.29 \\ \chi_{Z}&
    0.0179
                &B[6,2,{5/3}]&            0.3082\\
           h_{11}/10^{-6}&  0.0277         &C[8,2,1]&{      4.55,    166.93,    168.04 }\\
           h_{22}/10^{-4}& -0.0081    &D[3,2,{7/ 3}]&{      3.93,     95.60,    117.42 }\\
                   h_{33}&  0.0034     &E[3,2,{1/3}]&{      0.37,      3.21,      5.35 }\\
 f_{11}/10^{-6}&
  0.0242+  0.1635i
                      &&{     5.346,    138.97,    160.78 }\\
\hline\end{array}
 $$
\caption{\small{ II-4-a: Column 1 contains values   of the NMSGUT-SUGRY-NUHM  parameters at $M_X$
  derived from an  accurate fit to all 18 fermion data and compatible with RG constraints.
 See caption to Table 2 for explanation.}} \label{ II-4-a} \end{table}
 \begin{table}
 $$
 \begin{array}{|c|c|c|c|c|}
 \hline
 &&&&\\
 {\rm  Parameter }&Target =\bar O_i &Uncert.= \delta_i    &Achieved= O_i &Pull =(O_i-\bar O_i)/\delta_i\\
 \hline
    y_u/10^{-6}&  2.169056&  0.828579&  2.169068&  0.000015\\
    y_c/10^{-3}&  1.057321&  0.174458&  1.057338&  0.000102\\
            y_t&  0.390644&  0.015626&  0.390643& -0.000072\\
    y_d/10^{-5}&  6.611935&  3.854758&  6.614046&  0.000548\\
    y_s/10^{-3}&  1.255361&  0.592530&  1.256890&  0.002581\\
            y_b&  0.452379&  0.234785&  0.453431&  0.004481\\
\hline
 \end{array}
 $$
  \caption{\small{II-4-b: Fit   with $\chi_X=\sqrt{ \sum_{i=1}^{17}
 (O_i-\bar O_i)^2/\delta_i^2}=
    0.0061
 $. Target values,  at $M_X$ of the fermion yukawa
 couplings and mixing parameters, together with the estimated uncertainties, achieved values and pulls.
  See caption to Table 3 for explanation.}}
 \label{ II-4-b}\end{table}
 \begin{table}
 $$
 \begin{array}{|c|c|c|c|}
 \hline &&&\\ {\rm  Parameter }&SM(M_Z) & m^{GUT}(M_Z) & m^{MSSM}=(m+\Delta m)^{GUT}(M_Z) \\
 \hline
    m_d/10^{-3}&   2.90000&   0.60092&   2.89886\\
    m_s/10^{-3}&  55.00000&  11.41936&  55.02309\\
            m_b&   2.90000&   2.98616&   2.90724\\
    m_e/10^{-3}&   0.48657&   0.48126&   0.48642\\
         m_\mu &   0.10272&   0.10163&   0.10271\\
         m_\tau&   1.74624&   1.74575&   1.74469\\
    m_u/10^{-3}&   1.27000&   1.12845&   1.25425\\
            m_c&   0.61900&   0.55007&   0.61140\\
            m_t& 172.50000& 150.12439& 172.45759\\
 \hline
 \end{array}
 $$

 \caption{\small{II-4-c: Values of standard model
 fermion masses in GeV at $M_Z$ compared with the masses obtained from
 values of GUT derived  yukawa couplings  run down from $M_X^0$ to
 $M_Z$  both before and after threshold corrections.
  Fit with $\chi_Z=\sqrt{ \sum_{i=1}^{9} (m_i^{MSSM}- m_i^{SM})^2/ (m_i^{MSSM})^2} =
0.0179$.}}
 \label{ II-4-c } \end{table}

 \begin{table}
 $$
 \begin{array}{|c|c|c|c|}
 \hline
 {\rm  Parameter}  &Value&  {\rm  Parameter}&Value \\
 \hline
                       M_{1}&             35.00&   M_{{\tilde {\bar {u}}_1}}&           1355.00\\
                       M_{2}&            110.00&   M_{{\tilde {\bar {u}}_2}}&           1354.97\\
                       M_{3}&             82.43&   M_{{\tilde {\bar {u}}_3}}&          15017.39\\
     M_{{\tilde {\bar l}_1}}&           1668.31&               A^{0(l)}_{11}&         -37671.35\\
     M_{{\tilde {\bar l}_2}}&           1686.32&               A^{0(l)}_{22}&         -37622.42\\
     M_{{\tilde {\bar l}_3}}&          12622.76&               A^{0(l)}_{33}&         -23300.14\\
        M_{{\tilde {L}_{1}}}&           2873.66&               A^{0(u)}_{11}&         -42600.14\\
        M_{{\tilde {L}_{2}}}&           2878.88&               A^{0(u)}_{22}&         -42599.84\\
        M_{{\tilde {L}_{3}}}&           9306.38&               A^{0(u)}_{33}&         -21238.50\\
     M_{{\tilde {\bar d}_1}}&            362.45&               A^{0(d)}_{11}&         -37604.10\\
     M_{{\tilde {\bar d}_2}}&            363.78&               A^{0(d)}_{22}&         -37603.79\\
     M_{{\tilde {\bar d}_3}}&          17343.17&               A^{0(d)}_{33}&         -16155.25\\
          M_{{\tilde {Q}_1}}&           2110.14&                   \tan\beta&             51.00\\
          M_{{\tilde {Q}_2}}&           2110.24&                    \mu(M_Z)&          44372.34\\
          M_{{\tilde {Q}_3}}&          16267.76&                      B(M_Z)&
          2.6446 \times 10^{   8}
 \\
 M_{\bar {H}}^2&
         -1.9734 \times 10^{   9}
 &M_{H}^2&
         -2.1327 \times 10^{   9}
 \\
 \hline
 \end{array}
 $$
 \caption{ \small {II-4-d : Values (GeV) in  of the soft Susy parameters  at $M_Z$
 (evolved from the soft SUGRY-NUHM parameters at $M_X$).
 See caption to Table 5 for explanation. }}
  \label{ II-4-d }\end{table}

 \begin{table}
 $$
 \begin{array}{|c|c|}
 \hline {\mbox {Field } }&Mass(GeV)\\
 \hline
                M_{\tilde{G}}&             82.43\\
               M_{\chi^{\pm}}&            109.99,          44372.49\\
       M_{\chi^{0}}&             35.00,            109.99,          44372.43    ,          44372.44\\
              M_{\tilde{\nu}}&          2872.898,          2878.121,          9306.146\\
                M_{\tilde{e}}&           1668.92,           2874.07,           1683.95   ,           2881.03,           9294.77,          12631.49  \\
                M_{\tilde{u}}&           1354.50,           2109.42,           1354.39   ,           2109.57,          15008.78,          16276.95  \\
                M_{\tilde{d}}&            363.38,           2111.02,            364.49   ,           2111.16,          16229.75,          17378.88  \\
                        M_{A}&         116157.09\\
                  M_{H^{\pm}}&         116157.12\\
                    M_{H^{0}}&         116157.08\\
                    M_{h^{0}}&            127.41\\
 \hline
 \end{array}
 $$
\caption{\small{II-4-e: Spectra of supersymmetric partners calculated ignoring generation mixing effects.
  See caption to Table 6 for explanation.
  }} \label{ II-4-e }\end{table}

 \begin{table}
 $$
 \begin{array}{|c|c|}
 \hline {\mbox {Field } }&Mass(GeV)\\
 \hline
                M_{\tilde{G}}&             82.49\\
               M_{\chi^{\pm}}&            110.02,          44370.87\\
       M_{\chi^{0}}&             35.01,            110.02,          44370.81    ,          44370.82\\
              M_{\tilde{\nu}}&           2872.91,           2878.14,          9306.355\\
                M_{\tilde{e}}&           1668.80,           1681.53,           2874.11   ,           2881.63,           9295.61,          12631.53  \\
                M_{\tilde{u}}&           1354.49,           1354.60,           2109.47   ,           2123.10,          15008.73,          16275.15  \\
                M_{\tilde{d}}&            363.58,            364.64,           2111.08   ,           2124.64,          16228.06,          17378.66  \\
                        M_{A}&         116199.01\\
                  M_{H^{\pm}}&         116199.03\\
                    M_{H^{0}}&         116199.00\\
                    M_{h^{0}}&            127.43\\
 \hline
 \end{array}
 $$
 \caption{\small{II-4-f: Spectra of supersymmetric partners calculated including  generation mixing effects.
 See caption to Table 7 for explanation. }}\label{ II-4-f}\end{table}

\clearpage

 \begin{table}
 $$
 \begin{array}{|c|c|c|c|}
 \hline
 {\rm Parameter }&Value &{\rm  Field }&\hspace{10mm} Masses\\
 &&{\rm}[SU(3),SU(2),Y]&\hspace{10mm}( Units\,\,of 10^{16} GeV)\\ \hline
       \chi_{X}&  0.0143           &A[1,1,4]&  16007.39 \\ \chi_{Z}&
    0.0066
                &B[6,2,{5/3}]&            0.3294\\
           h_{11}/10^{-6}& -0.4946         &C[8,2,1]&{    671.99,   2566.01,   3875.58 }\\
           h_{22}/10^{-4}&  0.3413    &D[3,2,{7/ 3}]&{   1412.23,   9043.81,  10626.05 }\\
                   h_{33}&  0.0143     &E[3,2,{1/3}]&{      0.74,    488.82,    488.82 }\\
 f_{11}/10^{-6}&
 -0.1019-  0.0216i
                      &&{  1600.508,   4925.64,   7365.53 }\\
\hline\end{array}
 $$
 \caption{\small{ III-2-a : Column 1 contains values   of the NMSGUT-SUGRY-NUHM  parameters at $M_X$
  derived from an  accurate fit to all 18 fermion data and compatible with RG constraints.
  See caption to Table 2 for explanation.}} \label{ III-2-a}\end{table}
 \begin{table}
 $$
 \begin{array}{|c|c|c|c|c|}
 \hline
 &&&&\\
 {\rm  Parameter }&Target =\bar O_i &Uncert.= \delta_i    &Achieved= O_i &Pull =(O_i-\bar O_i)/\delta_i\\
 \hline
    y_u/10^{-6}&  2.114293&  0.807660&  2.114551&  0.000320\\
    y_c/10^{-3}&  1.030627&  0.170053&  1.030427& -0.001178\\
            y_t&  0.388741&  0.015550&  0.388744&  0.000205\\
    y_d/10^{-5}&  5.990999&  3.492753&  6.024454&  0.009578\\
    y_s/10^{-3}&  1.132757&  0.534661&  1.128624& -0.007730\\
            y_b&  0.546568&  0.283669&  0.545182& -0.004887\\
\hline
 \end{array}
 $$
 \caption{\small{III-2-b: Fit   with $\chi_X=\sqrt{ \sum_{i=1}^{17}
 (O_i-\bar O_i)^2/\delta_i^2}=
    0.0143
 $. Target values,  at $M_X$ of the fermion yukawa
 couplings and mixing parameters, together with the estimated uncertainties, achieved values and pulls.
  See caption to Table 3 for explanation.}}
  \label{ III-2-b}\end{table}
 \begin{table}
 $$
 \begin{array}{|c|c|c|c|}
 \hline &&&\\ {\rm  Parameter }&SM(M_Z) & m^{GUT}(M_Z) & m^{MSSM}=(m+\Delta m)^{GUT}(M_Z) \\
 \hline
    m_d/10^{-3}&   2.90000&   0.52468&   2.91331\\
    m_s/10^{-3}&  55.00000&   9.82943&  54.75545\\
            m_b&   2.90000&   3.24471&   2.89957\\
    m_e/10^{-3}&   0.48657&   0.47433&   0.48705\\
         m_\mu &   0.10272&   0.09592&   0.10272\\
         m_\tau&   1.74624&   1.74370&   1.74662\\
    m_u/10^{-3}&   1.27000&   1.10824&   1.26917\\
            m_c&   0.61900&   0.54005&   0.61848\\
            m_t& 172.50000& 148.32422& 172.51404\\
 \hline
 \end{array}
 $$

 \caption{\small{III-2-c: Values of standard model
 fermion masses in GeV at $M_Z$ compared with the masses obtained from
 values of GUT derived  yukawa couplings  run down from $M_X^0$ to
 $M_Z$  both before and after threshold corrections.
  Fit with $\chi_Z=\sqrt{ \sum_{i=1}^{9} (m_i^{MSSM}- m_i^{SM})^2/ (m_i^{MSSM})^2} =
0.0066$.}}
 \label{ III-2-c }\end{table}

 \begin{table}
 $$
 \begin{array}{|c|c|c|c|}
 \hline
 {\rm  Parameter}  &Value&  {\rm  Parameter}&Value \\
 \hline
                       M_{1}&             98.92&   M_{{\tilde {\bar {u}}_1}}&           3553.79\\
                       M_{2}&            341.43&   M_{{\tilde {\bar {u}}_2}}&           3550.53\\
                       M_{3}&            232.18&   M_{{\tilde {\bar {u}}_3}}&          20687.29\\
     M_{{\tilde {\bar l}_1}}&           1868.65&               A^{0(l)}_{11}&        -110036.65\\
     M_{{\tilde {\bar l}_2}}&            204.32&               A^{0(l)}_{22}&        -109904.91\\
     M_{{\tilde {\bar l}_3}}&          11397.10&               A^{0(l)}_{33}&         -66704.24\\
        M_{{\tilde {L}_{1}}}&           6369.22&               A^{0(u)}_{11}&        -132899.38\\
        M_{{\tilde {L}_{2}}}&           6233.07&               A^{0(u)}_{22}&        -132898.56\\
        M_{{\tilde {L}_{3}}}&          10527.00&               A^{0(u)}_{33}&         -66235.28\\
     M_{{\tilde {\bar d}_1}}&            712.29&               A^{0(d)}_{11}&        -109811.44\\
     M_{{\tilde {\bar d}_2}}&            705.83&               A^{0(d)}_{22}&        -109810.71\\
     M_{{\tilde {\bar d}_3}}&          41913.64&               A^{0(d)}_{33}&         -37711.03\\
          M_{{\tilde {Q}_1}}&           5387.22&                   \tan\beta&             50.00\\
          M_{{\tilde {Q}_2}}&           5385.73&                    \mu(M_Z)&         111532.64\\
          M_{{\tilde {Q}_3}}&          33269.50&                      B(M_Z)&
          2.1254 \times 10^{   9}
 \\
 M_{\bar {H}}^2&
         -1.1842 \times 10^{  10}
 &M_{H}^2&
         -1.3659 \times 10^{  10}
 \\
 \hline
 \end{array}
 $$
 \caption{ \small {III-2-d: Values (GeV) in  of the soft Susy parameters  at $M_Z$
 (evolved from the soft SUGRY-NUHM parameters at $M_X$).
  See caption to Table 5 for explanation.}}
  \label{ III-2-d}\end{table}

 \begin{table}
 $$
 \begin{array}{|c|c|}
 \hline {\mbox {Field } }&Mass(GeV)\\
 \hline
                M_{\tilde{G}}&            232.18\\
               M_{\chi^{\pm}}&            341.42,         111532.70\\
       M_{\chi^{0}}&             98.92,            341.42,         111532.68    ,         111532.68\\
              M_{\tilde{\nu}}&          6368.872,          6232.714,         10526.788\\
                M_{\tilde{e}}&           1869.20,           6369.40,            190.06   ,           6233.87,          10327.33,          11578.52  \\
                M_{\tilde{u}}&           3553.60,           5386.93,           3550.29   ,           5385.48,          20684.11,          33272.06  \\
                M_{\tilde{d}}&            712.77,           5387.56,            706.24   ,           5386.09,          33261.88,          41919.73  \\
                        M_{A}&         326055.13\\
                  M_{H^{\pm}}&         326055.14\\
                    M_{H^{0}}&         326055.11\\
                    M_{h^{0}}&            112.76\\
 \hline
 \end{array}
 $$
 \caption{\small{III-2-e: Spectra of supersymmetric partners calculated ignoring generation mixing effects.
  See caption to Table 6 for explanation.
  }}\label{  III-2-e}\end{table}

\begin{table}
 $$
 \begin{array}{|c|c|}
 \hline {\mbox {Field } }&Mass(GeV)\\
 \hline
                M_{\tilde{G}}&            232.36\\
               M_{\chi^{\pm}}&            341.50,         111528.02\\
       M_{\chi^{0}}&             98.94,            341.50,         111528.00    ,         111528.00\\
              M_{\tilde{\nu}}&           6232.83,           6368.97,         10529.123\\
                M_{\tilde{e}}&            342.67,           1868.73,           6223.30   ,           6369.45,          10391.14,          11540.07  \\
                M_{\tilde{u}}&           3550.99,           3554.30,           5367.30   ,           5387.36,          20677.92,          33271.88  \\
                M_{\tilde{d}}&            709.00,            715.54,           5367.93   ,           5387.99,          33261.71,          41917.16  \\
                        M_{A}&         326179.32\\
                  M_{H^{\pm}}&         326179.33\\
                    M_{H^{0}}&         326179.31\\
                    M_{h^{0}}&            112.78\\
 \hline
 \end{array}
 $$
\caption{\small{III-2-f: Spectra of supersymmetric partners calculated including  generation mixing effects.
 See caption to Table 7 for explanation.}} \label{  III-2-f}\end{table}

\clearpage

 \begin{table}
 $$
 \begin{array}{|c|c|c|c|}
 \hline
 {\rm Parameter }&Value &{\rm  Field }&\hspace{10mm} Masses\\
 &&{\rm}[SU(3),SU(2),Y]&\hspace{10mm}( Units\,\,of 10^{16} GeV)\\ \hline
       \chi_{X}&  0.0031           &A[1,1,4]&    689.26 \\ \chi_{Z}&
    0.0012
                &B[6,2,{5/3}]&            0.2457\\
           h_{11}/10^{-6}&  0.0205         &C[8,2,1]&{     38.27,    330.90,    346.79 }\\
           h_{22}/10^{-4}&  0.0364    &D[3,2,{7/ 3}]&{     36.96,    399.26,    408.77 }\\
                   h_{33}&  0.0027     &E[3,2,{1/3}]&{      0.40,     27.84,     27.84 }\\
 f_{11}/10^{-6}&
  0.0774-  0.1295i
                      &&{    30.102,    433.12,    477.24 }\\
\hline\end{array}
 $$
 \caption{\small{III-3-a  : Column 1 contains values   of the NMSGUT-SUGRY-NUHM  parameters at $M_X$
  derived from an  accurate fit to all 18 fermion data and compatible with RG constraints.
See caption to Table 2 for explanation. }}\label{ III-3-a} \end{table}

 \begin{table}
 $$
 \begin{array}{|c|c|c|c|c|}
 \hline
 &&&&\\
 {\rm  Parameter }&Target =\bar O_i &Uncert.= \delta_i    &Achieved= O_i &Pull =(O_i-\bar O_i)/\delta_i\\
 \hline
    y_u/10^{-6}&  2.121322&  0.810345&  2.121299& -0.000028\\
    y_c/10^{-3}&  1.034062&  0.170620&  1.033951& -0.000648\\
            y_t&  0.392382&  0.015695&  0.392383&  0.000076\\
    y_d/10^{-5}&  7.238271&  4.219912&  7.243490&  0.001237\\
    y_s/10^{-3}&  1.372396&  0.647771&  1.371099& -0.002002\\
            y_b&  0.512784&  0.266135&  0.513218&  0.001628\\
\hline
 \end{array}
 $$
 \caption{\small{III-3-b: Fit   with $\chi_X=\sqrt{ \sum_{i=1}^{17}
 (O_i-\bar O_i)^2/\delta_i^2}=
    0.0031
 $. Target values,  at $M_X$ of the fermion yukawa
 couplings and mixing parameters, together with the estimated uncertainties, achieved values and pulls.
  See caption to Table 3 for explanation.}}
 \label{ III-3-b} \end{table}
 \begin{table}
 $$
 \begin{array}{|c|c|c|c|}
 \hline &&&\\ {\rm  Parameter }&SM(M_Z) & m^{GUT}(M_Z) & m^{MSSM}=(m+\Delta m)^{GUT}(M_Z) \\
 \hline
    m_d/10^{-3}&   2.90000&   0.62377&   2.90229\\
    m_s/10^{-3}&  55.00000&  11.80718&  54.95706\\
            m_b&   2.90000&   3.08261&   2.90002\\
    m_e/10^{-3}&   0.48657&   0.47238&   0.48647\\
         m_\mu &   0.10272&   0.09671&   0.10272\\
         m_\tau&   1.74624&   1.74579&   1.74575\\
    m_u/10^{-3}&   1.27000&   1.10616&   1.26997\\
            m_c&   0.61900&   0.53916&   0.61901\\
            m_t& 172.50000& 149.41400& 172.44284\\
 \hline
 \end{array}
 $$

 \caption{\small{III-3-c: Values of standard model
 fermion masses in GeV at $M_Z$ compared with the masses obtained from
 values of GUT derived  yukawa couplings  run down from $M_X^0$ to
 $M_Z$  both before and after threshold corrections.
  Fit with $\chi_Z=\sqrt{ \sum_{i=1}^{9} (m_i^{MSSM}- m_i^{SM})^2/ (m_i^{MSSM})^2} =
0.0012$.}}
 \label{ III-3-c }\end{table}

 \begin{table}
 $$
 \begin{array}{|c|c|c|c|}
 \hline
 {\rm  Parameter}  &Value&  {\rm  Parameter}&Value \\
 \hline
                       M_{1}&             98.22&   M_{{\tilde {\bar {u}}_1}}&           4257.81\\
                       M_{2}&            273.31&   M_{{\tilde {\bar {u}}_2}}&           4256.95\\
                       M_{3}&            386.79&   M_{{\tilde {\bar {u}}_3}}&          15648.62\\
     M_{{\tilde {\bar l}_1}}&           1055.61&               A^{0(l)}_{11}&         -66907.20\\
     M_{{\tilde {\bar l}_2}}&            149.79&               A^{0(l)}_{22}&         -66822.57\\
     M_{{\tilde {\bar l}_3}}&          12367.95&               A^{0(l)}_{33}&         -39462.25\\
        M_{{\tilde {L}_{1}}}&           5350.98&               A^{0(u)}_{11}&         -79457.41\\
        M_{{\tilde {L}_{2}}}&           5300.00&               A^{0(u)}_{22}&         -79456.87\\
        M_{{\tilde {L}_{3}}}&          10339.21&               A^{0(u)}_{33}&         -39415.43\\
     M_{{\tilde {\bar d}_1}}&           2146.46&               A^{0(d)}_{11}&         -67055.77\\
     M_{{\tilde {\bar d}_2}}&           2145.58&               A^{0(d)}_{22}&         -67055.16\\
     M_{{\tilde {\bar d}_3}}&          25986.72&               A^{0(d)}_{33}&         -25007.87\\
          M_{{\tilde {Q}_1}}&           4117.71&                   \tan\beta&             51.50\\
          M_{{\tilde {Q}_2}}&           4117.04&                    \mu(M_Z)&          69115.80\\
          M_{{\tilde {Q}_3}}&          21502.97&                      B(M_Z)&
          7.5656 \times 10^{   8}
 \\
 M_{\bar {H}}^2&
         -4.7132 \times 10^{   9}
 &M_{H}^2&
         -5.2004 \times 10^{   9}
 \\
 \hline
 \end{array}
 $$
 \caption{ \small {III-3-d: Values (GeV) in  of the soft Susy parameters  at $M_Z$
 (evolved from the soft SUGRY-NUHM parameters at $M_X$).
  See caption to Table 5 for explanation. }}
  \label{ III-3-d}\end{table}

 \begin{table}
 $$
 %
 $$
 \caption{\small{III-3-e: Spectra of supersymmetric partners calculated ignoring generation mixing effects.
  See caption to Table 6 for explanation.
  }}\label{ III-3-e }\end{table}

 \begin{table}
 $$
 %
 $$
 \caption{\small{III-3-f : Spectra of supersymmetric partners calculated including  generation mixing effects.
 See caption to Table 7 for explanation. }}\label{ III-3-f }\end{table}
\clearpage

\begin{table}
 $$
 %
 $$
\caption{\small{III-4-a  : Column 1 contains values   of the NMSGUT-SUGRY-NUHM  parameters at $M_X$
  derived from an  accurate fit to all 18 fermion data and compatible with RG constraints.
  See caption to Table 2 for explanation.}} \label{ III-4-a} \end{table}
 \begin{table}
 $$
 %
 $$
  \caption{\small{III-4-b: Fit   with $\chi_X=\sqrt{ \sum_{i=1}^{17}
 (O_i-\bar O_i)^2/\delta_i^2}=    0.0469
 $. See caption to Table 3 for explanation. }}
\label{ III-4-b} \end{table}
 \begin{table}
 $$
 %
 $$

 \caption{\small{III-4-c: Values of standard model
 fermion masses in GeV at $M_Z$ compared with the masses obtained from
 values of GUT derived  yukawa couplings  run down from $M_X^0$ to
 $M_Z$  both before and after threshold corrections.
  Fit with $\chi_Z=\sqrt{ \sum_{i=1}^{9} (m_i^{MSSM}- m_i^{SM})^2/ (m_i^{MSSM})^2} =
0.0139$.}}
  \label{III-4-c }\end{table}

 \begin{table}
 $$
 %
 $$
  \caption{ \small {III-4-d: Values (GeV) in  of the soft Susy parameters  at $M_Z$
 (evolved from the soft SUGRY-NUHM parameters at $M_X$).
  See caption to Table 5 for explanation.}}
 \label{III-4-d }\end{table}

 \begin{table}
 $$
 %
 $$
 \caption{\small{III-4-e: Spectra of supersymmetric partners calculated ignoring generation mixing effects.
  See caption to Table 6 for explanation.
  }}\label{III-4-e}\end{table}

  \begin{table}
 $$
 %
 $$
 \caption{\small{Spectra of supersymmetric partners calculated including  generation mixing effects.
 See caption to Table 7 for explanation.}}\label{III-4-f}\end{table}

\end{document}